\documentclass[aps,prd,twocolumn,showpacs,floats,nofootinbib,longbibliography,10pt]{revtex4-1}

\usepackage{graphicx}
\usepackage{amsmath,amsfonts}
\usepackage{dcolumn}
\usepackage{hyperref}
\usepackage{dcolumn}

\newcolumntype{.}{D{.}{.}{-1}}
\newcolumntype{d}[1]{D{.}{.}{#1}}

\newcommand{\swY}[4][]{{}_{{}_{#2}}\!Y^{#1}_{#3}(#4)}
\newcommand{\swSH}[5][]{{}_{{}_{#2}}S^{#1}_{#3}(#4;#5)}

\newcommand{\YSH}[3][]{\mathcal{A}^{#1}_{#2}(#3)}
\newcommand{\YSHn}[2][]{\mathcal{A}^{#1}_{#2}}

\begin{document}

\title{Aspects of multimode {K}err ringdown fitting}

\author{Gregory B. Cook}\email{cookgb@wfu.edu}
\affiliation{Department of Physics, Wake Forest University,
		 Winston-Salem, North Carolina 27109}

\date{\today}

\begin{abstract}
A black hole that is ringing down to quiescence emits gravitational radiation of a very specific nature that can inform us of its mass and angular momentum, test the no-hair theorem for black holes, and perhaps even give us additional information about its progenitor system.  This paper provides a detailed description of, and investigation into the behavior of, multimode fitting of the ringdown signal provided by numerical simulations.  We find that there are at least three well-motivated multimode fitting schemes that can be used.  These methods are tested against a specific numerical simulation to allow for comparison to prior work.
\end{abstract}

\maketitle

\section{Introduction}
\label{sec:introduction}

Gravitational wave ringdown signals will be a generic feature of any dynamical scenario that ends in an isolated compact object.  The experimental study of such ringdown signals has already begun\cite{LIGO-grtests-2016,Isi-etal-2019,carullo-pozzo-veitch-2019}, and while this is certainly the most exciting new area of study, the study of ringdown signals in numerical simulations remains an important area of research\cite{babak-taracchini-buonanno-2017,bhagwat-etal-2018,giesler-etal-2019,huges-etal-2019a}.  In this paper, we will explore the multimode fitting of numerically simulated Kerr ringdown signals.

The term multimode can have different meanings.  First, any ringdown signal can be fit using some linear combination of fitting functions.  For example, these could be the quasinormal modes (QNMs) of a Kerr black hole\cite{BCP-2007,londonShoemakerhealy-2014,kamaretsos-etal-2012a,giesler-etal-2019}.  For numerical simulations, the gravitational-wave signal is typically decomposed as a time series in terms of spin-weighted spherical harmonics.  These modes can be analyzed individually, but a given gravitational QNM will contribute to multiple modes from a numerical simulation.  Fitting each simulation mode separately to one or more QNMs can result in multiple estimates for the amplitude of a given QNM's contribution to the gravitational wave signal\cite{londonShoemakerhealy-2014}.  Thus, another notion of multimode fitting is to fit multiple simulation modes simultaneously to the same set of QNMs\cite{huges-etal-2019b}.  This is the notion of multimode fitting we will explore in this paper.

The multimode fitting approach used in this paper is fundamentally based on linear least-squares fitting.  If we know the mass and angular momentum of a black hole, then a unique, but infinite, set of QNMs exist for that black hole, and we can choose a linear combination of them as a fitting function.  On the other hand, if we do not know the mass and angular momentum of the black hole, then determining these parameters by fitting makes the problem nonlinear.  A useful approach to handle both aspects of the problem is through the overlap integral between the gravitational waveform and the fitting function.  Extremizing the overlap yields a generalized eigenvalue problem\cite{Zimmerman-Chen-2011,Thesis:Zalutskiy}.  For fixed values of the mass and angular momentum of the black hole, the unique nonvanishing eigenvalue is the maximum overlap value, and the associated eigenvector yields the linear least-squares solution to the linear fitting problem.  The nonlinear problem of extracting the mass and angular momentum from the gravitational wave signal becomes a three-dimensional search where we maximize the value of the overlap.

Recently, Geisler {\em et al.}\cite{giesler-etal-2019} demonstrated the importance of QNM overtones up to $n=7$ in fitting ringdown signals accurately to a time early in the ringdown waveform.  They used a restricted fitting model that did not allow for multimode fitting.  As part of the work reported in this paper, we extend their exploration to full multimode fitting.  In doing so, we find that there are subtleties associated with computing the overlap that must be considered in order to fairly compare the different approaches.  These considerations lead to two closely related, but distinct, linear fitting approaches and three different ways to compute the overlap.  The different approaches converge when the fitting function and simulated waveform are fully compatible.

When multimode fitting is used, many of the QNM expansion coefficients can be fit with good consistency over a large range of fitting start times.  When used with simulation waveforms, multimode fitting can yield QNM expansion coefficients that will be useful in determining any general relationships between the expansion coefficients and the progenitor system such as those explored in Refs.~\cite{kamaretsos-etal-2012a,kamaretsos-et-al-2012b}.  They may also be useful in refining surrogate models for simulation waveforms\cite{babak-taracchini-buonanno-2017,varma-etal-2019}, and perhaps in uncovering systematic effects in simulation waveforms\cite{baibhav-etal-2018}.

The outline of this paper is as follows.  Section~\ref{sec:methods} presents a detailed description of the methods used in this paper, including relevant conventions and definitions.  Section~\ref{sec:sxsbbh0305} presents the results of numerous fits of the $m=2$ modes from the simulation waveform denoted SXS:BBH:0305 in the Simulating eXtreme Spacetimes catalog~\cite{SXS-waveforms,SXS-catalog-2013}.  We begin with a direct comparison of the overlap with the $\ell=2$, $m=2$ results in Geisler {\em et al.}\cite{giesler-etal-2019}.  We then explore the inclusion of $\ell=3$ and $4$ QNMs.  Finally, we explore full multimode fitting by simultaneously fitting the $m=2$, $\ell=2$, $3$, and $4$ simulation modes with a large set of QNMs.  All of these fits are performed using fixed values for the mass and angular momentum of the black hole.  Next, we explore the behavior of multimode fitting when used to also determine the mass and angular momentum.  Finally, we explore how the QNM fitting coefficients behave under various multimode fitting situations.  Section~\ref{sec:discussion} presents a discussion of the results described in the previous section.

\section{Methods}
\label{sec:methods}

\subsection{Conventions and definitions}
\label{sec:conventions}

The two independent polarizations of a gravitational waveform, usually referred to as $h_+$ and $h_\times$, can be conveniently represented by a complex scalar $h$ called the gravitational wave strain.  If the gravitational wave is propagating outward in the radial direction, then, at large distances, a standard convention is to define
\begin{subequations}
\begin{align}
	h_+ &= \frac12\left(h_{\hat\theta\hat\theta}-h_{\hat\phi\hat\phi}\right), \\
	h_\times &= h_{\hat\theta\hat\phi},
\end{align}
\end{subequations}
where $h_{\mu\nu}$ is the metric perturbation defined as the deviation of the metric from Minkowski space ($g_{\mu\nu}=\eta_{\mu\nu}+h_{\mu\nu}$).  $\hat\theta=\frac1r\partial_\theta$ and $\hat\phi=\frac1{r\sin\theta}\partial_\phi$ denote the usual orthonormal basis vectors on the surface of a sphere of radius $r$.  If we define the complex null tetrad
\begin{subequations}
\begin{align}
	k^\mu &= \frac1{\sqrt2}(\hat{t}^\mu + \hat{r}^\mu), \\
	\ell^\mu &= \frac1{\sqrt2}(\hat{t}^\mu - \hat{r}^\mu), \\
	\label{eqn:complexnulldiad}
	m^\mu &= \frac1{\sqrt2}(\hat\theta^\mu + i\hat\phi^\mu),
\end{align}
\end{subequations}
where $\hat{t}$ and $\hat{r}$ are the timelike and radial orthonormal basis vectors, then the gravitational strain is given by
\begin{equation}\label{eqn:strain}
	h = h_{\mu\nu}\bar{m}^\mu\bar{m}^\nu = h_+ - i h_\times,
\end{equation}
where $\bar{m}^\mu$ denotes the complex conjugate of $m^\mu$ (often also denoted with an asterisk ${}^*$).  In addition to the gravitational strain $h$, gravitational wave information is often presented in terms of the Newman-Penrose scalar
\begin{equation}\label{eqn:Psi4}
\Psi_4 = C_{\alpha\beta\gamma\delta}\ell^\alpha\bar{m}^\beta\ell^\gamma\bar{m}^\delta = -\ddot{h}.
\end{equation}
Here, $C_{\alpha\beta\gamma\delta}$ is the Weyl tensor, and $\dot{\ }$ denotes a derivative with respect to retarded time.  Note that there are differing conventions in defining the Weyl tensor and the Weyl scalars, and in the definition of $m^\mu$.  In this work, we will follow the conventions of Ref.~\cite{SXS-catalog-2019}.  Finally, the gravitational wave information can be presented in terms of the Bondi news function\cite{bondi-1962} $\mathcal{N}=\dot{h}$.   This approach is less widely used at this time and we will not present results based on $\mathcal{N}$, but as the fundamental quantity extracted by Cauchy-characteristic extraction\cite{bishop_etal96b,Handmer-etal-2016} methods, it may ultimately serve as the preferred means of representing the gravitational-wave information.

Notice that $h$, $\mathcal{N}$, and $\Psi_4$ are complex scalars with spin-weight $-2$.  Numerical relativity simulations use many techniques to extract gravitational wave information, but typically store and provide the data in terms of spin-weight $-2$ spherical harmonic modes.  We will assume that the modes are in a dimensionless form
\begin{equation}\label{eqn:psiNR}
\psi_{\rm NR} = \sum_{\ell{m}}{C_{\ell{m}}(t)\,\swY{-2}{\ell{m}}{\theta,\phi}},
\end{equation}
where $\psi_{\rm NR}= rh/M$, $r\mathcal{N}$, or $rM\Psi_4$ and $M$ is some mass parameter associated with the simulation.  Here, $\theta$ and $\phi$ are spherical coordinates associated with the numerical relativity code's extraction coordinate system, and $t$ is a retarded time.

In this paper, we are interested in fitting the ringdown waveform in terms of the gravitational quasinormal modes of the Kerr geometry for a black hole of mass $M_f$ and angular momentum $J_f$.  The mode frequencies, which are functions of the angular momentum parameter $a=J_f/M_f$, consist of two families of modes, $\omega^+_{\ell{m}n}$ and  $\omega^-_{\ell{m}n}$ related by
\begin{equation}\label{eqn:QNMfamilies}
\omega^+_{\ell{m}n} = -\omega^{-*}_{\ell{(-m)}n} \equiv \omega_{\ell{m}n}.
\end{equation}
Because of this relationship between the two families of QNMs, QNM data are typically only stored for the $\omega^+_{\ell{m}n}$ modes.  When we wish to highlight the use of a specific family of modes, we will use the notation $\omega^\pm_{\ell{m}n}$, but when we are writing expressions in a form useful for computation, we will use the form omitting the $\pm$ superscript in which case the expression has been transformed to use only the $+$ family of modes.  Whether we are considering data extracted in the form of $h$, $\mathcal{N}$, or $\Psi_4$, we can express the ringdown gravitational-wave signal in terms of QNMs as\cite{berticardosowill-2006}
\begin{eqnarray}\label{eqn:QNMexpansion1}
\psi &=& \sum_{\ell{m}n}\Bigl\{C^+_{\ell{m}n}e^{-i\omega^+_{\ell{m}n} (t-r^*)}
	\swSH{-2}{\ell{m}}{\theta^\prime,\phi^\prime}{a\omega^+_{\ell{m}n}} \\
	&&\mbox{}\qquad
	+ C^-_{\ell{m}n}e^{-i\omega^-_{\ell{m}n} (t-r^*)}
	\swSH{-2}{\ell{m}}{\theta^\prime,\phi^\prime}{a\omega^-_{\ell{m}n}}
	\Bigr\},\nonumber
\end{eqnarray}
where again $\psi$ is either $rh/M$, $r\mathcal{N}$, or $rM\Psi_4$.  The angular behavior is expressed in terms of the spin-weight $-2$ spheroidal harmonics $\swSH{-2}{\ell{m}}{\theta^\prime,\phi^\prime}{c}$.  In this case, $\theta^\prime$ and $\phi^\prime$ are angular spheroidal coordinates associated with the remnant Kerr black hole, where the $z^\prime$ axis is aligned with the spin-axis of the black hole, and the spheroidal parameter $c=a\omega^\pm_{\ell{m}n}$.  In the exponentials, $t-r_*$ is the retarded time expressed in terms of the tortoise-coordinate $r^*$.

Rewriting Eq.~(\ref{eqn:QNMexpansion1}) in a form more suitable for use with just the $+$ family of QNM modes, we find
\begin{subequations}
\begin{eqnarray}
\psi &=& \sum_{\ell{m}n}\Bigl\{C^+_{\ell{m}n}e^{-i\omega_{\ell{m}n}t}
	\swSH{-2}{\ell{m}}{\theta^\prime,\phi^\prime}{a\omega_{\ell{m}n}} \\
	&&\mbox{}\qquad
	+ C^-_{\ell{m}n}e^{i\omega^*_{\ell{(-m)}n}t}
	\swSH{-2}{\ell{m}}{\theta^\prime,\phi^\prime}{-a\omega^*_{\ell{(-m)}n}}
	\Bigr\},\nonumber \\ \label{eqn:QNMexpansion2}
	&=& \sum_{\ell{m}n}\Bigl\{C^+_{\ell{m}n}e^{-i\omega_{\ell{m}n}t}
	\swSH{-2}{\ell{m}}{\theta^\prime,\phi^\prime}{a\omega_{\ell{m}n}} \\
	&&\mbox{}
	+ (-1)^\ell C^-_{\ell{(-m)}n}e^{i\omega^*_{\ell{m}n}t}
	\swSH[*]{-2}{\ell{m}}{\pi-\theta^\prime,\phi^\prime}{a\omega_{\ell{m}n}}
	\Bigr\},\nonumber
\end{eqnarray}
\end{subequations}
where we now redefine $t$ as the retarded time to simplify the expressions.

The spin-weighted spheroidal harmonics can be expanded in terms of the spin-weighted spherical harmonics\cite{teukolsky-1973,cook-zalutskiy-2014}
\begin{equation} \label{eqn:swSH:expan}
\swSH{-2}{\ell{m}}{\theta^\prime,\phi^\prime}{c} = \sum_{\acute\ell} 
	\YSH{\acute\ell\ell{m}}{c} \swY{-2}{\acute\ell{m}}{\theta^\prime,\phi^\prime}.
\end{equation}
The spheroidal-harmonic expansion coefficients $\YSH{\acute\ell\ell{m}}{c}$ are also functions of the angular momentum parameter $a$ through the spheroidal parameter $c$.  The second spheroidal-harmonic function in Eq.~(\ref{eqn:QNMexpansion2}) can also be expressed in terms of the spherical harmonics, using various symmetry properties\cite{cook-zalutskiy-2014}, as
\begin{eqnarray} \label{eqn:swSconj:expan}
\swSH[*]{-2}{\ell{m}}{\pi-\theta^\prime,\phi^\prime}{c} &&\; = \\
&&\mbox\!\!\!\! \sum_{\acute\ell} (-1)^{\acute\ell}
	\YSH[*]{\acute\ell\ell{m}}{c} \swY{-2}{\acute\ell(-m)}{\theta^\prime,\phi^\prime}. \nonumber
\end{eqnarray}

\begin{figure}
\begin{picture}(180,220)(0,0)
\put(0,0){\includegraphics[width=2.5in,clip]{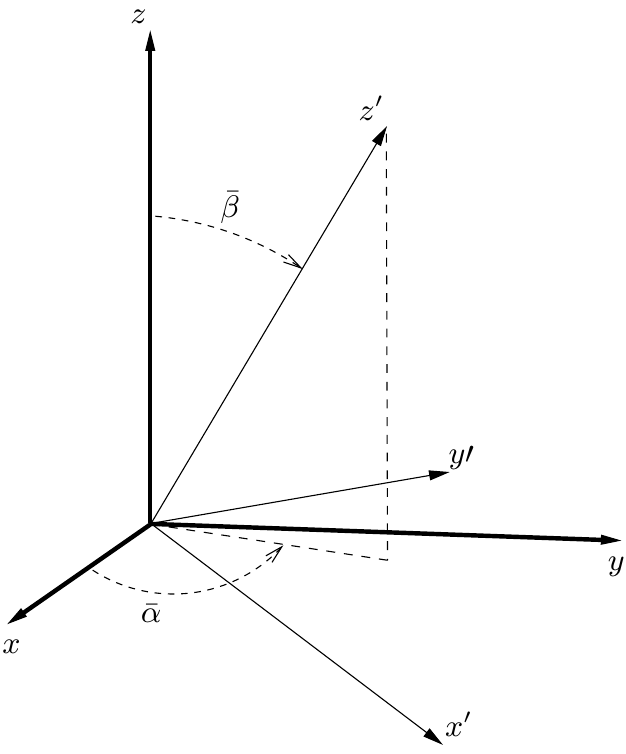}}
% \put(0,27){$x$}
% \put(175,51){$y$}
% \put(37,209){$z$}
% \put(128,3){${x^\prime}$}
% \put(129,82){${y\prime}$}
% \put(103,181){${z^\prime}$}
% \put(40,36){$\bar\alpha$}
% \put(63,153){$\bar\beta$}
\end{picture}
\caption{\label{fig:coordrot} The unprimed $\{x,y,z\}$ coordinate frame corresponds to the numerical relativity frame in which the data for $\psi_{\rm NR}$ are extracted.  The primed $\{x^\prime,y^\prime,z^\prime\}$ coordinate frame has its $z$-axis aligned with the angular momentum vector of the black hole.  $\bar\beta$ represents the angle of the black hole's angular momentum vector relative to the numerical-relativity $z$-axis.}
\end{figure}

If we equate the numerical relativity waveform $\psi_{\rm NR}$ from Eq.~(\ref{eqn:psiNR}) with the ringdown expansion $\psi$ from Eq.~(\ref{eqn:QNMexpansion1}), and if we assume that the waveform extraction coordinates coincide with the QNM coordinates aligned with the black hole's spin, then we can express the wave-form expansion coefficients $C_{\ell{m}}$ in terms of the QNM expansion coefficients $C^\pm_{\ell{m}n}$
\begin{eqnarray}
C_{\acute\ell{m}} &=& \sum_{\ell{n}}\Bigl\{C^+_{\ell{m}n}e^{-i\omega_{\ell{m}n}t}
	\YSH{\acute\ell\ell{m}}{a\omega_{\ell{m}n}} \\
	&&\mbox{}
	+ (-1)^{\ell+\acute\ell} C^-_{\ell{m}n}e^{i\omega^*_{\ell{(-m)}n}t}
	\YSH[*]{\acute\ell\ell{(-m)}}{a\omega_{\ell{(-m)}n}}
	\Bigr\}.\nonumber
\end{eqnarray}
Of course, the two coordinate systems will not necessarily be aligned.  They will, in general, be related by a Lorentz transformation.  When the transformation includes a boost, transforming between the two frames can be quite complicated\cite{boyle-2016,Thesis:Zalutskiy}.  However, when the coordinate systems simply differ by a rotation as illustrated in Fig.~\ref{fig:coordrot}, the transformation is straightforward.  In general, we find
\begin{eqnarray}\label{eqn:Clm_expansion}
C_{\acute\ell\acute{m}} &=& \sum_{\ell{m}n}
	D^{\acute\ell}_{\acute{m}{m}}(\bar\alpha,\bar\beta,0)
	\Bigl\{C^+_{\ell{m}n}e^{-i\omega_{\ell{m}n}t}
	\YSHn{\acute\ell\ell{m}n} \\
	&&\mbox{}\qquad
	+ (-1)^{\ell+\acute\ell} C^-_{\ell{m}n}e^{i\omega^*_{\ell{(-m)}n}t}
	\YSHn[*]{\acute\ell\ell{(-m)}n}
	\Bigr\},\nonumber
\end{eqnarray}
where we have simplified the notation by defining
\begin{equation}\label{eqn:swSH:simp}
\YSH{\acute\ell\ell{m}}{a\omega_{\ell{m}n}} \equiv
\YSHn{\acute\ell\ell{m}n},
\end{equation}
and  $D^\ell_{\acute{m}{m}}(\bar\alpha,\bar\beta,\bar\gamma)$ is the Wigner rotation matrix.  We follow the conventions as defined in Ref.~\cite{cook-zalutskiy-2014}.
% \begin{equation}
%   D^\ell_{mn}(\alpha,\beta,\gamma) \equiv 
%        e^{-im\alpha} d^\ell_{mn}(\beta) e^{-in\gamma}.
% \end{equation}

Equation~(\ref{eqn:Clm_expansion}) can be used as the foundation for fitting for the QNM expansion coefficients $C^\pm_{\ell{m}n}$ and also for the remnant mass $M_f$, the dimensionless angular momentum parameter $\chi_f=a/M_f$, and the inclination angle $\bar\beta$.  Note that we cannot fit for the azimuthal angle $\bar\alpha$ because this parameter simply corresponds to a constant phase change for each of the expansion coefficients $C^\pm_{\ell{m}n}$.  If we assume that the basic remnant parameters $M_f$, $\chi_f$, and $\bar\beta$ are provided by the numerical relativity simulation, then the expansion coefficients $C^\pm_{\ell{m}n}$ can be determined by linear-least-squares fitting\cite{MSThesis:Zalutskiy,Thesis:Zalutskiy}.  Alternatively, any subset of the basic remnant parameters can also be determined by some form of nonlinear fitting\cite{BCP-2007,MSThesis:Zalutskiy,Thesis:Zalutskiy}

In order to perform a fit of the ringdown portion of simulated gravitational-wave data, we must fix some choices.  First, we must choose what simulation data will be fit.  In Ref.~\cite{BCP-2007}, the real and imaginary parts of individual $C_{\ell{m}}(t)$ modes were fit separately, but it is straightforward to fit the full complex signal for individual modes.  However, because QNM modes are defined with respect to the spin-weighted spheroidal harmonics, many different $C_{\ell{m}}$ mode amplitudes depend on the same $C^\pm_{\ell{m}n}$ mode amplitudes.  Because of this, it may be useful to fit a set of $C_{\ell{m}}$ mode amplitudes against a chosen set of QNMs.

Introducing some new notation to simplify the expressions, we will denote the numerical-relativity waveform (\ref{eqn:psiNR}) to which we are fitting as
\begin{equation}\label{eqn:psi_NR_lm}
\psi_{\rm NR} = \sum_{\{\ell{m}\}\in\{{\rm NR}\}}{C_{\ell{m}}|\ell{m}\rangle},
\end{equation}
where $\{{\rm NR}\}$ denotes the chosen set of numerical relativity modes against which we are fitting.  For a ringdown signal, our fitting function is given by Eq.~(\ref{eqn:QNMexpansion1}) which we now write as
\begin{equation}\label{eqn:psi_fit_k}
\psi_{\rm fit} = \sum_{k\in\{{\rm QNM}\}}{C_k\psi_k}.
\end{equation}
where
\begin{align}
\psi_k &= e^{-i\omega_kt}\swSH{-2}{\ell{m}}{\theta,\phi}{a\omega_k}, \\
&\equiv e^{-i\omega_kt}|k\rangle.
\end{align}
Here, $k$ denotes a general QNM triplet ${\ell{m}n}$ which can represent modes from either of the $\pm$ families, and $\{{\rm QNM}\}$ denotes the full set of QNMs being used in the fitting function.  When it is important to distinguish between the $\pm$ families, we will use the notation $|k\pm\rangle$.  So, Eq.~(\ref{eqn:swSH:expan}) can be expressed for the two families as
\begin{eqnarray}
|k+\rangle = |\ell{m}n+\rangle 
	&=& \sum_{\acute\ell}{\YSHn{\acute\ell\ell{m}n}|\acute\ell{m}\rangle} ,\\
|k-\rangle = |\ell{m}n-\rangle 
	&=& \sum_{\acute\ell}{(-1)^{\acute\ell+\ell}\YSHn[*]{\acute\ell\ell{(-m)}n}|\acute\ell{m}\rangle}.
\end{eqnarray}

\subsection{Waveform fitting}
\label{sec:waveformfitting}
To handle both the linear and nonlinear aspects of multimode ringdown fitting, it is useful to express the quality of the fit in terms of the overlap $\rho$ between the ringdown waveform $\psi_{\rm NR}$ and the fitting function $\psi_{\rm fit}$\cite{Zimmerman-Chen-2011},
\begin{equation}\label{eqn:rho2}
\rho^2 = \frac{\left|\langle\psi_{\rm fit}|\psi_{\rm NR}\rangle\right|^2}
	{\langle\psi_{\rm NR}|\psi_{\rm NR}\rangle
	\langle\psi_{\rm fit}|\psi_{\rm fit}\rangle},
\end{equation}
where the inner product between any two complex functions is defined as
\begin{equation}\label{eqn:innerproduct}
\langle\psi_1|\psi_2\rangle \equiv
\int_{t_i}^{t_e}{dt\oint{d\Omega\psi^*_1(t,\Omega)\psi_2(t,\Omega)}}.
\end{equation}
Below, we adapt and expand upon the approach outlined in Sec.~IV.A of Ref.~\cite{Zimmerman-Chen-2011} and Sec.~5.6 of Ref.~\cite{Thesis:Zalutskiy}.  In terms of our fitting function in Eq.~(\ref{eqn:psi_fit_k}) we find
\begin{equation}\label{eqn:rho2explicit}
\rho^2 = \frac{\left|\sum_k{C^*_kA_k}\right|^2}
	{\langle\psi_{\rm NR}|\psi_{\rm NR}\rangle\sum_{i,j}{C^*_iB_{ij}C_j}},
\end{equation}
where
\begin{eqnarray}
A_k &\equiv& \langle\psi_k|\psi_{\rm NR}\rangle, \\
B_{ij} &\equiv& \langle\psi_i|\psi_j\rangle.
\end{eqnarray}

If we extremize $\rho^2$ with respect to $C^*_k$, we find
\begin{equation}
\frac1{\langle\psi_{\rm NR}|\psi_{\rm NR}\rangle}\sum_k{A_iA^*_kC_k}
	= \rho^2\sum_k{B_{ik}C_k},
\end{equation}
which is a generalized eigenvalue problem that we can write in standard matrix notation as
\begin{equation}\label{eqn:generalized_EVP}
\frac1{\langle\psi_{\rm NR}|\psi_{\rm NR}\rangle}
	\vec{A}\otimes\vec{A}^\dag\cdot\vec{C}_n
	= \rho^2_n{\mathbb B}\cdot\vec{C}_n.
\end{equation}

Both of the matrices, $\vec{A}\otimes\vec{A}^\dag$ and ${\mathbb B}$, are clearly Hermitian.  Therefore, the eigenvalues $\rho_n^2$ will be real, and the eigenvectors $\vec{C}_n$ will be (or can be made) orthogonal with respect to ${\mathbb B}$.  Let us assume that $\rho^2_1\ne0$, then for $n\ne1$ we have
\begin{equation}\label{eqn:orthog_cond}
\frac1{\langle\psi_{\rm NR}|\psi_{\rm NR}\rangle}
	\vec{C}^\dag_n\cdot\vec{A}\otimes\vec{A}^\dag\cdot\vec{C}_1
	= \rho^2_1\vec{C}^\dag_n\cdot{\mathbb B}\cdot\vec{C}_1 = 0.
\end{equation}
Clearly, from Eq.~(\ref{eqn:generalized_EVP}), $\vec{A}^\dag\cdot\vec{C}_1\ne0$ so long as $\rho^2_1\ne0$.  But, because the left-hand side of Eq.~(\ref{eqn:orthog_cond}) must vanish for $n\ne1$, it must be true that $\vec{A}^\dag\cdot\vec{C}_n=0$.  Now, dotting Eq.~(\ref{eqn:generalized_EVP}) from the left with $\vec{C}^\dag_n$ and noting that $\vec{C}^\dag_n\cdot{\mathbb B}\cdot\vec{C}_n\ne0$, we see that $\rho^2_n=0$ for $n\ne1$.  Thus, there is only one nonvanishing eigenvalue, $\rho^2_1\equiv\rho^2_{\rm max}$.  It follows immediately that
\begin{align}\label{eqn:rho2max}
\rho^2_{\rm max} &= \frac1{\langle\psi_{\rm NR}|\psi_{\rm NR}\rangle}
	{\rm Tr}({\mathbb B}^{-1}\cdot\vec{A}\otimes\vec{A}^\dag), \nonumber \\
	&= \frac1{\langle\psi_{\rm NR}|\psi_{\rm NR}\rangle}
	\vec{A}^\dag\cdot{\mathbb B}^{-1}\cdot\vec{A}.
\end{align}
The associated eigenvector is easily seen to be proportional to
\begin{equation}\label{eqn:least-squares_coefs}
\vec{C} = {\mathbb B}^{-1}\cdot\vec{A}.
\end{equation}
In fact, Eq.~(\ref{eqn:least-squares_coefs}) is the correctly normalized linear least-squares solution (assuming fixed values of the remnant parameters $M_f$, $\chi_f$, and $\bar\beta$) as can be easily verified from
\begin{equation}\label{eqn:chi2full}
	\chi^2_{{}_{ev}} = \Bigl\langle\psi_{\rm NR}-\sum_jC_j\psi_j\Big|
	\psi_{\rm NR}-\sum_kC_k\psi_k\Bigr\rangle
\end{equation}
by extremizing $\chi^2_{{}_{ev}}$ with respect to $\vec{C}^\dag$.

At this point, it is important to make explicit a few subtleties of linear least-squares fitting of the ringdown.  A straightforward implementation of linear least-squares fitting of Eq.~(\ref{eqn:psiNR}) for the
complex coefficients $C_{\ell{m}}$ would be based upon minimization of 
\begin{equation}\label{eqn:chi2ls}
	\chi^2_{{}_{ls}} = \sum_{t,\{\rm NR\}}\Bigl|\psi_{\rm NR}-\sum_kC_k\tilde\psi_k\Bigr|^2.
\end{equation}
The sum is over all values of $t$ over which the fit is being performed, and for each simulation mode in $\{\rm NR\}$.  More importantly, $\tilde\psi_i$ is the fitting basis function projected onto the subspace of spin-weighted spherical harmonic modes covered by $\psi_{\rm NR}$.  That is,
\begin{equation}\label{eqn:proj_psi_k}
	\tilde\psi_k = \!\!\!\!\sum_{\{\ell{m}\}\in\{{\rm NR\}}}\!\!\!\!|\ell{m}\rangle\langle\ell{m}|\psi_k.
\end{equation}

Let us consider the explicit construction of the least-squares solution for Eq.~(\ref{eqn:chi2ls}).  The numerical data represented by $\psi_{\rm NR}$ consist of a set of complex coefficients $C_{\ell{m}}(t)$ obtained at a large number $N_t$ of retarded times.  Assuming that the set of waveform modes $\{{\rm NR}\}$ contains $N_{\rm NR}$ $(\ell,m)$ modes, then we are fitting to $M=N_t\times N_{\rm NR}$ complex data points.  And, if the set of QNMs $\{{\rm QNM}\}$ contains $N_{\rm QNM}$ modes, then we are fitting for $N_{\rm QNM}$ complex coefficients $C_k$.  Each of the $M$ linear equations is represented as
\begin{equation}\label{eqn:leastsquareseqn}
\sum_{k\in\{{\rm QNM}\}}{e^{-i\omega_kt}\langle\ell{m}|k\rangle C_k}=C_{\ell{m}}(t),
\end{equation}
which can be expressed in matrix notation as
\begin{equation}\label{eqn:lsmatrixnotedm}
{\mathbb L}\cdot\vec{C}=\vec{R}.
\end{equation}
Here, ${\mathbb L}$ is a $M\times N_{\rm QNM}$ matrix usually referred to\cite{numrec_c++} as the design matrix of the least-squares fitting problem.

Explicit forms for the components of ${\mathbb L}$ can be derived if we first note that
the inner products of the spin-weighted spheroidal and spherical harmonics, including the rotation between their frames, can be written as 
\begin{eqnarray}\label{eqn:sphericalspheroida+}
\langle\acute\ell\acute{m}|\ell{m}n+\rangle &=&
	\YSHn{\acute\ell\ell{m}n}
	D^{\acute\ell}_{\acute{m}{m}}(\bar\alpha,\bar\beta,0), \\
\label{eqn:sphericalspheroida-}
\langle\acute\ell\acute{m}|\ell{m}n-\rangle &=&
	(-1)^{\acute\ell+\ell}\!\YSHn[*]{\acute\ell\ell{(-m)}n}
	D^{\acute\ell}_{\acute{m}{m}}(\bar\alpha,\bar\beta,0).
\end{eqnarray}
These components can also be read off from Eq.~(\ref{eqn:Clm_expansion}).

This same system can be solved in terms of what are usually referred to\cite{numrec_c++} as the normal equations.  In this case, the system that is solved is
\begin{equation}\label{eqn:lsmatrixnotenormal}
\left({\mathbb L}^\dagger\cdot{\mathbb L}\right)\cdot\vec{C} = {\mathbb L}^\dagger\cdot\vec{R}.
\end{equation}
Equation~(\ref{eqn:lsmatrixnotenormal}) looks very much like the equation that gives rise to the eigenvector of Eq.~(\ref{eqn:least-squares_coefs}) that is equivalent to the least-squares solution of Eq.~(\ref{eqn:chi2full}).  In fact, it is easy to see that the right-hand-side vector ${\mathbb L}^\dagger\cdot\vec{R}$ is essentially equal to $\vec{A}$.  However, as we will see below, the normal matrix ${\mathbb L}^\dagger\cdot{\mathbb L}$ is not, in general, the same as ${\mathbb B}$.

The components of $\vec{A}$ are given by
\begin{align}
\label{eqn:Acomp+}
\langle\psi_{\ell{m}n+}|\psi_{\rm NR}\rangle &= \\ \nonumber
	\int_{t_i}^{t_e}
	\!\!dt&\,e^{i\omega^*_{\ell{m}n}t}\!\!\!\!\!\!\sum_{\{\acute\ell\acute{m}\}\in\{{\rm NR}\}}
	{\!\!\!\!\!\!C_{\acute\ell\acute{m}}(t)\YSHn[*]{\acute\ell\ell{m}n}
	D^{\acute\ell*}_{\acute{m}{m}}(\bar\alpha,\bar\beta,0)}, \\
\label{eqn:Acomp-}
\langle\psi_{\ell{m}n-}|\psi_{\rm NR}\rangle &= \\ \nonumber
	\int_{t_i}^{t_e}
	\!\!dt&\,e^{-i\omega_{\ell{(-m)}n}t}\times \\\nonumber
	\sum_{\{\acute\ell\acute{m}\}\in\{{\rm NR}\}}&
	{\!\!\!\!\!\!(-1)^{\acute\ell+\ell}C_{\acute\ell\acute{m}}(t)
	\YSHn{\acute\ell\ell{(-m)}n}
	D^{\acute\ell*}_{\acute{m}{m}}(\bar\alpha,\bar\beta,0)}.
\end{align}
The dot-product in ${\mathbb L}^\dagger\cdot\vec{R}$ incorporates both the summation over $\{\acute\ell\acute{m}\}$ and the integration over $t$ in Eqs.~(\ref{eqn:Acomp+}) and (\ref{eqn:Acomp-}), so ${\mathbb L}^\dagger\cdot\vec{R}\simeq\vec{A}$ up to minor differences associated with the implementation of the numerical integral over $t$.

Now consider the components of ${\mathbb B}$.  These are given by
\begin{align}
\label{eqn:Bcomp++}
\langle\psi_{\ell{m}n+}|\psi_{\acute\ell\acute{m}\acute{n}+}\rangle &= \\\nonumber
	\delta_{m\acute{m}}&\int_{t_i}^{t_e}{
	\!\!dt\,e^{i(\omega^*_{\ell{m}n}-\omega_{\acute\ell\acute{m}\acute{n}})t}
	\sum_{\breve\ell}
	\YSHn[*]{\breve\ell\ell{m}n}
	\YSHn{\breve\ell\acute\ell\acute{m}\acute{n}}}, \\
\label{eqn:Bcomp+-}
\langle\psi_{\ell{m}n+}|\psi_{\acute\ell\acute{m}\acute{n}-}\rangle &= \\\nonumber
	\delta_{m\acute{m}}&\int_{t_i}^{t_e}{
	\!\!dt\,e^{i(\omega^*_{\ell{m}n}+\omega^*_{\acute\ell(-\acute{m})\acute{n}})t}\times}
	\\\nonumber & \qquad\qquad\sum_{\breve\ell}(-1)^{\breve\ell+\acute\ell}
	\YSHn[*]{\breve\ell\ell{m}n}
	\YSHn[*]{\breve\ell\acute\ell(-\acute{m})\acute{n}} , \\
\label{eqn:Bcomp--}
\langle\psi_{\ell{m}n-}|\psi_{\acute\ell\acute{m}\acute{n}-}\rangle &= \\\nonumber
	\delta_{m\acute{m}}&\int_{t_i}^{t_e}{
	\!\!dt\,e^{-i(\omega_{\ell(-m)n}-\omega^*_{\acute\ell(-\acute{m})\acute{n}})t}\times}
	\\\nonumber &\qquad\quad\sum_{\breve\ell}(-1)^{\ell+\acute\ell}
	\YSHn{\breve\ell\ell{(-m)}n}
	\YSHn[*]{\breve\ell\acute\ell(-\acute{m})\acute{n}}.
\end{align}
There are two differences between ${\mathbb L}^\dagger\cdot{\mathbb L}$ and ${\mathbb B}$.  The most obvious is connected to the fact that the Wigner rotation matrix $D^{\acute\ell}_{\acute{m}m}(\alpha,\beta,\gamma)$ does not appear in any of the components of ${\mathbb B}$.  The second difference is associated with the summation over $\breve\ell$.  If the set of simulations modes $\{{\rm RN}\}$ is large enough, then the Wigner rotation matrices can be eliminated based on their unitarity\cite{Rotations-in-QM-1987}
\begin{equation}\label{eqn:WignerDUnitarity}
\sum_{\breve{m}}{D^{\acute\ell*}_{\breve{m}m}(\alpha,\beta,\gamma)D^{\acute\ell}_{\breve{m}\acute{m}}(\alpha,\beta,\gamma)}=\delta_{m\acute{m}}.
\end{equation}
The sum over $\breve\ell$ comes from the expansion of the spin-weighted spheroidal harmonics in terms of the spin-weighted spherical harmonics in Eq.~(\ref{eqn:swSH:expan}).  The sum is formally over all allowed values of $\breve\ell$, but is practically only over those for which $\YSHn{\breve\ell\ell{m}n}$ is significant.  Again, if the set of simulation modes $\{{\rm RN}\}$ is large enough, then this sum will be effectively represented.  But this will not be the case in general.

Not surprisingly, the standard least-squares approach can be reformulated in terms of extremizing the overlap by using the projected fitting functions of Eq.~(\ref{eqn:proj_psi_k}).  That is, we define a new fitting function
\begin{equation}\label{eqn:proj_psi_fit_k}
	\tilde\psi_{\rm fit} = \!\!\!\!\sum_{\{\ell{m}\}\in\{{\rm NR\}}}\!\!\!\!|\ell{m}\rangle\langle\ell{m}|\psi_{\rm fit} = \sum_k{C_k\tilde\psi_k}.
\end{equation}
The new fitting function $\tilde\psi_{\rm fit}$ is simply the original fitting function projected into the subspace of spin-weighted spherical harmonic modes included in the numerical-relativity waveform (\ref{eqn:psi_NR_lm}) to which we are fitting.  The change in the fitting function directly affects only the components of the matrix $\mathbb{B}$\footnote{The components of $\vec{A}$ are unchanged since $\psi_{\rm fit}$ is already projected into the subspace of modes covered by $\psi_{\rm NR}$ by the inner product that defines $\vec{A}$.}.  These are now given by\footnote{Note that the arguments $(\alpha,\beta,\gamma)$ of the Wigner matrices have been omitted for brevity.  In all cases, these are given by $(\bar\alpha,\bar\beta,0)$.}
\begin{align}
\label{eqn:mlBcomp++}
\langle\psi_{\ell{m}n+}|\psi_{\acute\ell\acute{m}\acute{n}+}\rangle &= 
	\int_{t_i}^{t_e}{
	\!\!dt\,e^{i(\omega^*_{\ell{m}n}-\omega_{\acute\ell\acute{m}\acute{n}})t}} \times \\\nonumber
	&\!\!\!\!\!\sum_{\{\breve\ell\breve{m}\}\in\{{\rm NR}\}}\!\!\!\!\!
	\YSHn[*]{\breve\ell\ell{m}n}D^{\breve\ell*}_{\breve{m}{m}}
	\YSHn{\breve\ell\acute\ell\acute{m}\acute{n}}D^{\breve\ell}_{\breve{m}\acute{m}}, \\
\label{eqn:mlBcomp+-}
\langle\psi_{\ell{m}n+}|\psi_{\acute\ell\acute{m}\acute{n}-}\rangle &=
	\int_{t_i}^{t_e}{
	\!\!dt\,e^{i(\omega^*_{\ell{m}n}+\omega^*_{\acute\ell(-\acute{m})\acute{n}})t}\times}
	\\\nonumber & \!\!\!\!\!\sum_{\{\breve\ell\breve{m}\}\in\{{\rm NR}\}}\!\!\!\!\!
	(-1)^{\breve\ell+\acute\ell}
	\YSHn[*]{\breve\ell\ell{m}n}D^{\breve\ell*}_{\breve{m}{m}}
	\YSHn[*]{\breve\ell\acute\ell(-\acute{m})\acute{n}}D^{\breve\ell}_{\breve{m}\acute{m}} , \\
\label{eqn:mlBcomp--}
\langle\psi_{\ell{m}n-}|\psi_{\acute\ell\acute{m}\acute{n}-}\rangle &= 
	\int_{t_i}^{t_e}{
	\!\!dt\,e^{-i(\omega_{\ell(-m)n}-\omega^*_{\acute\ell(-\acute{m})\acute{n}})t}\times}
	\\\nonumber & \!\!\!\!\!\!\!\!\!\!\sum_{\{\breve\ell\breve{m}\}\in\{{\rm NR}\}}\!\!\!\!\!\!\!
	(-1)^{\ell+\acute\ell}
	\YSHn{\breve\ell\ell{(-m)}n}D^{\breve\ell*}_{\breve{m}{m}}
	\YSHn[*]{\breve\ell\acute\ell(-\acute{m})\acute{n}}D^{\breve\ell}_{\breve{m}\acute{m}}.
\end{align}
We will denote the matrix $\mathbb{B}$ computed using Eqs.~(\ref{eqn:mlBcomp++})--(\ref{eqn:mlBcomp--}) as $\mathbb{B}_{\rm ml}$ because it is computed from $\tilde\psi_{\rm fit}$, the ``mode-limited'' version of the fitting function.  We find that $\mathbb{L}^\dagger\cdot\mathbb{L}\simeq\mathbb{B}_{\rm ml}$ up to minor differences associated with the implementation of the numerical integral over $t$.

We see that, given fixed values for the remnant parameters $M_f$, $\chi_f$, and $\bar\beta$, there are two linear least-squares problems that can be used to determine the ringdown expansion coefficients $\vec{C}$.  The first version, based on $\psi_{\rm fit}$, is represented by Eqs.~(\ref{eqn:chi2full}), (\ref{eqn:least-squares_coefs}), and (\ref{eqn:rho2max}).  When we compute the ringdown expansion coefficients or the overlap using these equations, we will denote them as being determined by the ``Eigenvalue Method'' (EV).  The second version, based on $\tilde\psi_{\rm fit}$, is fundamentally represented by Eq.~(\ref{eqn:chi2ls}), but the expansion coefficients and overlap can be computed in slightly different ways.  We will be primarily interested in the approach that follows the EV method but where $\mathbb{B}\to\mathbb{B}_{\rm ml}$ in Eqs.~(\ref{eqn:least-squares_coefs}) and (\ref{eqn:rho2max}).  We will refer to this approach as the ``mode-limited Eigenvalue Method'' (mlEV).  We can also compute the ringdown expansion coefficients using either Eq.~(\ref{eqn:lsmatrixnotedm}) or (\ref{eqn:lsmatrixnotenormal}).  We will refer to this as the ``Least-Squares Method'' (LS).  We can also compute the overlap associated with a fit determined by the LS method, but there are two ways that it can be computed.  Both are expressed by Eq.~(\ref{eqn:rho2explicit}), but we can use either $\mathbb{B}$ or $\mathbb{B}_{\rm ml}$.  By default, we will assume that we are computing $\rho^2$ using the $\psi_{\rm fit}$ (ie. using $\mathbb{B}$).  However, we can also compute $\rho^2$ using $\tilde\psi_{\rm fit}$ (ie. using $\mathbb{B}_{\rm ml}$).  We will refer to the second approach as computing a ``mode-limited'' overlap.  Furthermore, we note that we can also compute a mode-limited overlap for the EV method.  Finally, we note that the mlEV and LS methods are essentially the same, differing only in how the time integration is accomplished.

\subsection{Implementation details}\label{sec:implementation}
Let us now consider the remaining details necessary to actually compute numerical
values for the overlap $\rho_{\rm max}$ and the QNM expansion coefficients $\vec{C}$.  First, the QNM data for $\omega_{\ell{m}n}(\chi_f)$ and $\YSH{\acute\ell\ell{m}n}{\chi_f}$ are those described in Ref.~\cite{cook-zalutskiy-2014}.  The data were computed with high numerical precision and should have an absolute accuracy of $10^{-11}$ or better.  These data sets are publicly accessible at \href{https://doi.org/10.5281/zenodo.2650358}{https://doi.org/10.5281/zenodo.2650358} \cite{QNMdata-cook-2019}.  The complex QNM frequencies are provided in dimensionless form as $\bar\omega_{\ell{m}n}(\chi_f)=M_f\omega_{\ell{m}n}(\chi_f)$.  The spheroidal harmonic expansion coefficients $\YSH{\acute\ell\ell{m}n}{\chi_f}$ are normalized so that
\begin{equation}
	\sum_{\acute\ell}{|\YSHn{\acute\ell\ell{m}n}|^2}=1,
\end{equation}
and the phase is chosen so that the coefficient with $\max_{\acute\ell}{|\YSHn{\acute\ell\ell{m}n}}|$ is real.

The numerical relativity waveforms are functions of retarded time $t$ that is usually given in dimensionless form $\bar{t}\equiv t/M$, where $M$ is some natural mass scale for the simulation.  The arguments of the exponentials in Eqs.~(\ref{eqn:QNMexpansion2}) and (\ref{eqn:Acomp+})--(\ref{eqn:Bcomp--}) are of the form,
\begin{align}
	i\omega_{\ell{m}n}(\chi_f)t &= i\left[M_f\omega_{\ell{m}n}(\chi_f)\right](t/M)(M/M_f) \nonumber \\
		&= i\bar\omega_{\ell{m}n}(\chi_f)\bar{t}/\delta, 
\end{align}
where $\delta\equiv M_f/M$ is the remnant mass ratio, defined in terms of the mass used in defining $\bar{t}$.

We define the remnant parameters as the set $\mathcal{R}=\{\delta,\chi_f,\bar\beta,\bar\alpha\}$.  With $\mathcal{R}$ fixed, we can compute $\rho_{\rm max}(\bar{t})$ and $\vec{C}(\bar{t})$.  To do this, we must first choose the set $\{\rm NR\}$ of numerical relativity modes $C_{\ell{m}}(\bar{t})$ that will constitute $\psi_{\rm NR}$, and the set $\{\rm QNM\}$ of QNMs that will constitute $\psi_{\rm fit}$.  Now, for the EV method, we can construct $\vec{A}$ and $\mathbb{B}$ by evaluating the integrals of Eqs.~(\ref{eqn:Acomp+})--(\ref{eqn:Bcomp--}) with $t_i=M\bar{t}_i$ and $t_e$ set to some desired time in the tail of the ringdown signal.  For the mlEV method, we can construct $\vec{A}$ and $\mathbb{B}_{\rm ml}$ by evaluating the integrals of Eqs.~(\ref{eqn:Acomp+}, (\ref{eqn:Acomp-}), and (\ref{eqn:mlBcomp++})--(\ref{eqn:mlBcomp--}) with $t_i=M\bar{t}_i$ and $t_e$ set as above.  Explicit forms for the components of $\vec{R}$ and $\mathbb{L}$ needed for the LS method are not given in the text, but can be easily inferred from Eqs.~(\ref{eqn:leastsquareseqn})--(\ref{eqn:sphericalspheroida-}).  The matrix inverses ${\mathbb B}^{-1}$, ${\mathbb B}_{\rm ml}^{-1}$, and $({\mathbb L}^\dag\cdot{\mathbb L})^{-1}$ are computed through Singular-Value Decomposition (SVD)\cite{numrec_c++} allowing us to set a tolerance on the size of the singular values associated with each matrix and effectively remove QNMs that are (at least numerically) irrelevant to the fit.  Finally, we note that
\begin{equation}\label{eqn:psiNRpsiNR}
\langle\psi_{\rm NR}|\psi_{\rm NR}\rangle = 
\int_{t_i}^{t_e}{dt\!\!\!\!\!\!\!\sum_{\{\ell{m}\}\in\{\rm NR\}}\!\!\!\!\!\!\left|C_{\ell{m}}(t)\right|^2}.
\end{equation}

\section{Exploration of SXS:BBH:0305}\label{sec:sxsbbh0305}

Recently, Giesler {\em et.~al.}~\cite{giesler-etal-2019}, hereafter referred to as GIST, studied the importance of overtones in fitting the ringdown portion of the gravitational wave signal.  The gravitational waveform that they investigated was from a simulation of a binary-black-hole system similar to that responsible for the first gravitational-wave observation, GW150914\cite{GW150914-2016}.  The numerical simulation, labeled SXS:BBH:0305, is from the Simulating eXtreme Spacetimes (SXS) catalog~\cite{SXS-waveforms,SXS-catalog-2013}.  Below, we will explore the same waveform.  More precisely, we will use the numerical results from level 6, with second-order extrapolation, and center-of-mass corrected.  Furthermore we will denote as a reference time, $t_{\rm peak}=3692.7198494029326M$ which is the interpolated peak of $|h^{\rm NR}_{22}|$, the $\ell=2$, $m=2$ spherical harmonic mode of the complex strain as computed in this numerical simulation.  The remnant parameters used for this model are $\mathcal{R}_{\rm NR}=\{0.952032939704,0.692085186818,0,0\}$, and were obtained directly from the numerical simulation.  This simulation was chosen to facilitate direct comparisons with the results in GIST.

In the remainder of this section, we will extensively explore the ringdown portion of this data set.  We will consider many different combinations of the set of numerical relativity modes to which fits will be made, sets of QNMs with which the fits will be made, and several different relationships between these modes.  Unless otherwise noted, the fitting will be based on the Eigenvalue Method.  For clarity within the text, figures, and tables, we will use the shorthand notation of Table~\ref{table:cases} to distinguish the different principal fitting combinations.

\subsection{Comparison with GIST}\label{sec:GISTcomparison}
In their study, GIST considered a restricted fitting model.  For the numerical data, they fit only to the $C_{22}$ mode of the complex strain $h$.   Furthermore, they fit the data only to the $\omega^+_{22n}$ QNMs with $n=0,1,\ldots,7$.  Finally, they simplified the fitting model by assuming that the spheroidal-harmonic expansion coefficients [see Eqs.~(\ref{eqn:swSH:expan}) and (\ref{eqn:swSH:simp})] were given by $\YSHn{\acute\ell\ell{m}n}=\delta_{\acute\ell\ell}$.  In the notation of this paper, the GIST version of Eq.~(\ref{eqn:Clm_expansion}) can be written
\begin{equation}\label{eqn:GISTfitmodel}
	C_{22}=\sum_{n=0}^{N}{C^+_{22n}}e^{-i\omega_{22n}t}.
\end{equation}
This fitting combination will be referred to as case $\{22\mbox{-$\mathcal{A}$}\}$.  It is important to note that, when using the restricted fitting model of case $\{22\mbox{-$\mathcal{A}$}\}$, there is no difference between the EV method and the mlEV method.

\begin{figure}
\includegraphics[width=\linewidth,clip]{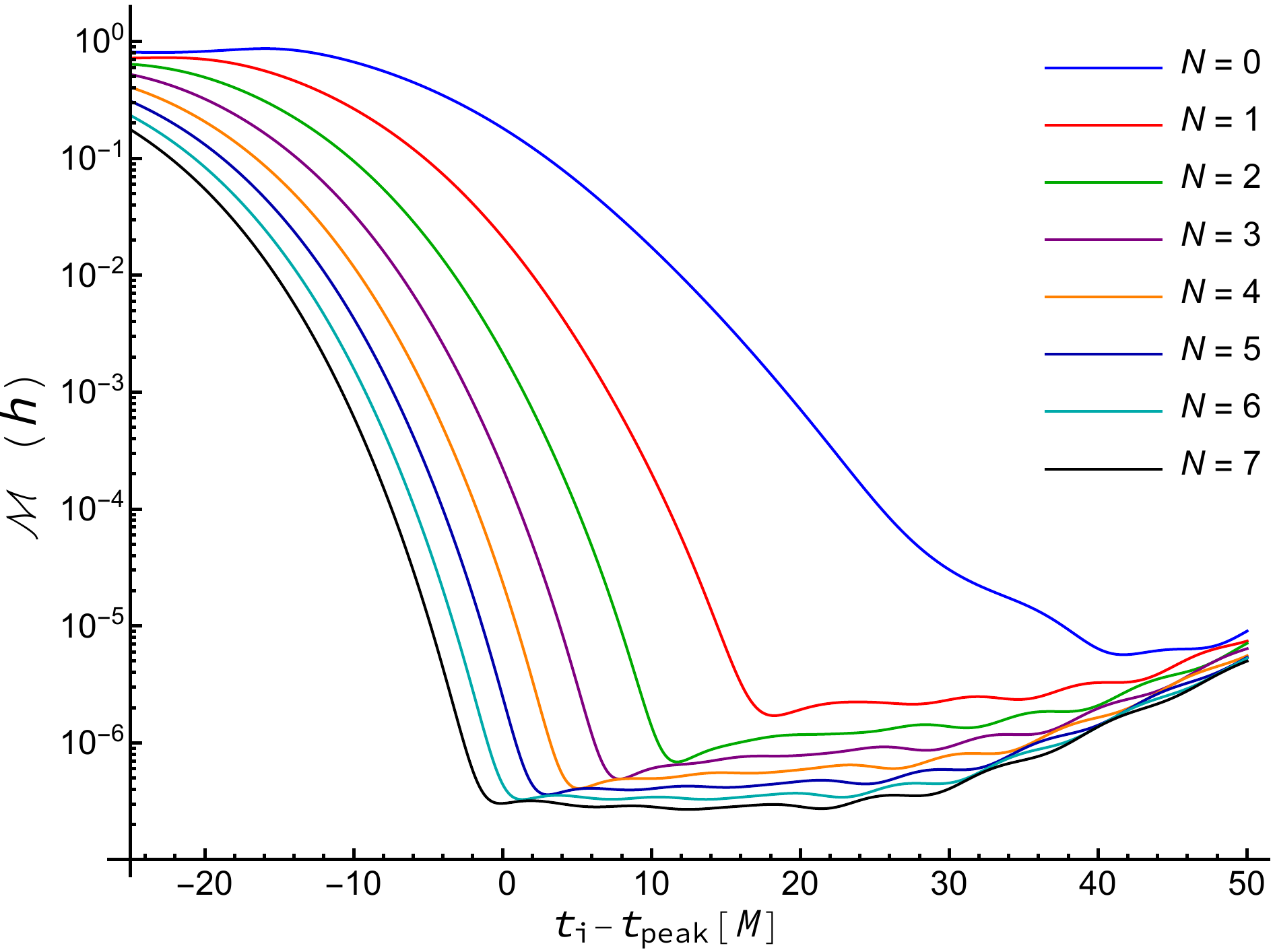}
 \caption{\label{fig:h7noamp} Mismatch $\mathcal{M}$ plotted as a function of $t_i-t_{\rm peak}$ for $h$ using fitting case $\{22\mbox{-$\mathcal{A}$}\}$ and SVD tolerance $\tau=0$.  The number $N$ associated with each line denotes the maximum value of the overtone index $n$ used in the fitting mode set $\{{\rm QNM}\}$.  This case uses $\YSHn{\acute\ell\ell{m}n}=\delta_{\acute\ell\ell}$ and can be compared directly with Fig.~1 of GIST.}
\end{figure}

In Fig.~\ref{fig:h7noamp} we recreate Fig.~1 of GIST showing the mismatch ${\cal M}=1-\rho$ from the gravitational strain $h$ for the $\{22\mbox{-$\mathcal{A}$}\}$ fitting model.  In computing the inner products, Eq.~(\ref{eqn:innerproduct}), we vary the start time of the integral, $t_i$, and fix the end time as $t_e=t_{\rm peak}+90M$.  Each line in the figure corresponds to a different number of overtones included in the sum for $\psi_{\rm fit}$, Eq.~(\ref{eqn:psi_fit_k}).  The uppermost line labeled $N=0$ has $\{{\rm QNM}\}=\{220+\}$, second from the top is the line labeled $N=1$ which has $\{{\rm QNM}\}=\{220+,221+\}$, and so forth.
The results are very similar to those in GIST, but not identical.  The main difference is a general trend of increasing value for ${\cal M}$ for $t_i-t_{\rm peak}\gtrsim30M$.  The differences with the results of GIST are very small, and are suggestive of small algorithmic differences.

\begin{figure}
\includegraphics[width=\linewidth,clip]{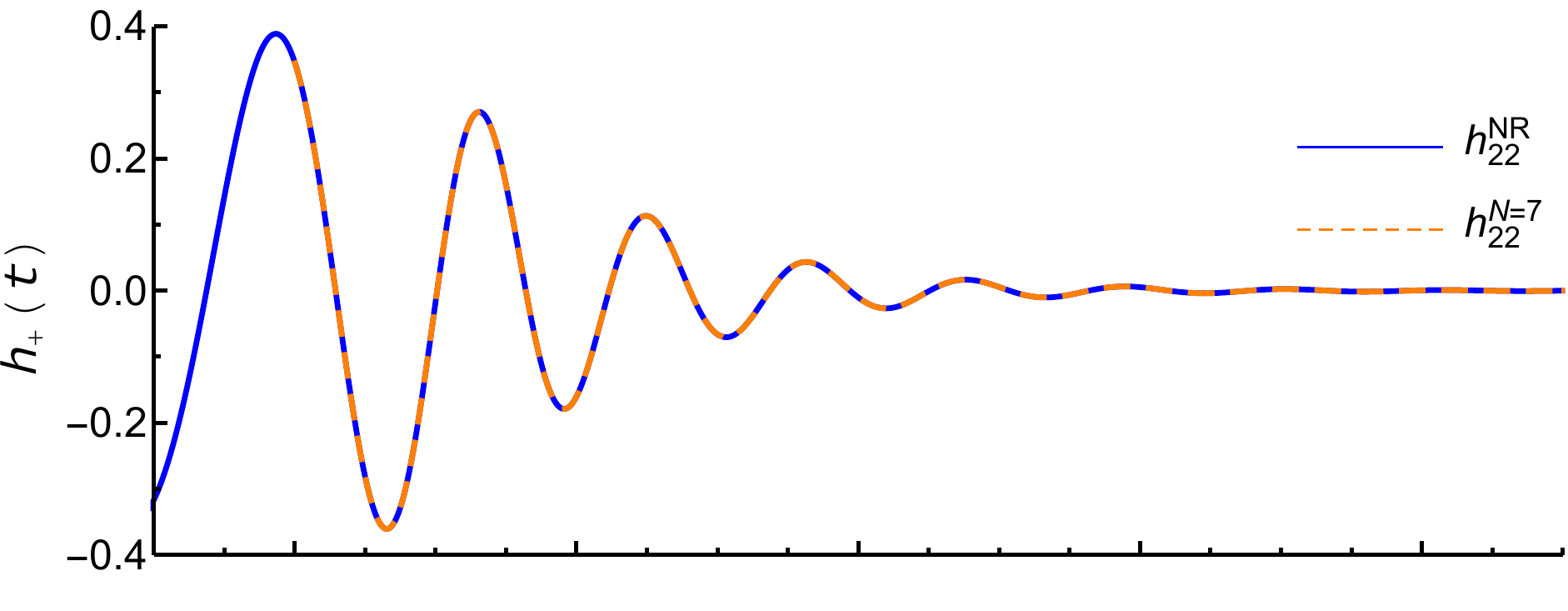}
\includegraphics[width=\linewidth,clip]{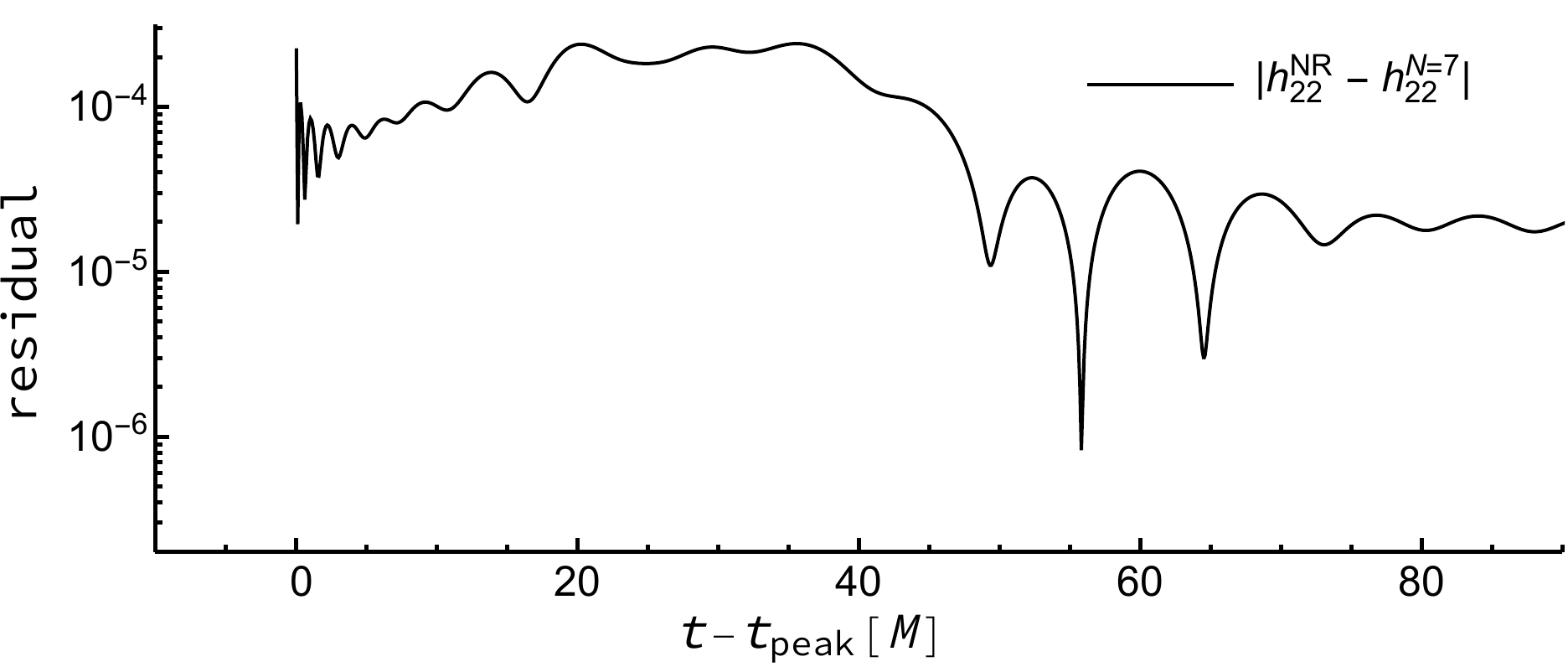}
 \caption{\label{fig:h7noampcmp} Comparison of the waveform and fit using case $\{22\mbox{-$\mathcal{A}$}\}$.  The upper panel shows $h_+$ from the numerical-relativity waveform and the $N=7$ version of the fit with $t_i=t_{\rm peak}$.  The lower panel shows the magnitude of the difference of the two complex signals.  This case uses $\YSHn{\acute\ell\ell{m}n}=\delta_{\acute\ell\ell}$ and can be compared directly with Fig.~2 of GIST.}
\end{figure}

In Fig.~\ref{fig:h7noampcmp} we recreate Fig.~2 of GIST comparing the waveform $h$ with the fit at $t_i=t_{\rm peak}$ and $N=7$.  The upper plot in the figure directly compares the real part of the waveform $h_+$, while the lower plot shows the magnitude of the difference of the two complex waveforms.
Again, there are small differences with the results in GIST.  While the upper plot in both cases shows excellent agreement between the simulation and the fit, there appears to be a phase shift between the data used in this work and that shown in GIST.  This suggests that we may not be comparing results from exactly the same input data.  However, as we will see, the differences are quite small for all quantities that can be accurately determined.  In the lower plot of the figure, we find that we have good agreement except for $t_i-t_{\rm peak}\gtrsim50M$ where the residuals in this work are slightly larger than those reported by GIST.

\begin{table}
	\begin{tabular}{l|ll|c}
	Case & $\{{\rm NR}\}$ & $\{{\rm QNM}\}$ & Fit equation \\
	\hline
	\hline
		$\{22\}$ & $\{22\}$ & $\{22n+\}$ & (\ref{eqn:Clm_expansion}) \\
		$\{22\mbox{-$\mathcal{A}$}\}$ & $\{22\}$ & $\{22n+\}$ & (\ref{eqn:GISTfitmodel}) \\
		$\{22\mbox{+3}\}$ & $\{22\}$ & $\{22n+,32n+\}$ & (\ref{eqn:Clm_expansion}) \\
		$\{22\mbox{+4}\}$ & $\{22\}$ & $\{22n+,32n+,42n+\}$ & (\ref{eqn:Clm_expansion}) \\
		$\{22,32\}$ & $\{22,32\}$ & $\{22n+,32n+\}$ & (\ref{eqn:Clm_expansion}) \\
		$\{22,32,42\}$ & $\{22,32,42\}$ & $\{22n+,32n+,42n+\}$ & (\ref{eqn:Clm_expansion}) \\
		$\{22,32,42*\}$ & $\{22,32,42\}$ & $\{220+,221+,222+,$\\
		&&\qquad\qquad$320+,420+\}$ & (\ref{eqn:Clm_expansion})
	\end{tabular}
	\caption{Notation for the different fitting cases explored in this paper.  The first column is the shorthand notation used in this paper.  The second column denotes the set of numerical relativity modes against which we are fitting as described in Eq.~(\ref{eqn:psi_NR_lm}).  The third column denotes the QNM ringdown modes used in the fitting function as defined in Eq.~(\ref{eqn:psi_fit_k}).  In all cases, $n\in\{0,1,\ldots,7\}$.  The fourth column lists the equation defining the fundamental relation between the numerical relativity modes and the QNMs.}
	\label{table:cases}
\end{table}

\begin{figure}
\includegraphics[width=\linewidth,clip]{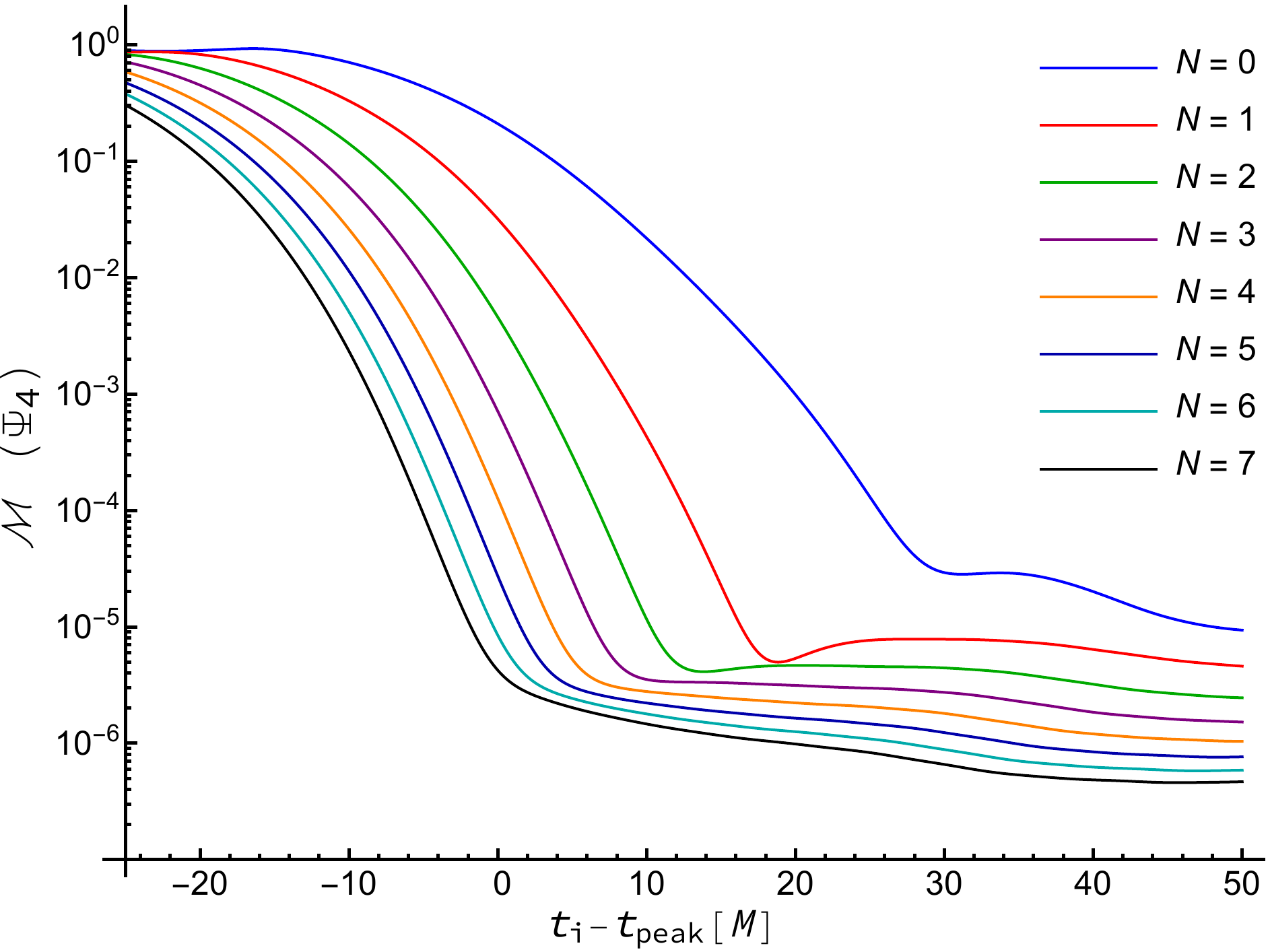}
 \caption{\label{fig:Psi7noamp} Mismatch $\mathcal{M}$ plotted as a function of $t_i-t_{\rm peak}$ for $\Psi_4$ using fitting case $\{22\mbox{-$\mathcal{A}$}\}$ and SVD tolerance $\tau=0$.  The number $N$ associated with each line denotes the maximum value of the overtone index $n$ used in the fitting mode set $\{{\rm QNM}\}$.}
\end{figure}

\begin{figure}
\includegraphics[width=\linewidth,clip]{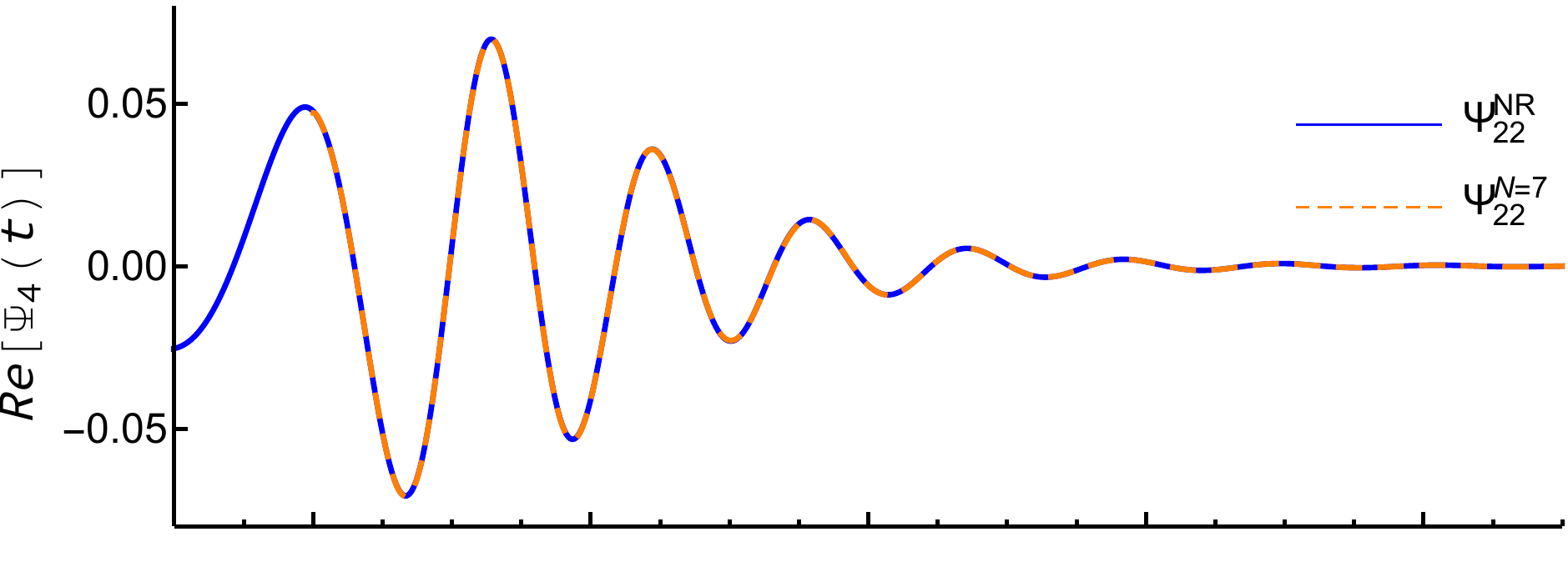}
\includegraphics[width=\linewidth,clip]{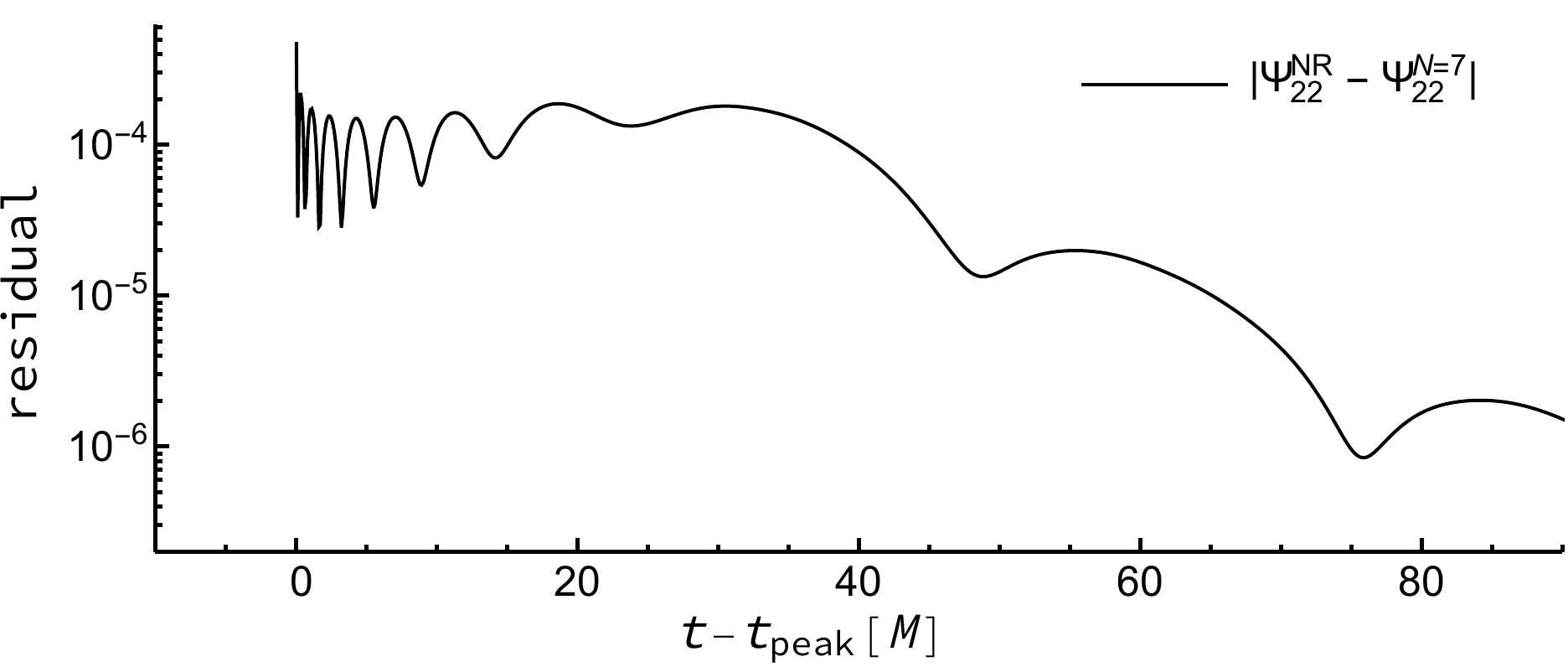}
 \caption{\label{fig:Psi7noampcmp} Comparison of the waveform and fit using case $\{22\mbox{-$\mathcal{A}$}\}$.  The upper panel shows $\rm{Re}[\Psi_4]=-\ddot{h}_+$ from the numerical-relativity waveform and the $N=7$ version of the fit with $t_i=t_{\rm peak}$.  The lower panel shows the magnitude of the difference of the two complex signals.}
\end{figure}

Ringdown fitting can be performed using either the strain $h$, news function $\mathcal{N}$, or the Weyl scalar $\Psi_4$.  In Figs.~\ref{fig:Psi7noamp} and \ref{fig:Psi7noampcmp} we plot the same results from Figs.~\ref{fig:h7noamp} and \ref{fig:h7noampcmp} but where the $\Psi_4$ ringdown waveform has been used for fitting.
The quality of the fits is very similar to that obtained by fitting to $h$.  The main difference seems to be that both the mismatch and the fit residual seems to be smaller at large values of $t_i-t_{\rm peak}$ when fitting to $\Psi_4$.

Tables~\ref{table:h7noA}--\ref{table:h7topsi7noA} display the least-squares fit amplitudes and phases for the fits using all eight overtones (i.e.\ for $N=7$) for $t_i=t_{\rm peak}$.  Table~\ref{table:h7noA} displays the results for fitting to $h$ and the results can be compared to the last row in Table~I of GIST.  The agreement is quite good, especially for the lower overtones, with differences increasing with the overtone number.  Table~\ref{table:Psi7noA} shows the same results for the fits to $\Psi_4$.  The complex amplitudes for fitting to $h$ and $\Psi_4$ are, of course, related by
\begin{equation}\label{eqn:htoPsi4}
C^{(\Psi_4)}_{\ell{m}n}=\omega^2_{\ell{m}n}C^{(h)}_{\ell{m}n}.
\end{equation}
\begin{table}
\begin{tabular}{c|d{5}d{4}|d{7}d{7}}
 Mode & \multicolumn{1}{c}{Amplitude} & \multicolumn{1}{c}{Phase/$\pi$} & \multicolumn{1}{c}{$\sigma$(Amp)} & \multicolumn{1}{c}{$\sigma$(Phase)/$\pi$} \\
 \hline\hline
 $C^+_{220}$ & 0.971 & 0.475 & 0.000083 & 0.000027 \\
 $C^+_{221}$ & 4.21 & -0.208 & 0.0027 & 0.00021 \\
 $C^+_{222}$ & 11.4 & 0.918 & 0.031 & 0.00086 \\
 $C^+_{223}$ & 23.1 & -0.0807 & 0.16 & 0.0022 \\
 $C^+_{224}$ & 33.4 & 0.849 & 0.44 & 0.0042 \\
 $C^+_{225}$ & 30.1 & -0.243 & 0.61 & 0.0064 \\
 $C^+_{226}$ & 14.7 & 0.671 & 0.41 & 0.0088 \\
 $C^+_{227}$ & 3.05 & -0.387 & 0.11 & 0.011
\end{tabular}
 \caption{\label{table:h7noA} The magnitude and phase of the QNM amplitudes from fitting $h$ with fitting case $\{22\mbox{-$\mathcal{A}$}\}$ with $N=7$ at $t_i=t_{\rm peak}$ and using SVD tolerance $\tau=0$.  Also displayed are the $1\sigma$ uncertainties.}
\end{table}
\begin{table}
\begin{tabular}{c|d{5}d{4}|d{7}d{7}}
 Mode & \multicolumn{1}{c}{Amplitude} & \multicolumn{1}{c}{Phase/$\pi$} & \multicolumn{1}{c}{$\sigma$(Amp)} & \multicolumn{1}{c}{$\sigma$(Phase)/$\pi$} \\
 \hline\hline
 $C^+_{220}$ & 0.306 & 0.380 & 0.000068 & 0.000070 \\
 $C^+_{221}$ & 1.52 & -0.481 & 0.0022 & 0.00047 \\
 $C^+_{222}$ & 5.25 & 0.516 & 0.025 & 0.0015 \\
 $C^+_{223}$ & 14.2 & -0.554 & 0.13 & 0.0030 \\
 $C^+_{224}$ & 26.3 & 0.351 & 0.36 & 0.0044 \\
 $C^+_{225}$ & 28.5 & -0.740 & 0.49 & 0.0055 \\
 $C^+_{226}$ & 16.0 & 0.185 & 0.33 & 0.0066 \\
 $C^+_{227}$ & 3.70 & -0.859 & 0.089 & 0.0076
\end{tabular}
 \caption{\label{table:Psi7noA} The magnitude and phase of the QNM amplitudes from fitting $\Psi_4$ with fitting case $\{22\mbox{-$\mathcal{A}$}\}$ with $N=7$ at $t_i=t_{\rm peak}$ and using SVD tolerance $\tau=0$.  Also displayed are the $1\sigma$ uncertainties.}
\end{table}
Table~\ref{table:h7topsi7noA} uses Eq.~(\ref{eqn:htoPsi4}) to transform the complex expansion coefficients shown in Table~\ref{table:h7noA} so that they can be compared with the results in Table~\ref{table:Psi7noA}.
\begin{table}
\begin{tabular}{c|d{5}d{5}}
 Mode & \multicolumn{1}{c}{Amplitude} & \multicolumn{1}{c}{Phase/$\pi$} \\
 \hline\hline
 $C^+_{220}$ & 0.307 & 0.378 \\
 $C^+_{221}$ & 1.52 & -0.490 \\
 $C^+_{222}$ & 5.24 & 0.476 \\
 $C^+_{223}$ & 14.3 & -0.653 \\
 $C^+_{224}$ & 28.2 & 0.181 \\
 $C^+_{225}$ & 34.1 & -0.972 \\
 $C^+_{226}$ & 22.6 & -0.0984 \\
 $C^+_{227}$ & 6.25 & 0.812 \\
\end{tabular}
 \caption{\label{table:h7topsi7noA} The magnitude and phase of the QNM amplitudes for $\Psi_4$ from fitting $h$ with fitting case $\{22\mbox{-$\mathcal{A}$}\}$ with $N=7$ at $t_i=t_{\rm peak}$ and using SVD tolerance $\tau=0$.  The coefficients used to produce Table~\ref{table:h7noA} were converted to coefficients for $\Psi_4$ using Eq.~(\ref{eqn:htoPsi4}).  The results in this table can be directly compared to those in Table~\ref{table:Psi7noA}.}
\end{table}
In general we see good agreement between modes fit using $h$ and $\Psi_4$, with the agreement being best for the lowest overtones (small $n$) and with the differences increasing as the overtone number increases.

\subsection{Singular-value decomposition}

As mentioned in Sec.~\ref{sec:implementation} we make use of SVD to construct the matrix inverse of ${\mathbb B}$ (or ${\mathbb B}_{\rm ml}$ or ${\mathbb L}^\dag\cdot{\mathbb L}$).  This approach is especially useful in fitting models where the inner product of some of the fitting modes is very small.  In such cases, the matrices ${\mathbb B}$, ${\mathbb B}_{\rm ml}$, and ${\mathbb L}^\dag\cdot{\mathbb L}$ may be singular, or nearly singular.  By using SVD, we gain explicit control of these modes by setting a tolerance on the size of the singular values of the matrix that will be included in the inverse.  When the ratio of a given singular value to the largest singular value is below this specified tolerance, then the inverse of that singular value is set to zero when constructing the inverse metric (see Sec.~2.6.2 of Ref.~\cite{numrec_c++}).  This process creates what is called a pseudoinverse of the matrix.

\begin{figure}
\includegraphics[width=\linewidth,clip]{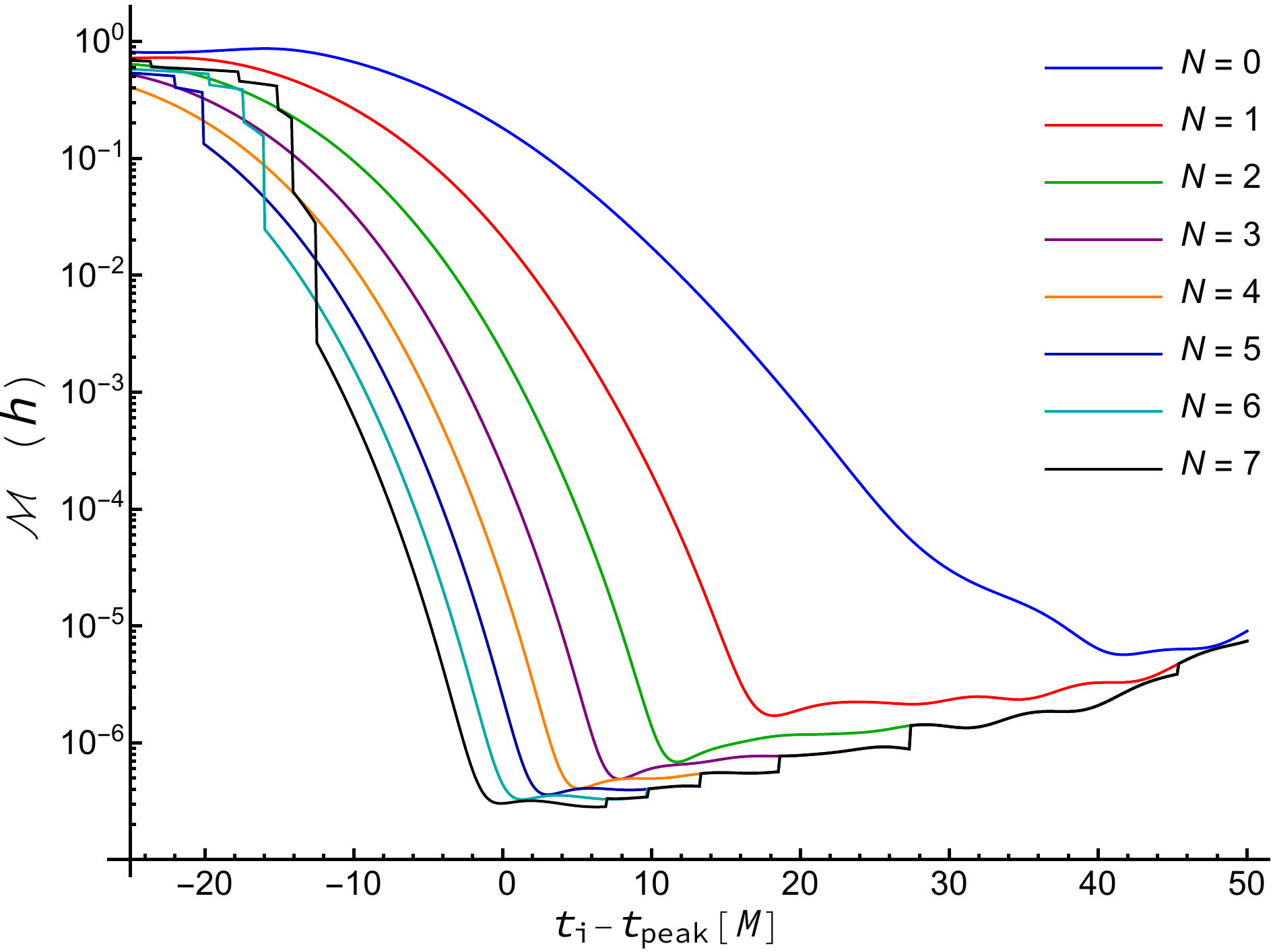}
 \caption{\label{fig:h7noamptol} This figure plots the same case as Fig.~\ref{fig:h7noamp} but with SVD tolerance $\tau=10^{-16}$.}
\end{figure}

The matrix inverses used in computing the mismatch curves shown in Figs.~\ref{fig:h7noamp} and \ref{fig:Psi7noamp} used a tolerance of $0$, meaning that the true inverse was used.  In both figures, as $N$ approached $7$, roundoff error became noticeable and extended precision was used to alleviate this and make the lines relatively smooth.  Sensitivity to roundoff error is an indication that the matrix being inverted is ill-conditioned.  Figure~\ref{fig:h7noamptol} shows the same mismatch curves displayed in Fig.~\ref{fig:h7noamp}, but with the tolerance set to $10^{-16}$.
Examining the curves for $t_i-t_{\rm peak}>0$ clearly shows the effect of setting a nonvanishing tolerance.  Near $t_i-t_{\rm peak}\sim7M$, the $N=7$ curve changes discontinuously to the same level as the $N=6$ curve.  The reason for this is that the decay rate for the $\omega_{227}$ mode is very large and this mode quickly becomes so small that it should not contribute significantly to the fit.  However, if the SVD tolerance is set too low, then the linear fitting process will try to increase the amplitude of this mode so that it can contribute.  In this case the large amplitude for this mode is not physical.  The decay rate for each of the modes decreases with decreasing $n$, and one can clearly see where the nonvanishing SVD tolerance is causing successive modes to be ignored in the fitting process.  A similar ``stepping'' in the mismatch curves can also be seen in many of the curves for $t_i-t_{\rm peak}<0$.  In this case, it is typically the modes with small $n$ which are subdominant and are being filtered out via the pseudoinverse.

\subsection{Full fitting of $C_{22}$ using $\omega_{22n}$}\label{sec:fullC22}

Now consider the same fitting, but using the full fitting model of Eq.~(\ref{eqn:Clm_expansion}) instead of the restricted model of Eq.~(\ref{eqn:GISTfitmodel}).  This fitting combination will be referred to as case $\{22\}$.  In the numerical simulation being studied, the total angular momentum of the final system is aligned with the $z$-axis of the simulation, so the Wigner rotation matrix $D^\ell_{\acute{m}{m}}(\bar\alpha,\bar\beta,\bar\gamma)\to\delta_{\acute{m}{m}}$ and plays no role.  The difference is that we will now correctly include the spheroidal-harmonic expansion coefficients $\YSHn{
 \acute\ell\ell{m}n}$.  As pointed out by GIST, these coefficients are nearly $1$ for the dominant mode $C_{22}$ implying that there is very little mixing with the $C_{\ell2}$ modes with $\ell>2$.
\begin{table}
\begin{tabular}{c|c}
 Mode & Amplitude \\
 \hline\hline
 $\YSHn{2220}$ & 0.997554 \\
 $\YSHn{2221}$ & 0.997197 \\
 $\YSHn{2222}$ & 0.996422 \\
 $\YSHn{2223}$ & 0.995151 \\
 $\YSHn{2224}$ & 0.993368 \\
 $\YSHn{2225}$ & 0.990998 \\
 $\YSHn{2226}$ & 0.987693 \\
 $\YSHn{2227}$ & 0.983360 \\
\end{tabular}
\caption{\label{table:spexpcoefs}Spheroidal expansion coefficients for $\chi_f=0.692085$ }
\end{table}
Table~\ref{table:spexpcoefs} shows the values of these expansion coefficients for $\chi_f=0.692085$.

\begin{figure}
\includegraphics[width=\linewidth,clip]{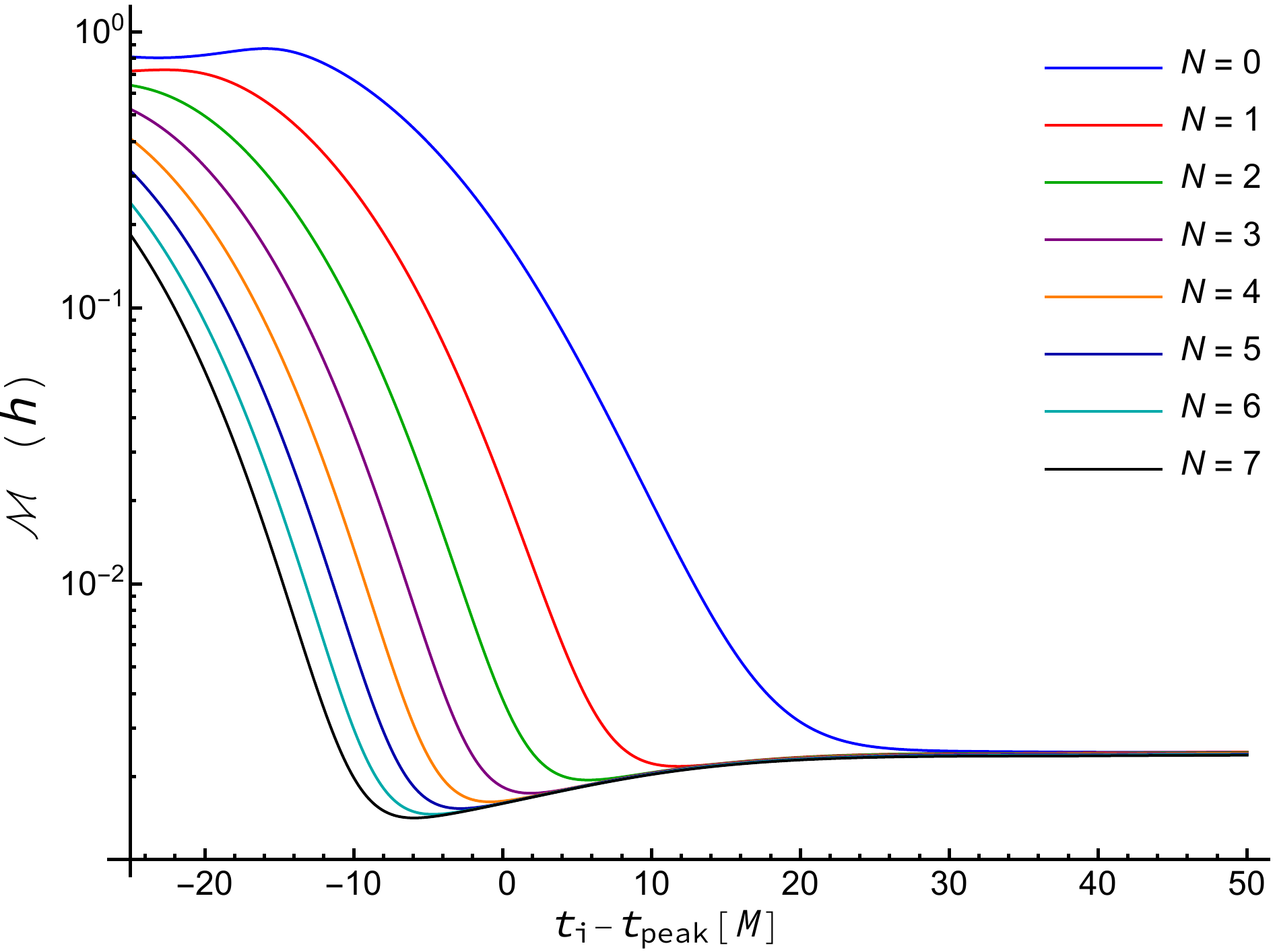}
 \caption{\label{fig:h7_C22} Mismatch $\mathcal{M}$ plotted as a function of $t_i-t_{\rm peak}$ for $h$ using fitting case $\{22\}$, the EV method, and SVD tolerance $\tau=0$.  The number $N$ associated with each line denotes the maximum value of the overtone index $n$ used in the fitting mode set $\{{\rm QNM}\}$.}
\end{figure}

\begin{figure}
\includegraphics[width=\linewidth,clip]{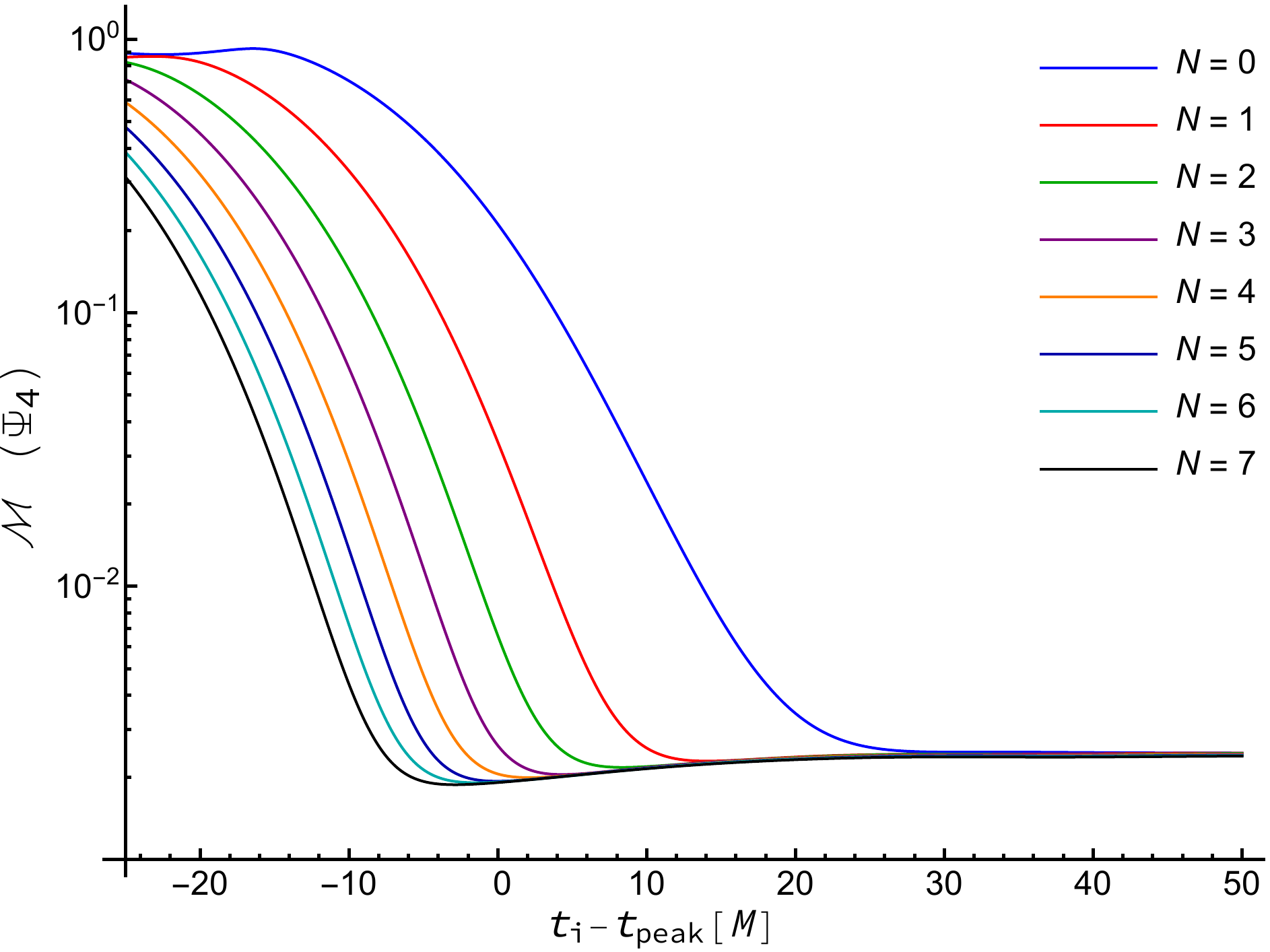}
 \caption{\label{fig:Psi7_C22} Mismatch $\mathcal{M}$ plotted as a function of $t_i-t_{\rm peak}$ for $\Psi_4$ using fitting case $\{22\}$, the EV method, and SVD tolerance $\tau=0$.  The number $N$ associated with each line denotes the maximum value of the overtone index $n$ used in the fitting mode set $\{{\rm QNM}\}$.}
\end{figure}

Figures~\ref{fig:h7_C22} and \ref{fig:Psi7_C22} show the same mismatch curves as plotted in Figs.~\ref{fig:h7noamp} and \ref{fig:Psi7noamp} but using the correct spherical-harmonic expansion coefficients.  It is obvious that the mismatch is significantly larger.  The reason for this increase in the mismatch will be discussed in detail below.

Tables~\ref{table:h7}--\ref{table:h7topsi7} show the amplitudes and phases for each mode used in the $N=7$ fit at $t_i=t_{\rm peak}$ when full fitting is used.  As with the modes computed using the GIST restricted fitting, the modes with lower overtone number are obtained with good accuracy and with good agreement between the fits using $h$ and $\Psi_4$.
\begin{table}
\begin{tabular}{c|d{5}d{4}|d{6}d{6}}
 Mode & \multicolumn{1}{c}{Amplitude} & \multicolumn{1}{c}{Phase/$\pi$} & \multicolumn{1}{c}{$\sigma$(Amp)} & \multicolumn{1}{c}{$\sigma$(Phase)/$\pi$} \\
 \hline\hline
 $C^+_{220}$ & 0.970 & 0.476 & 0.00056 & 0.00018 \\
 $C^+_{221}$ & 4.31 & -0.210 & 0.018 & 0.0013 \\
 $C^+_{222}$ & 12.3 & 0.888 & 0.19 & 0.0050 \\
 $C^+_{223}$ & 28.3 & -0.176 & 1.0 & 0.011 \\
 $C^+_{224}$ & 51.9 & 0.681 & 2.6 & 0.016 \\
 $C^+_{225}$ & 62.9 & -0.448 & 3.5 & 0.018 \\
 $C^+_{226}$ & 41.6 & 0.459 & 2.3 & 0.018 \\
 $C^+_{227}$ & 11.2 & -0.593 & 0.60 & 0.017 \\
\end{tabular}
\caption{\label{table:h7} The magnitude and phase of the QNM amplitudes from fitting $h$ with fitting case $\{22\}$ and the EV method with $N=7$ at $t_i=t_{\rm peak}$ and using SVD tolerance $\tau=0$.}
\end{table}
\begin{table}
\begin{tabular}{c|d{5}d{4}|d{6}d{6}}
 Mode & \multicolumn{1}{c}{Amplitude} & \multicolumn{1}{c}{Phase/$\pi$} & \multicolumn{1}{c}{$\sigma$(Amp)} & \multicolumn{1}{c}{$\sigma$(Phase)/$\pi$} \\
 \hline\hline
 $C^+_{220}$ & 0.306 & 0.380 & 0.00013 & 0.00013 \\
 $C^+_{221}$ & 1.52 & -0.477 & 0.0041 & 0.00085 \\
 $C^+_{222}$ & 5.34 & 0.528 & 0.044 & 0.0026 \\
 $C^+_{223}$ & 14.8 & -0.528 & 0.23 & 0.0049 \\
 $C^+_{224}$ & 28.2 & 0.392 & 0.60 & 0.0068 \\
 $C^+_{225}$ & 31.0 & -0.683 & 0.81 & 0.0083 \\
 $C^+_{226}$ & 17.8 & 0.260 & 0.53 & 0.0095 \\
 $C^+_{227}$ & 4.17 & -0.766 & 0.14 & 0.010 \\
\end{tabular}
\caption{\label{table:Psi7} The magnitude and phase of the QNM amplitudes from fitting $\Psi_4$ with fitting case $\{22\}$ and the EV method with $N=7$ at $t_i=t_{\rm peak}$ and using SVD tolerance $\tau=0$.}
\end{table}
\begin{table}
\begin{tabular}{c|d{5}d{4}}
 Mode & \multicolumn{1}{c}{Amplitude} & \multicolumn{1}{c}{Phase/$\pi$} \\
 \hline\hline
 $C^+_{220}$ & 0.307 & 0.379 \\
 $C^+_{221}$ & 1.56 & -0.492 \\
 $C^+_{222}$ & 5.67 & 0.446 \\
 $C^+_{223}$ & 17.6 & -0.749 \\
 $C^+_{224}$ & 43.7 & 0.0131 \\
 $C^+_{225}$ & 71.3 & 0.823 \\
 $C^+_{226}$ & 3.79 & -0.310 \\
 $C^+_{227}$ & 23.0 & 0.606 \\
\end{tabular}
 \caption{\label{table:h7topsi7} The magnitude and phase of the QNM amplitudes for $\Psi_4$ from fitting $h$ with fitting case $\{22\}$ and the EV method with $N=7$ at $t_i=t_{\rm peak}$ and using SVD tolerance $\tau=0$.  The coefficients used to produce Table~\ref{table:h7} were converted to coefficients for $\Psi_4$ using Eq.~(\ref{eqn:htoPsi4}).  The results in this table can be directly compared to those in Table~\ref{table:Psi7}.}
\end{table}
It is also clear that the modes with lower overtone number are in good agreement when comparing the results of full fitting with those using the restricted GIST fitting model.  Furthermore, while the values for the modes with larger $n$ disagree between the two fitting models, the general trends in the size of the amplitudes agree.

Returning to the comparison of the mismatch between Figs.~\ref{fig:h7_C22} and \ref{fig:h7noamp}, the reason for the increase in the mismatch originates in the fact that when full fitting is used, $\psi_{\rm fit}$ is expanded in terms of spin-weighted spheroidal harmonics instead of spherical harmonics.  So, even in the simplest case where we only use $\{22n\pm\}$ QNMs to construct $\psi_{\rm fit}$, it contains contributions for $\ell=3$ and higher spin-weighted spherical harmonic modes.  When we construct $\langle\psi_{\rm fit}|\psi_{\rm fit}\rangle=\sum_{ij}C^*_iB_{ij}C_j$, we get a larger result than if we first projected out only the $\ell=2$ spherical harmonic contributions to $\psi_{\rm fit}$.  These higher order $\ell$ modes enter the calculation through the sum over $\breve\ell$ in Eqs.~(\ref{eqn:Bcomp++}), (\ref{eqn:Bcomp+-}), and (\ref{eqn:Bcomp--}) for the components of $\mathbb{B}$.

Does a larger mismatch imply that the full fit [case $\{22\})$] is significantly worse than the restricted fit [case $\{22\mbox{-$\mathcal{A}$}\}$]?  A more equal comparison between the results from the full fitting of $C_{22}$ and the restricted fitting can be obtained by recomputing the overlap using Eq.~(\ref{eqn:rho2explicit}) with the expansion coefficients $C^+_{22n}$ from case $\{22\}$, but with the matrix elements $B_{ij}$ recomputed with $\breve\ell=2$ instead of summed over all values.  That is, we us the mode-limited $\mathbb{B}_{\rm ml}$ [see Eqs~(\ref{eqn:mlBcomp++})--(\ref{eqn:mlBcomp--})] to compute the overlap in Eq.~(\ref{eqn:rho2explicit}).
\begin{figure}
\includegraphics[width=\linewidth,clip]{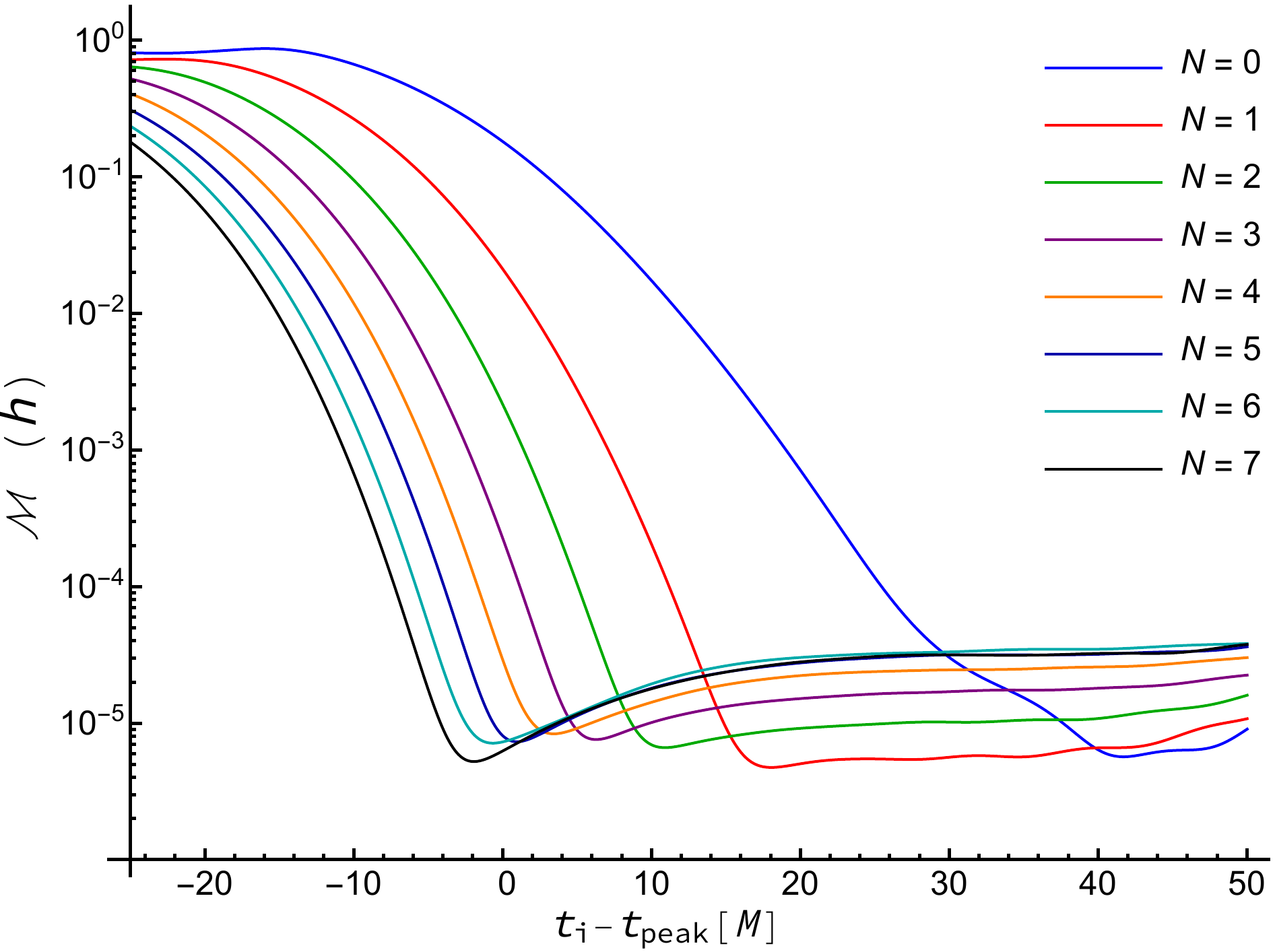}
 \caption{\label{fig:h7_C22rhorestrict} Same as Fig.~\ref{fig:h7_C22} but with the mismatch recomputed using $\mathbb{B}_{\rm ml}$ as described in the text.}
\end{figure}
The result is show in Fig.~\ref{fig:h7_C22rhorestrict}.  We see that the mismatch is much smaller than in Fig.~\ref{fig:h7_C22}.  Comparing Fig.~\ref{fig:h7_C22rhorestrict} with the results from case $\{22\mbox{-$\mathcal{A}$}\}$ seen in Fig.~\ref{fig:h7noamp} we see that the mismatch for the $N=0$ cases are identical, but the mismatch curves for $N>0$ show generally larger values than seen in the restricted fitting case.

We can also recompute the overlap using the mlEV method.  That is, using $\mathbb{B}_{\rm ml}$ in the computation of the expansion coefficients and the overlap.  The results for the mismatch are identical to those in Fig.~\ref{fig:h7noamp}.  The amplitudes of the expansion coefficients differ from those in Table~\ref{table:h7} (full fitting) but are, in fact, equal to the values from Table~\ref{table:h7noA} (restricted fitting), but divided by the appropriate spheroidal expansion coefficient from Table~\ref{table:spexpcoefs}.

The differences between the mismatches in Figs.~\ref{fig:h7noamp} and \ref{fig:h7_C22rhorestrict} most clearly illustrates the fundamental difference between the EV method and the mlEV(or LS) method.  For fitting case $\{22\}$, the mlEV method is the same as the LS method (aside from minor differences corresponding to the implementation of the time integral) and the fitting performed in GIST.  When the mismatch from the full EV method is recomputed using $\mathbb{B}_{\rm ml}$, the only difference in comparison with the mlEV method mismatch comes from the computed values of the expansion coefficients $C^+_{22n}$.

\subsection{Fitting $C_{22}$ with higher order modes}\label{sec:C22-higher-order}

\begin{figure}
\includegraphics[width=\linewidth,clip]{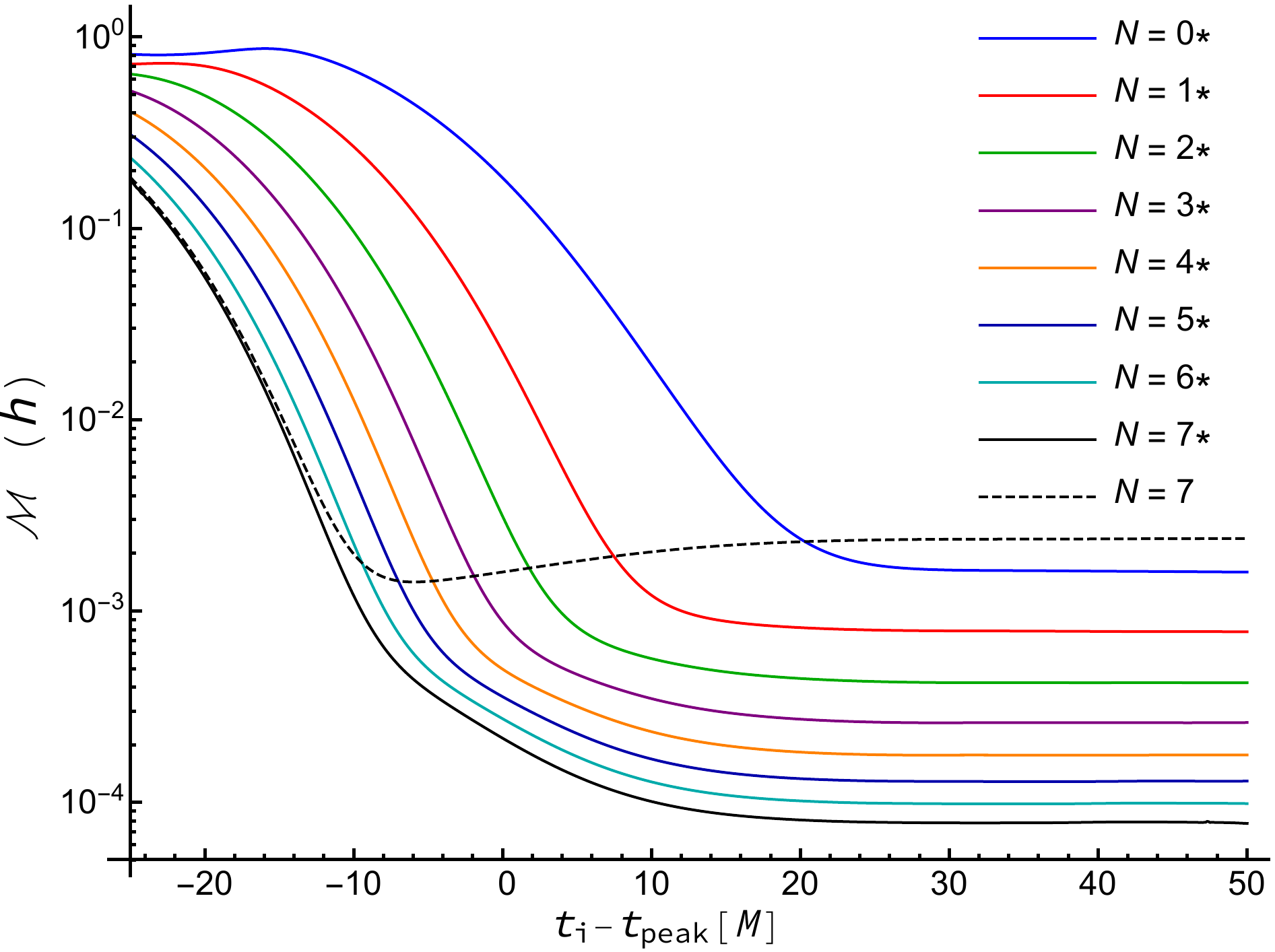}
 \caption{\label{fig:h7_C22a} Mismatch $\mathcal{M}$ plotted as a function of $t_i-t_{\rm peak}$ for $h$ using fitting case $\{22\mbox{+3}\}$, the EV method, and SVD tolerance $\tau=0$.  The number $N$ associated with each line denotes the maximum value of the overtone index $n$ used in the fitting mode set $\{{\rm QNM}\}$.}
\end{figure}

\begin{figure}
\includegraphics[width=\linewidth,clip]{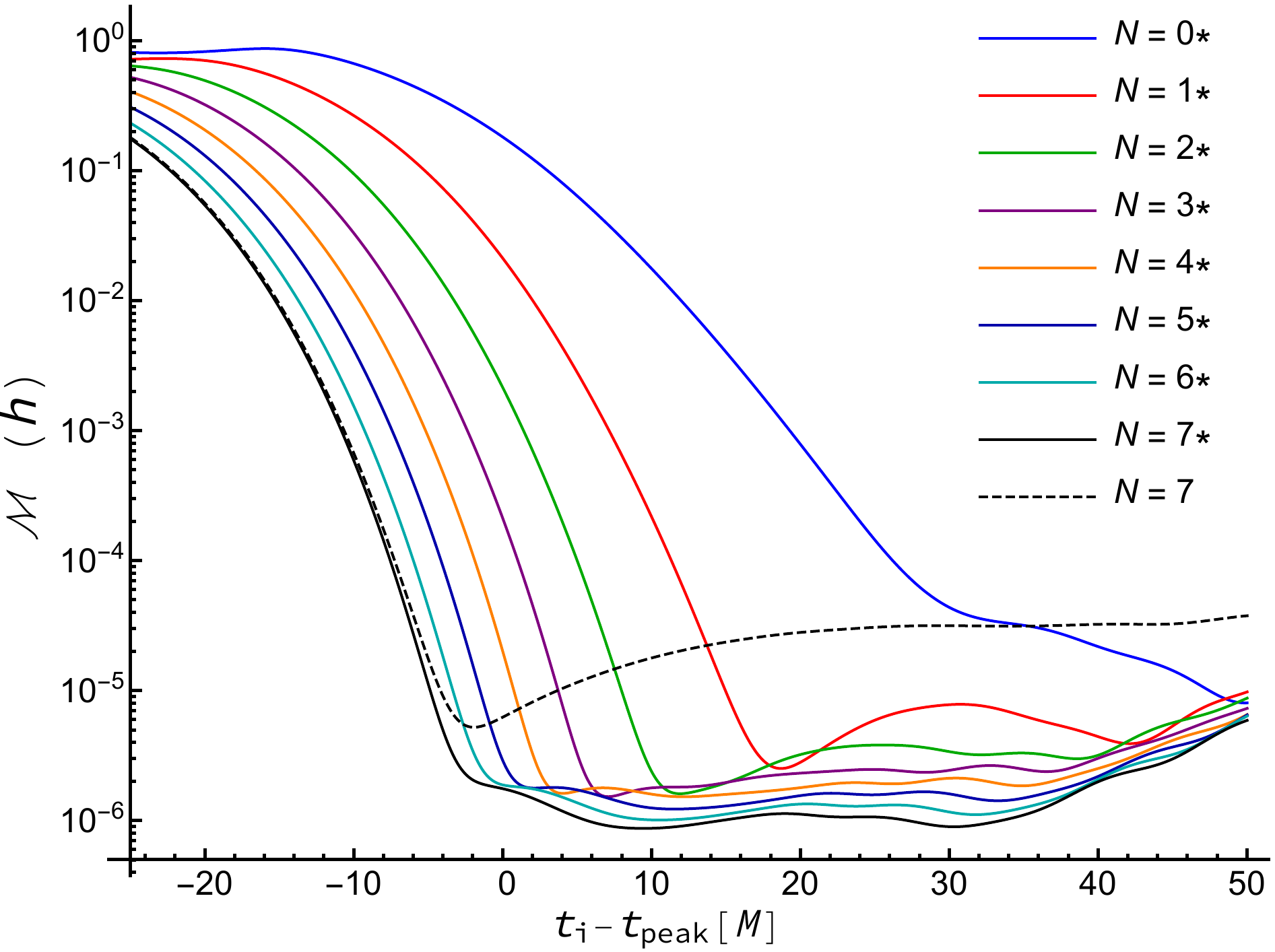}
 \caption{\label{fig:h7_C22arhorestrict} Same as Fig.~\ref{fig:h7_C22a}, but with the mismatch recomputed using $\mathbb{B}_{\rm ml}$ as described in the text.}
\end{figure}

In order to increase the overlap of the fits (decrease the mismatch), we must include higher order modes.  Here, we will explore the effects of adding the $\omega_{32n}$ and $\omega_{42n}$ modes to fitting the $C_{22}$ waveform with the EV method.  Figure~\ref{fig:h7_C22a} displays the results for fitting using the $\omega_{22n}$ and $\omega_{32n}$ modes with $n=0,1,\ldots,7$.  This fitting combination will be referred to as case $\{22\mbox{+3}\}$.  In order to further distinguish the different models, we label the mismatch curves that include modes $\omega^+_{22n}$ and $\omega^+_{32n}$ with overtones $n$ up to $N$ as $N\ast$.  Included in Fig.~\ref{fig:h7_C22a} for comparison is the $N=7$ mismatch curve from Fig.~\ref{fig:h7_C22}.
Clearly, including the $\ell=3$ QNMs significantly reduces the mismatch.  For comparison, Fig.~\ref{fig:h7_C22arhorestrict} presents the mismatch for the same fit as shown in Fig.~\ref{fig:h7_C22a} but recomputed using $\mathbb{B}_{\rm ml}$.

For Figs.~\ref{fig:h7_C22a} and \ref{fig:h7_C22arhorestrict} (and Figs.~\ref{fig:h7_C22} and \ref{fig:Psi7_C22}), the mismatch was computed using the EV method with the least-squares fitting of the expansion coefficients based on Eq.~(\ref{eqn:chi2full}).  Now, consider how the fitting behaves if we use the LS method with fitting based on Eq.~(\ref{eqn:chi2ls}) and use $\mathbb{B}$ to compute the mismatch.
\begin{figure}
\includegraphics[width=\linewidth,clip]{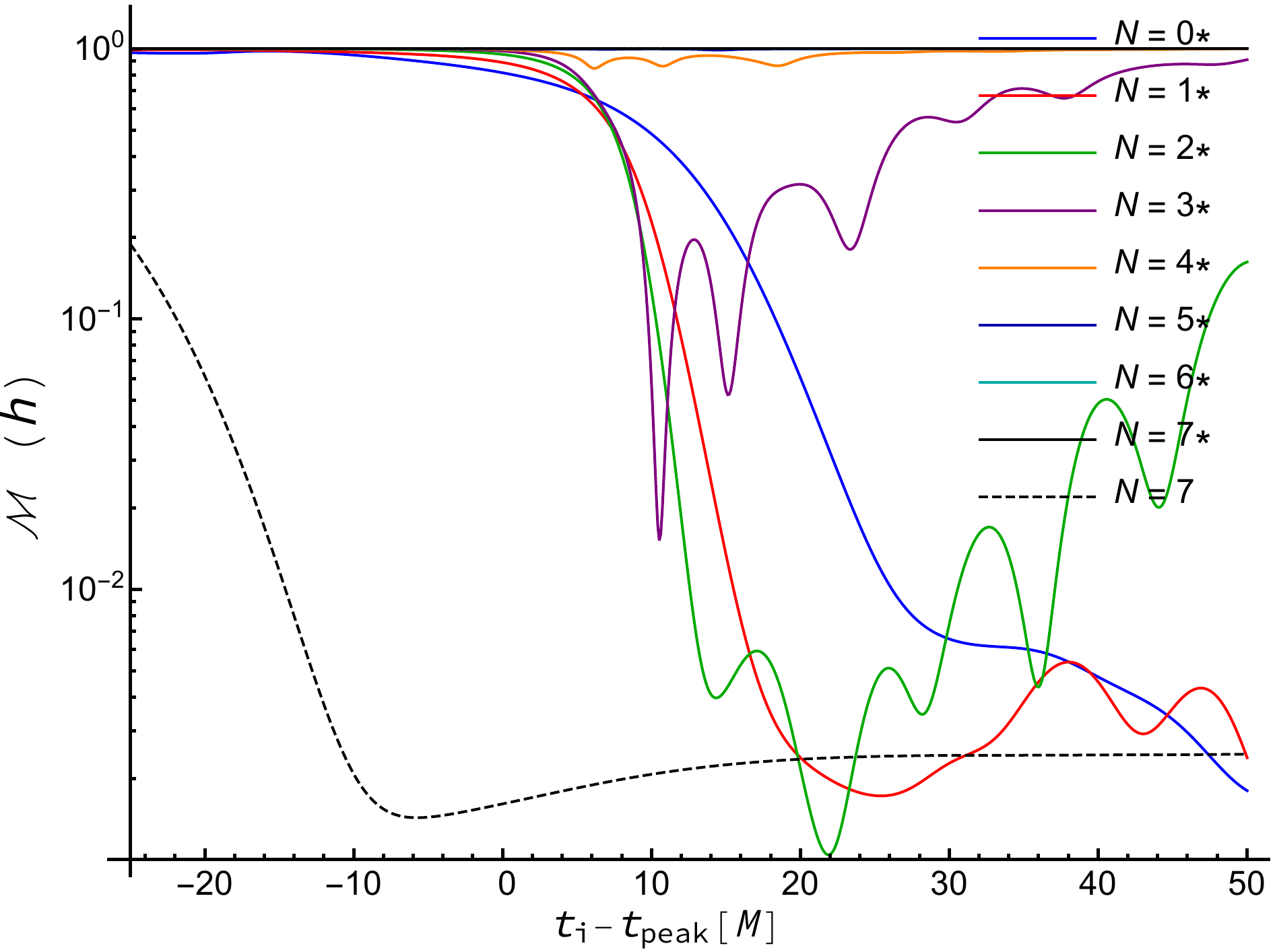}
 \caption{\label{fig:h7_C22NEa} Same as Fig.~\ref{fig:h7_C22a}, but fit with the LS method using least-squares fitting based on Eq.~(\ref{eqn:chi2ls}) instead of using the EV method.  As with the EV method, the mismatch is computed using $\mathbb{B}$.}
\end{figure}
The results are shown in Fig.~\ref{fig:h7_C22NEa}.  The mismatch seems to indicates a poor fit.  However, if we recompute the mismatch with $\mathbb{B}_{\rm ml}$, the mismatch as shown in Fig.~\ref{fig:h7_C22NEarhorestrict}, actually appears quite good for $N\le5$.
\begin{figure}
\includegraphics[width=\linewidth,clip]{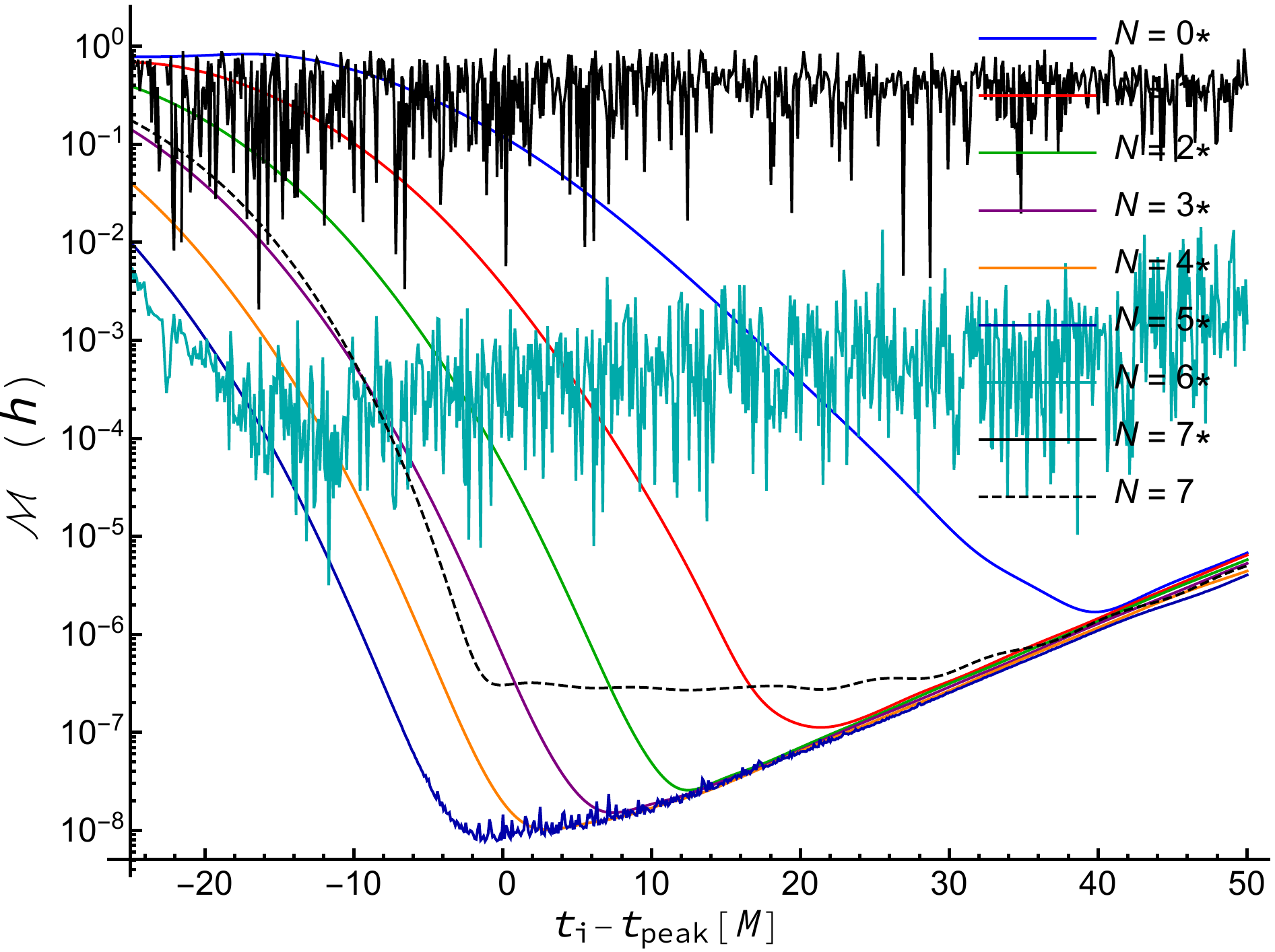}
 \caption{\label{fig:h7_C22NEarhorestrict} Same as Fig.~\ref{fig:h7_C22NEa}, but with the mismatch recomputed using $\mathbb{B}_{\rm ml}$ as described in the text.}
\end{figure}
The $N=6$ and $7$ cases show the effects of $\mathbb{L}^\dag\cdot\mathbb{L}$ becoming increasingly ill-conditioned.  The rather poor mismatch of the $N=6$ and $7$ cases can be remedied by using a nonvanishing SVD tolerance, but for the sake of comparison, Figs.~\ref{fig:h7_C22NEa} and \ref{fig:h7_C22NEarhorestrict} have been computed with $\tau=0$.

\begin{figure}
\includegraphics[width=\linewidth,clip]{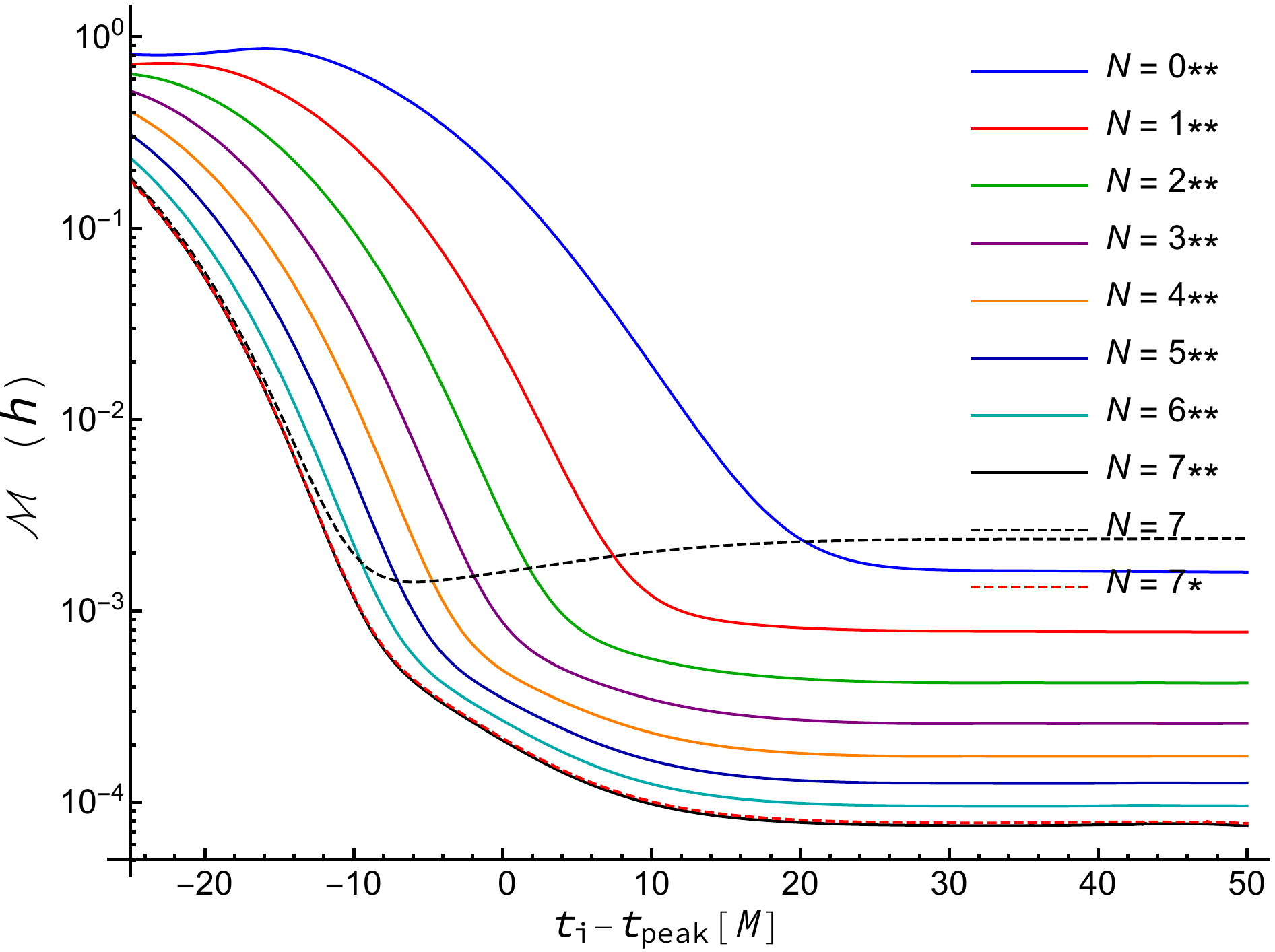}
 \caption{\label{fig:h7_C22b} Mismatch $\mathcal{M}$ plotted as a function of $t_i-t_{\rm peak}$ for $h$ using fitting case $\{22\mbox{+4}\}$, the EV method, and SVD tolerance $\tau=0$.  The number $N$ associated with each line denotes the maximum value of the overtone index $n$ used in the fitting mode set $\{{\rm QNM}\}$.}
\end{figure}

Returning to the EV method, we can extend the fit of $C_{22}$ to include the $\omega_{42n}$ modes.  Figure~\ref{fig:h7_C22b} shows the results for fitting using the $\omega^+_{22n}$, $\omega^+_{32n}$, and $\omega^+_{42n}$ modes with $n=0,1,\ldots,7$.  This fitting combination will be referred to as case $\{22\mbox{+4}\}$.  In order to further distinguish the different models, we label the mismatch curves that include modes $\omega^+_{22n}$, $\omega^+_{32n}$ and $\omega^+_{42n}$ with overtones $n$ up to $N$ as $N\!\ast\!\ast$.  Included in Fig.~\ref{fig:h7_C22b} for comparison is the $N=7$ mismatch curve from Fig.~\ref{fig:h7_C22} and the $N=7\ast$ curve from Fig.~\ref{fig:h7_C22a}.
It is apparent that adding the $\ell=4$ modes to the fit has very little effect on the mismatch.  Corresponding mismatch plots (not shown) using $\mathbb{B}_{\rm ml}$ or the LS method show a similar lack of improvement with the inclusion of $\ell=4$ QNMs.
Table~\ref{table:h722} presents the amplitude and phase for the fit denoted $N=7\!\ast\!\ast$ in Fig.~\ref{fig:h7_C22b} at time $t_i=t_{\rm peak}$.  Table~\ref{table:Psi722} shows the corresponding amplitudes and phases for fits to $\Psi_4$.  Table~\ref{table:h7topsi722} shows the amplitudes and phases from the fit to $h$ in Table~\ref{table:h722} converted to amplitudes and phases from $\Psi_4$ for comparison with the results in Table~\ref{table:Psi722}.
%\begin{figure}
%\includegraphics[width=\linewidth,clip]{MM0305_Psi4_o7C22b}
% \caption{\label{fig:Psi7_C22b} $\Psi_4$ fit to $C_{22}$, $C^+_{22(0-7)}$, $C^+_{32(0-7)}$, $C^+_{42(0-7)}$, $Tol=0$}
%\end{figure}
\begin{table}
\begin{tabular}{c|d{5}d{6}|d{6}d{7}}
 Mode & \multicolumn{1}{c}{Amplitude} & \multicolumn{1}{c}{Phase/$\pi$} & \multicolumn{1}{c}{$\sigma$(Amp)} & \multicolumn{1}{c}{$\sigma$(Phase)/$\pi$} \\
 \hline\hline
 $C^+_{220}$ & 0.967 & 0.475 & 0.00022 & 0.000072 \\
 $C^+_{221}$ & 4.20 & -0.199 & 0.0072 & 0.00055 \\
 $C^+_{222}$ & 12.3 & 0.959 & 0.082 & 0.0021 \\
 $C^+_{223}$ & 31.1 & 0.00553 & 0.44 & 0.0045 \\
 $C^+_{224}$ & 60.7 & 0.984 & 1.2 & 0.0063 \\
 $C^+_{225}$ & 73.0 & -0.0554 & 1.7 & 0.0072 \\
 $C^+_{226}$ & 47.1 & 0.914 & 1.1 & 0.0077 \\
 $C^+_{227}$ & 12.7 & -0.0936 & 0.31 & 0.0077 \\
 \hline
 $C^+_{320}$ & 0.0629 & -0.715 & 0.00022 & 0.0011 \\
 $C^+_{321}$ & 2.81 & 0.443 & 0.0073 & 0.00082 \\
 $C^+_{322}$ & 28.5 & -0.482 & 0.078 & 0.00087 \\
 $C^+_{323}$ & 125. & 0.549 & 0.38 & 0.00098 \\
 $C^+_{324}$ & 287. & -0.442 & 0.99 & 0.0011 \\
 $C^+_{325}$ & 364. & 0.556 & 1.4 & 0.0012 \\
 $C^+_{326}$ & 241. & -0.452 & 1.0 & 0.0014 \\
 $C^+_{327}$ & 65.3 & 0.539 & 0.31 & 0.0015 \\
 \hline
 $C^+_{420}$ & 0.00111 & 0.238 & 0.00023 & 0.066 \\
 $C^+_{421}$ & 0.0754 & -0.830 & 0.0074 & 0.031 \\
 $C^+_{422}$ & 1.22 & 0.142 & 0.079 & 0.021 \\
 $C^+_{423}$ & 7.82 & -0.875 & 0.38 & 0.016 \\
 $C^+_{424}$ & 24.1 & 0.109 & 0.96 & 0.013 \\
 $C^+_{425}$ & 38.6 & -0.907 & 1.3 & 0.011 \\
 $C^+_{426}$ & 30.9 & 0.0768 & 0.93\ & 0.0096 \\
 $C^+_{427}$ & 9.78 & -0.939 & 0.27 & 0.0087 \\
\end{tabular}
\caption{\label{table:h722} The magnitude and phase of the QNM amplitudes from fitting $h$ with fitting case $\{22\mbox{+4}\}$ and the EV method with $N=7$ at $t_i=t_{\rm peak}$ and using SVD tolerance $\tau=0$.}
\end{table}

\begin{table}
\begin{tabular}{c|d{5}d{5}|d{7}d{6}}
 Mode & \multicolumn{1}{c}{Amplitude} & \multicolumn{1}{c}{Phase/$\pi$} & \multicolumn{1}{c}{$\sigma$(Amp)} & \multicolumn{1}{c}{$\sigma$(Phase)/$\pi$} \\
 \hline\hline
 $C^+_{220}$ & 0.305 & 0.380 & 0.000099 & 0.00010 \\
 $C^+_{221}$ & 1.50 & -0.475 & 0.0033 & 0.00069 \\
 $C^+_{222}$ & 5.07 & 0.543 & 0.037 & 0.0023 \\
 $C^+_{223}$ & 13.2 & -0.490 & 0.20 & 0.0047 \\
 $C^+_{224}$ & 23.3 & 0.460 & 0.54 & 0.0074 \\
 $C^+_{225}$ & 23.9 & -0.582 & 0.75 & 0.010 \\
 $C^+_{226}$ & 12.7 & 0.401 & 0.51 & 0.013 \\
 $C^+_{227}$ & 2.84 & -0.573 & 0.14 & 0.016 \\
 \hline
 $C^+_{320}$ & 0.0181 & -0.793 & 0.00010 & 0.0018 \\
 $C^+_{321}$ & 0.773 & 0.368 & 0.0033 & 0.0014 \\
 $C^+_{322}$ & 7.65 & -0.564 & 0.035 & 0.0015 \\
 $C^+_{323}$ & 33.2 & 0.462 & 0.17 & 0.0017 \\
 $C^+_{324}$ & 76.3 & -0.533 & 0.45 & 0.0019 \\
 $C^+_{325}$ & 96.6 & 0.462 & 0.64 & 0.0021 \\
 $C^+_{326}$ & 64.0 & -0.547 & 0.47 & 0.0023 \\
 $C^+_{327}$ & 17.3 & 0.442 & 0.14 & 0.0025 \\
 \hline
 $C^+_{420}$ & 0.000322 & 0.0969 & 0.00010 & 0.10 \\
 $C^+_{421}$ & 0.0228 & -0.950 & 0.0034 & 0.047 \\
 $C^+_{422}$ & 0.365 & 0.0368 & 0.036 & 0.031 \\
 $C^+_{423}$ & 2.29 & -0.974 & 0.17 & 0.024 \\
 $C^+_{424}$ & 6.94 & 0.0136 & 0.44 & 0.020 \\
 $C^+_{425}$ & 11.0 & 0.999 & 0.60 & 0.017 \\
 $C^+_{426}$ & 8.69 & -0.0169 & 0.42 & 0.015 \\
 $C^+_{427}$ & 2.73 & 0.967 & 0.12 & 0.014 \\
\end{tabular}
\caption{\label{table:Psi722} The magnitude and phase of the QNM amplitudes from fitting $\Psi_4$ with fitting case $\{22\mbox{+4}\}$ and the EV method with $N=7$ at $t_i=t_{\rm peak}$ and using SVD tolerance $\tau=0$.}
\end{table}

\begin{table}
\begin{tabular}{c|d{8}d{4}}
 Mode & \multicolumn{1}{c}{Amplitude} & \multicolumn{1}{c}{Phase/$\pi$} \\
 \hline\hline
 $C^+_{220}$ & 0.306 & 0.379 \\
 $C^+_{221}$ & 1.52 & -0.480 \\
 $C^+_{222}$ & 5.64 & 0.516 \\
 $C^+_{223}$ & 19.3 & -0.567 \\
 $C^+_{224}$ & 51.2 & 0.317 \\
 $C^+_{225}$ & 82.8 & -0.784 \\
 $C^+_{226}$ & 72.2 & 0.145 \\
 $C^+_{227}$ & 26.0 & -0.894 \\
 \hline
 $C^+_{320}$ & 0.0402 & -0.786 \\
 $C^+_{321}$ & 1.94 & 0.234 \\
 $C^+_{322}$ & 22.6 & -0.819 \\
 $C^+_{323}$ & 121. & 0.0992 \\
 $C^+_{324}$ & 348. & -0.987 \\
 $C^+_{325}$ & 564. & -0.0667 \\
 $C^+_{326}$ & 477. & 0.864 \\
 $C^+_{327}$ & 163. & -0.193 \\
 \hline
 $C^+_{420}$ & 0.00116 & 0.182 \\
 $C^+_{421}$ & 0.0828 & -0.998 \\
 $C^+_{422}$ & 1.48 & -0.134 \\
 $C^+_{423}$ & 10.8 & 0.751 \\
 $C^+_{424}$ & 39.1 & -0.353 \\
 $C^+_{425}$ & 75.1 & 0.554 \\
 $C^+_{426}$ & 72.9 & -0.527 \\
 $C^+_{427}$ & 28.1 & 0.402 \\
\end{tabular}
 \caption{\label{table:h7topsi722} The magnitude and phase of the QNM amplitudes for $\Psi_4$ from fitting $h$ with fitting case $\{22\mbox{+4}\}$ and the EV method with $N=7$ at $t_i=t_{\rm peak}$ and using SVD tolerance $\tau=0$.  The coefficients used to produce Table~\ref{table:h722} were converted to coefficients for $\Psi_4$ using Eq.~(\ref{eqn:htoPsi4}).  The results in this table can be directly compared to those in Table~\ref{table:Psi722}.}
\end{table}

\subsection{Multimode fitting of the $m=2$ data}\label{sec:multimodefit}

\begin{figure}
\includegraphics[width=\linewidth,clip]{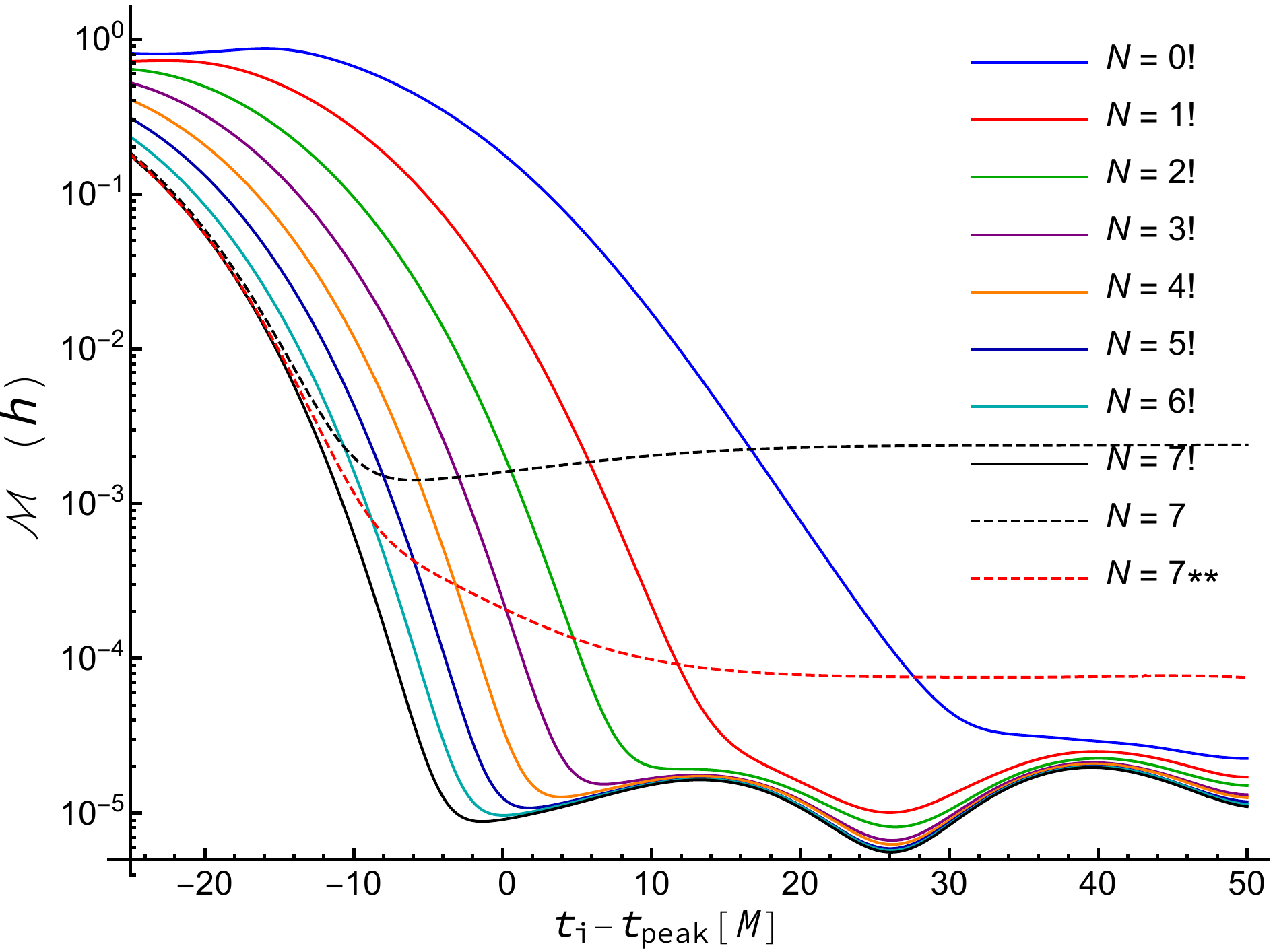}
 \caption{\label{fig:h7_C2232} Mismatch $\mathcal{M}$ plotted as a function of $t_i-t_{\rm peak}$ for $h$ using fitting case $\{22,32\}$, the EV method, and SVD tolerance $\tau=0$.  The number $N$ associated with each line denotes the maximum value of the overtone index $n$ used in the fitting mode set $\{{\rm QNM}\}$.}
\end{figure}

So far, we have only considered fitting a single mode of the numerical relativity waveform.  Now let us consider fitting the ringdown to the two most dominant $m=2$ modes: $C_{22}$ and $C_{32}$.  Figure~\ref{fig:h7_C2232} shows the mismatch from fitting these two modes of $h$ using the EV method and the $\omega^+_{22n}$ and $\omega^+_{32n}$ modes with $n=0,1,\ldots,7$.  This fitting combination will be referred to as case $\{22,32\}$.  In order to further distinguish the different models, we label the mismatch curves that include modes $\omega^+_{22n}$ and $\omega^+_{32n}$ with overtones $n$ up to $N$ as $N!$.  Included in Fig.~\ref{fig:h7_C2232} for comparison is the $N=7$ mismatch curve from Fig.~\ref{fig:h7_C22} and the $N=7\!\ast\!\ast$ curve from Fig.~\ref{fig:h7_C22b}.  

\begin{figure}
\includegraphics[width=\linewidth,clip]{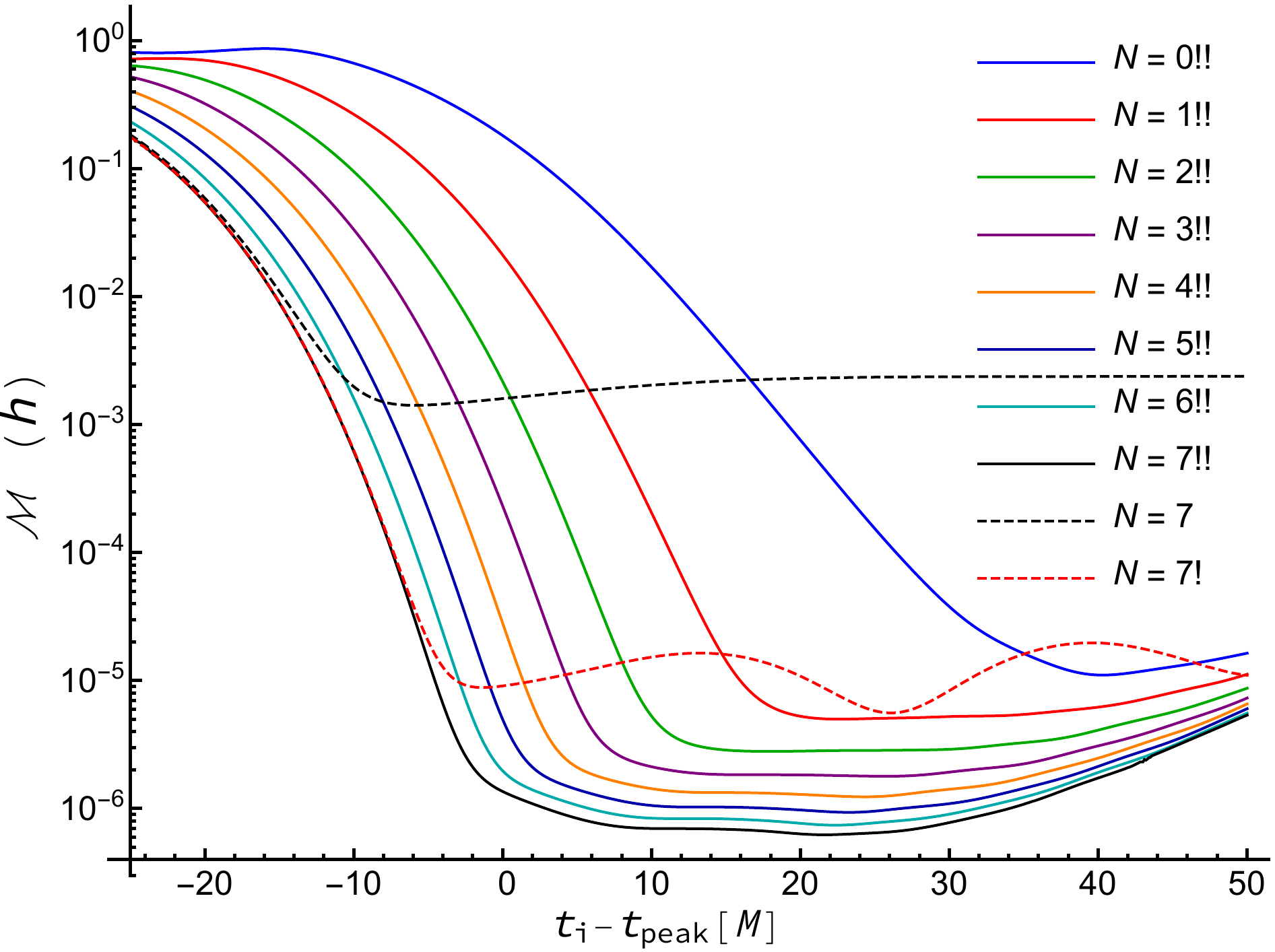}
 \caption{\label{fig:h7_C223242} Mismatch $\mathcal{M}$ plotted as a function of $t_i-t_{\rm peak}$ for $h$ using fitting case $\{22,32,42\}$, the EV method, and SVD tolerance $\tau=0$.  The number $N$ associated with each line denotes the maximum value of the overtone index $n$ used in the fitting mode set $\{{\rm QNM}\}$.}
\end{figure}

Clearly, including the information from the $C_{32}$ mode improves the mismatch by about an order of magnitude compared to fitting just $C_{22}$ against the same set of QNMs.  We can gain nearly another order of magnitude in the mismatch by fitting to the three most dominant $m=2$ modes:  $C_{22}$, $C_{32}$, and $C_{42}$.  Figure~\ref{fig:h7_C223242} shows the mismatch from fitting these three modes of $h$ using the EV method and the $\omega^+_{22n}$, $\omega^+_{32n}$, and $\omega^+_{42n}$ modes with $n=0,1,\ldots,7$.  This fitting combination will be referred to as case $\{22,32,42\}$.  In order to further distinguish the different models, we label the mismatch curves that include modes , $\omega^+_{32n}$, and $\omega^+_{42n}$ with overtones $n$ up to $N$ as $N!!$.  Included in Fig.~\ref{fig:h7_C223242} for comparison is the $N=7$ mismatch curve from Fig.~\ref{fig:h7_C22} and the $N=7!$ curve from Fig.~\ref{fig:h7_C2232}.

\begin{figure}
\includegraphics[width=\linewidth,clip]{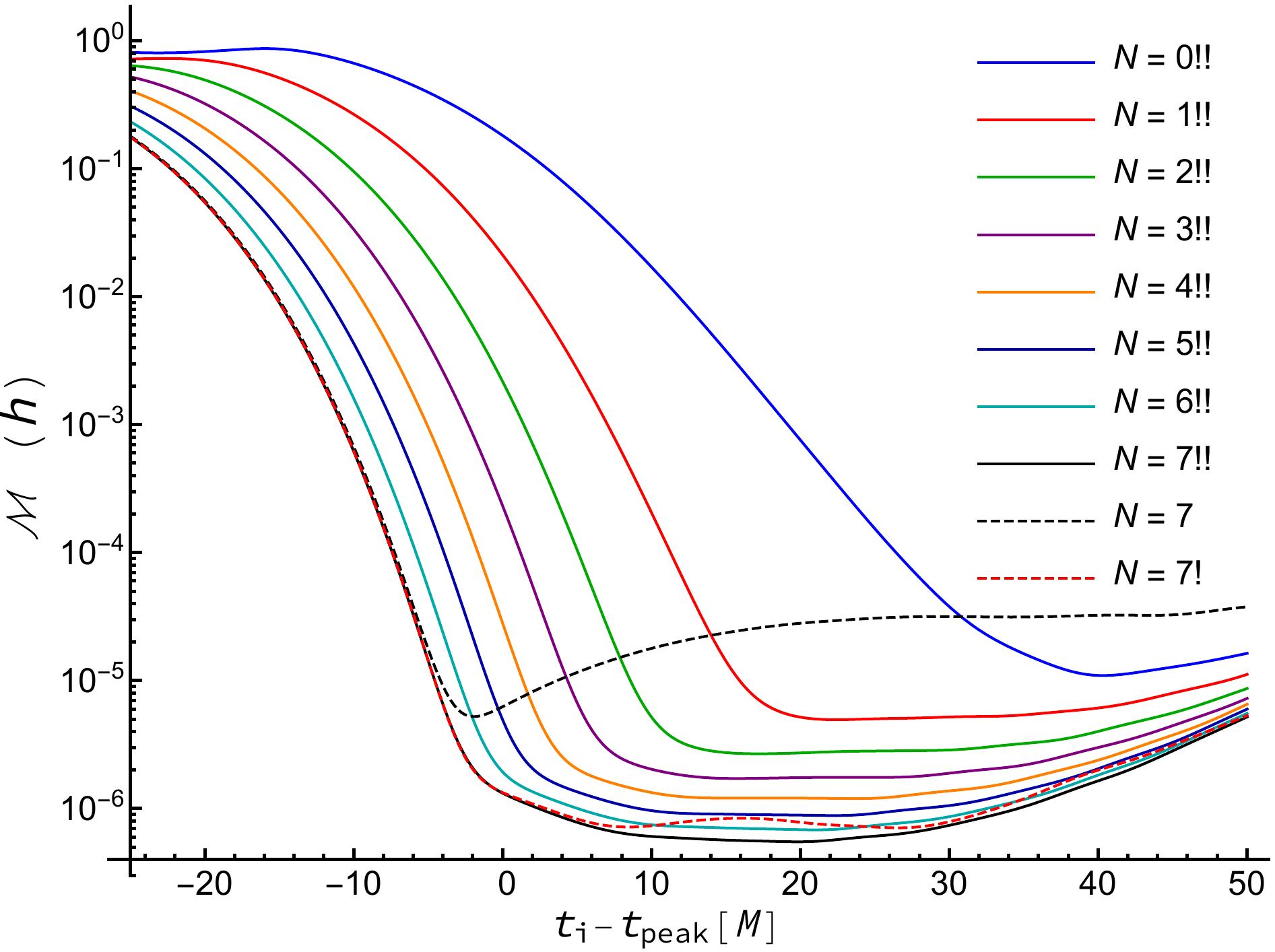}
 \caption{\label{fig:h7_C223242rhorestrict} Same as Fig.~\ref{fig:h7_C223242}, but with the mismatch recomputed using $\mathbb{B}_{\rm ml}$ as described in the text.}
\end{figure}

By including modes beyond $\ell=2$ in the set of simulation modes $\{{\rm NR}\}$, the mismatches computed directly by the EV method with $\mathbb{B}$ and from using $\mathbb{B}_{\rm ml}$ show increasingly less difference.  In fact, Fig.~\ref{fig:h7_C223242rhorestrict} which shows the mismatch from fitting case $\{22,32,42\}$ recomputed using $\mathbb{B}_{\rm ml}$ is very similar to Fig.~\ref{fig:h7_C223242}.
The primary differences are seen in the $N=7$ and $7!$ lines respectively associated with the $\{22\}$ and $\{22,32\}$ fitting cases.

\begin{figure}
\includegraphics[width=\linewidth,clip]{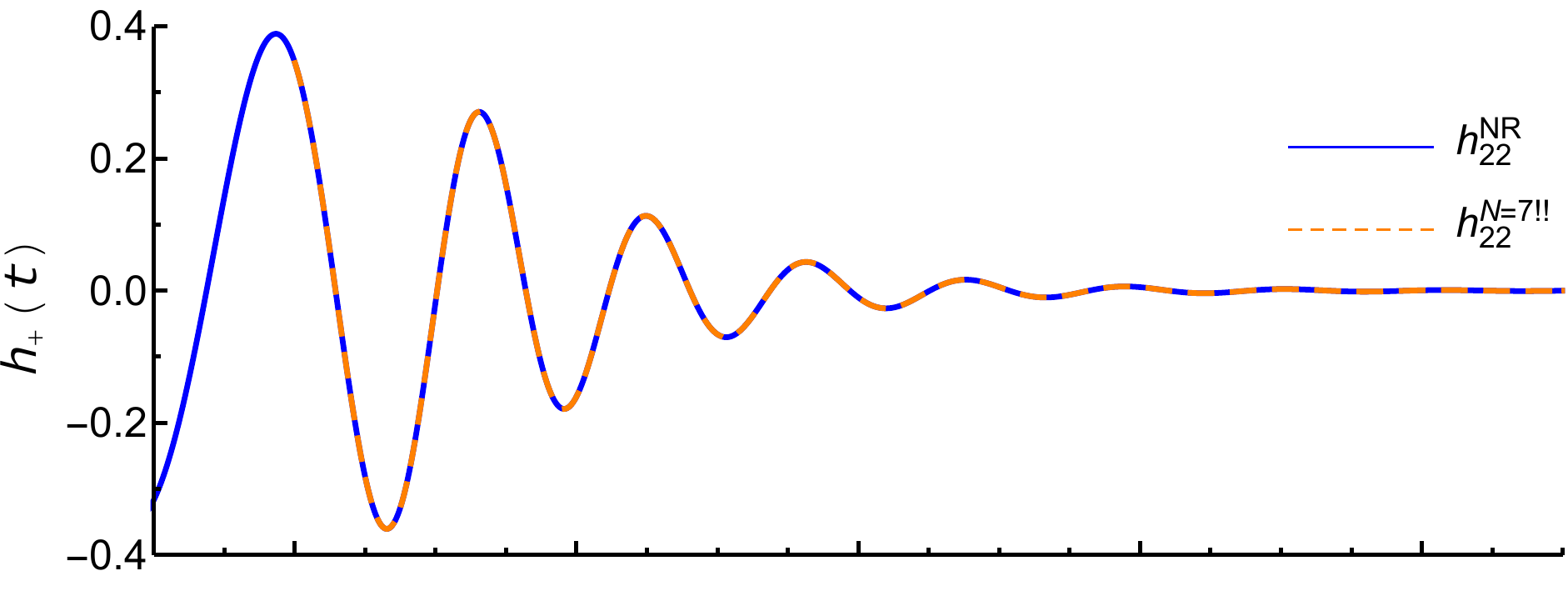}
\includegraphics[width=\linewidth,clip]{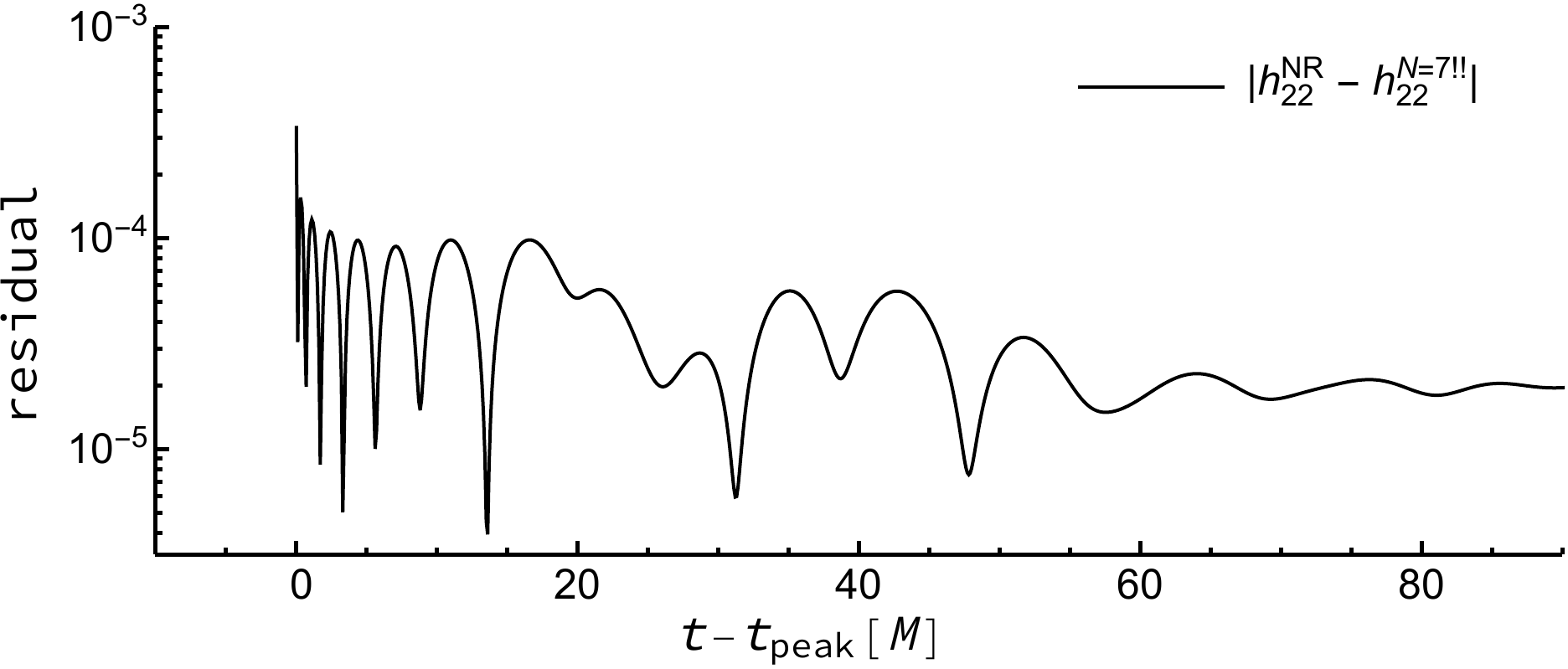}
 \caption{\label{fig:h22_o7C223242} Comparison of the $C_{22}$ waveform and fit using case $\{22,32,42\}$.  The upper panel shows the $(2,2)$ mode of $h_+$ from the numerical-relativity waveform and the $N=7$ version of its fit with $t_i=t_{\rm peak}$.  The lower panel shows the magnitude of the difference of the two complex signals.}
\end{figure}

\begin{figure}
\includegraphics[width=\linewidth,clip]{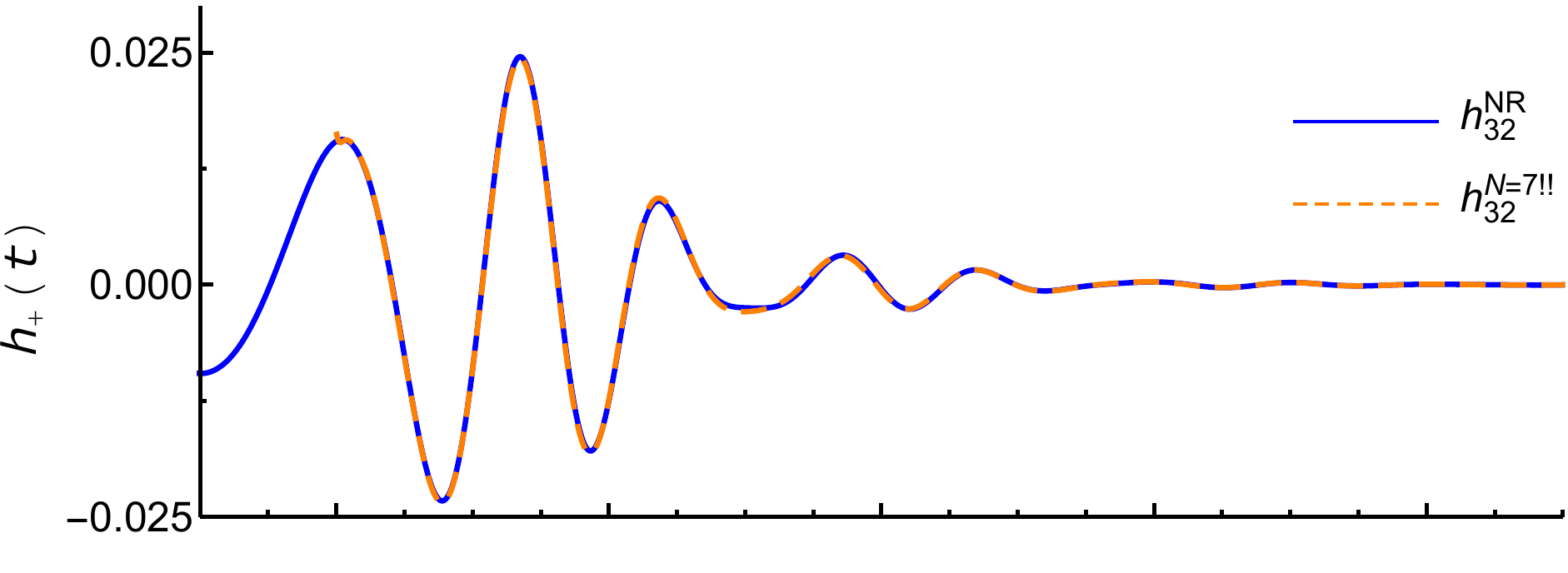}
\includegraphics[width=\linewidth,clip]{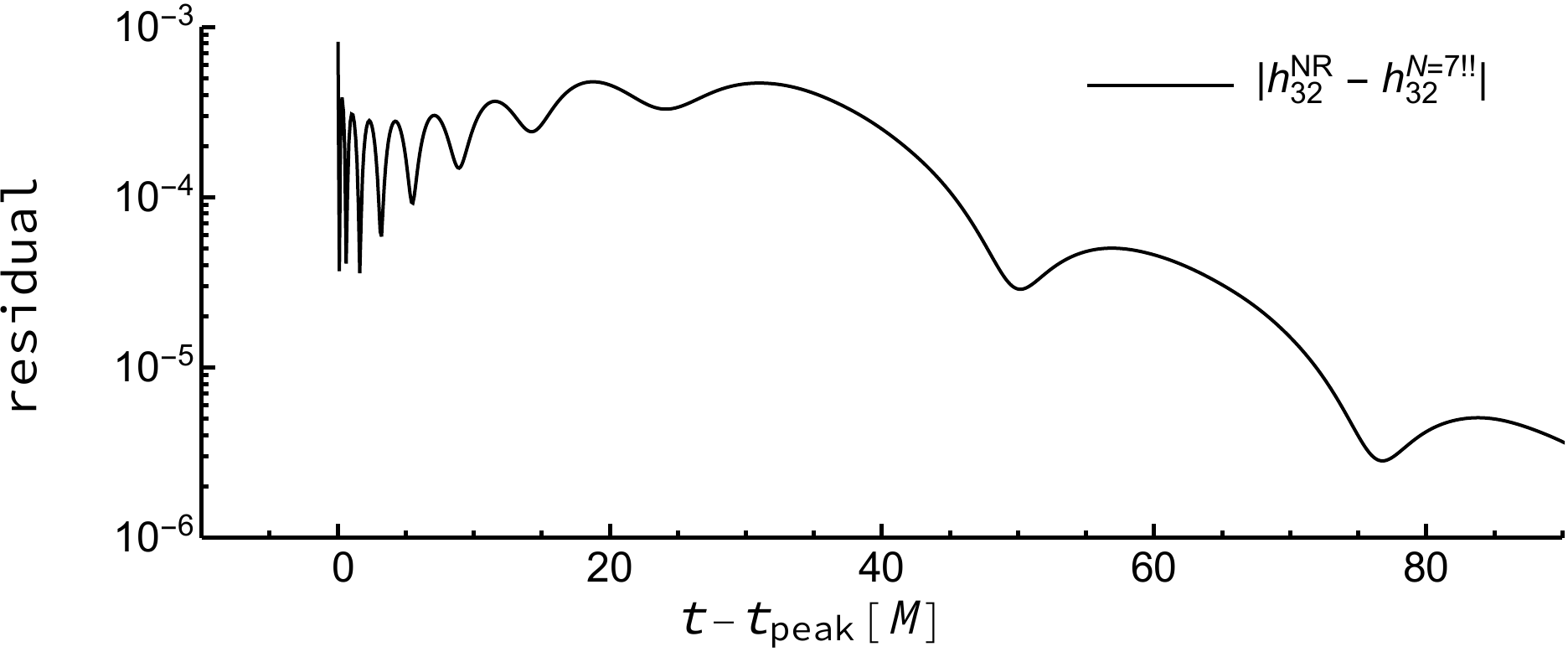}
 \caption{\label{fig:h32_o7C223242} Comparison of the $C_{32}$ waveform and fit using case $\{22,32,42\}$.  The upper panel shows the $(3,2)$ mode of $h_+$ from the numerical-relativity waveform and the $N=7$ version of its fit with $t_i=t_{\rm peak}$.  The lower panel shows the magnitude of the difference of the two complex signals.}
\end{figure}

\begin{figure}
\includegraphics[width=\linewidth,clip]{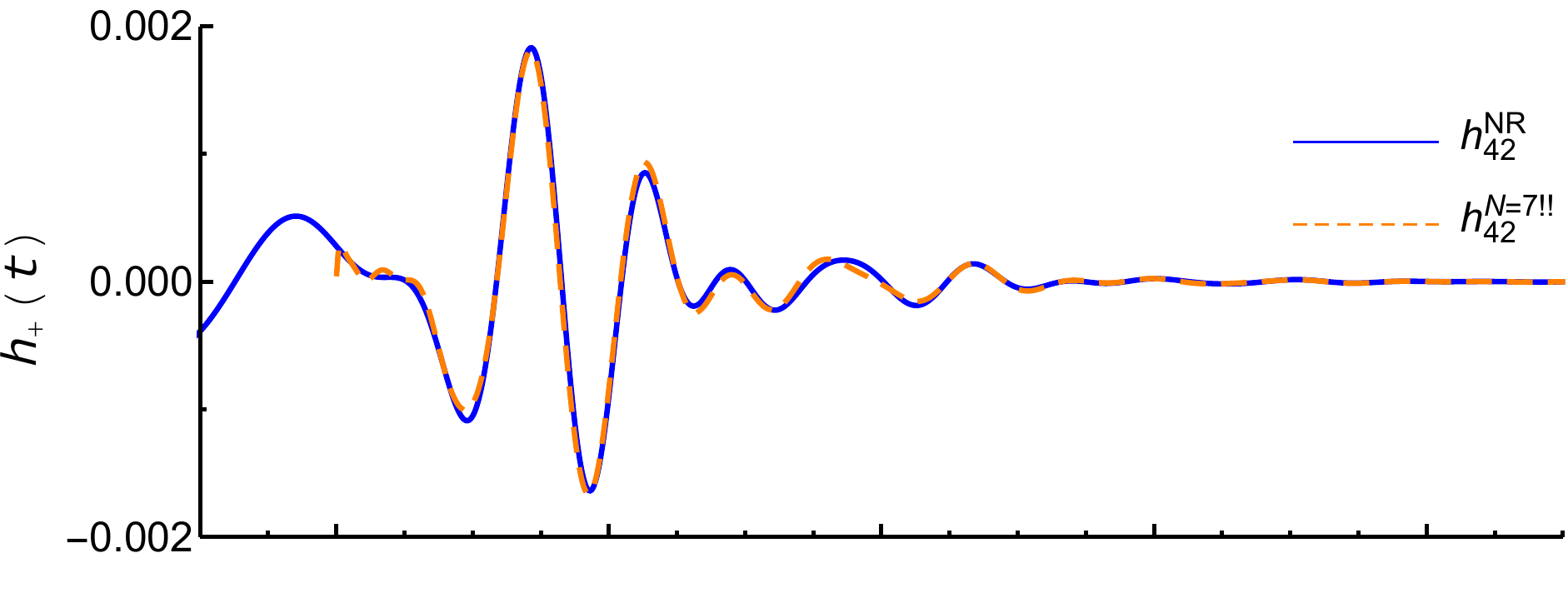}
\includegraphics[width=\linewidth,clip]{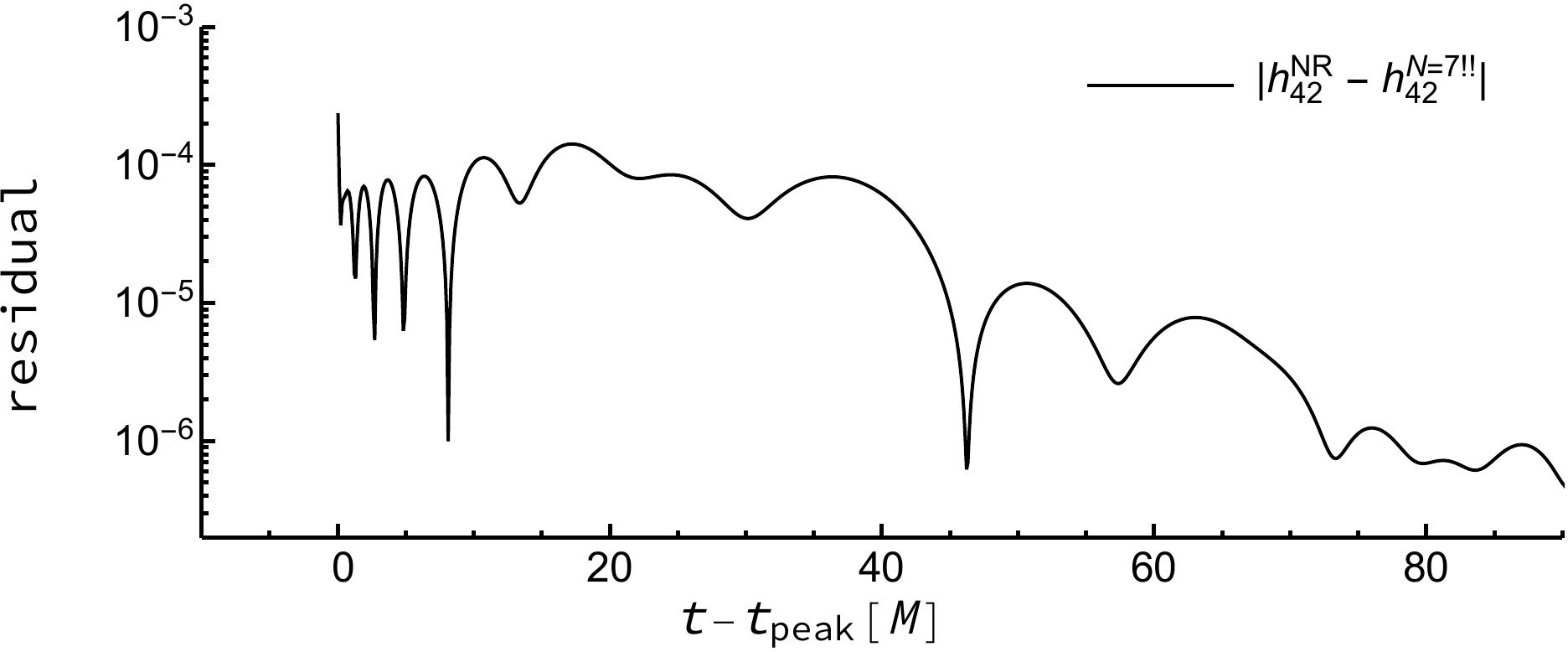}
 \caption{\label{fig:h42_o7C223242} Comparison of the $C_{42}$ waveform and fit using case $\{22,32,42\}$.  The upper panel shows the $(4,2)$ mode of $h_+$ from the numerical-relativity waveform and the $N=7$ version of its fit with $t_i=t_{\rm peak}$.  The lower panel shows the magnitude of the difference of the two complex signals.}
\end{figure}

In Figs.~\ref{fig:h22_o7C223242}--\ref{fig:h42_o7C223242} we compare the waveform $h$ with the $N=7$ version of fitting case $\{22,32,42\}$.  The upper plot in each of these figures compares the real part of the waveform $h_+$ with the fit, while the lower plot shows the magnitude of the difference of the complex data and the complex fit.  In each case, the values for the $C^+_{\ell2n}$ fit coefficients are taken at $t_i=t_{\rm peak}$.  Figure~\ref{fig:h22_o7C223242} shows this comparison for $\ell=2$.  That is for the $h_{22}$ data and fit.
Figure~\ref{fig:h32_o7C223242} shows the comparison for the $\ell=3$, $h_{32}$ data and fit.
Figure~\ref{fig:h42_o7C223242} shows the comparison for the $\ell=4$, $h_{42}$ data and fit.

\begin{figure}
\includegraphics[width=\linewidth,clip]{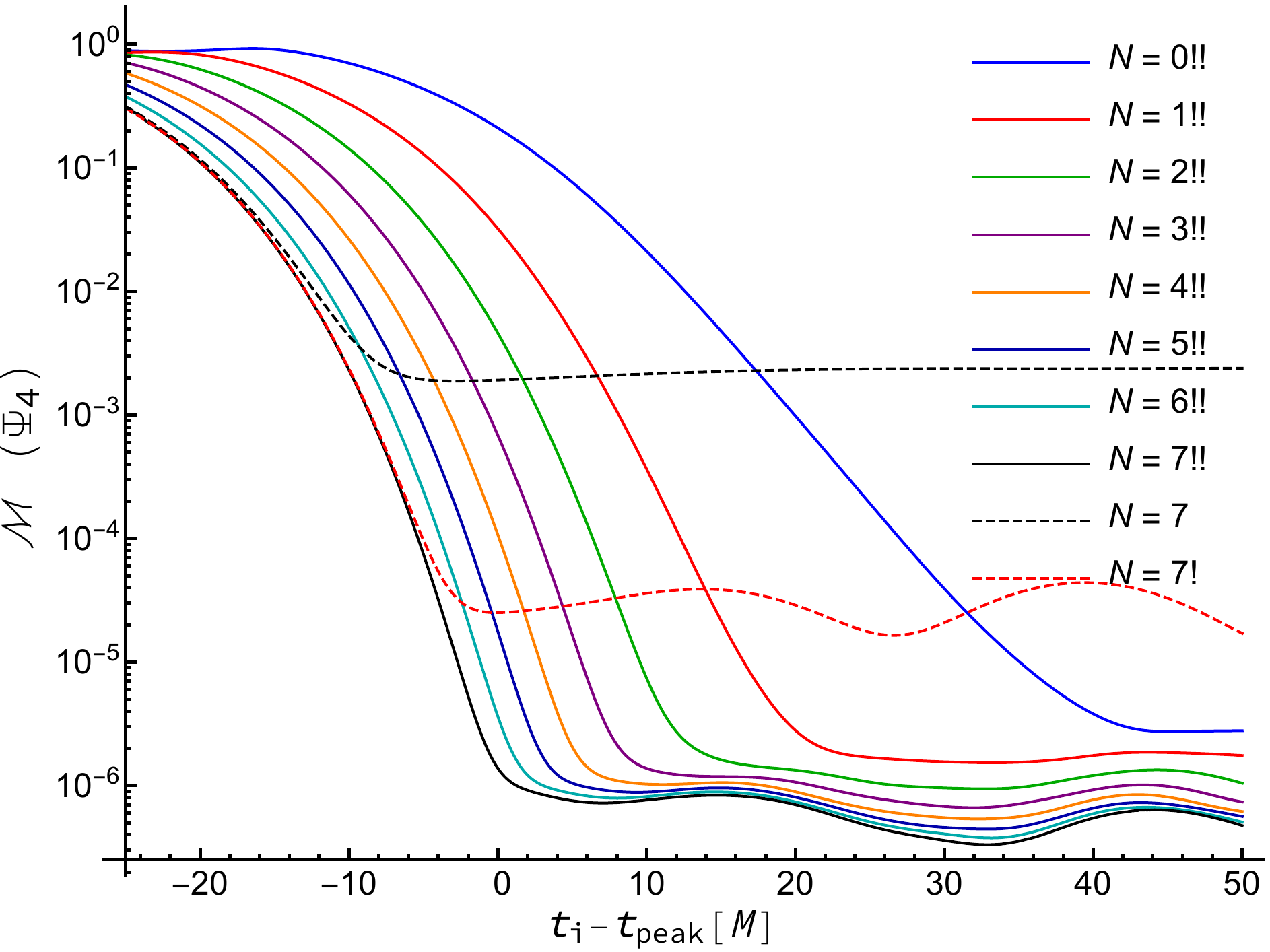}
 \caption{\label{fig:Psi7_C223242} Mismatch $\mathcal{M}$ plotted as a function of $t_i-t_{\rm peak}$ for $\Psi_4$ using fitting case $\{22,32,42\}$, the EV method, and SVD tolerance $\tau=0$.  The number $N$ associated with each line denotes the maximum value of the overtone index $n$ used in the fitting mode set $\{{\rm QNM}\}$.}
\end{figure}

Of course, the same multimode fitting can be performed using the waveform extracted through $\Psi_4$.  The full fitting of the $C_{22}$, $C_{32}$, and $C_{42}$ modes of the $\Psi_4$ data using the EV method and the $\omega^+_{22n}$, $\omega^+_{32n}$, and $\omega^+_{42n}$ modes with $n=0,1,\dots,7$ is shown in Fig.~\ref{fig:Psi7_C223242}.  The lines labeled as $N=7$ and $N=7!$ in this plot are from the $\Psi_4$ analogues of Figs.~\ref{fig:h7_C22} and \ref{fig:h7_C2232}, but these plots are not actually presented in this paper.

\begin{figure}
\includegraphics[width=\linewidth,clip]{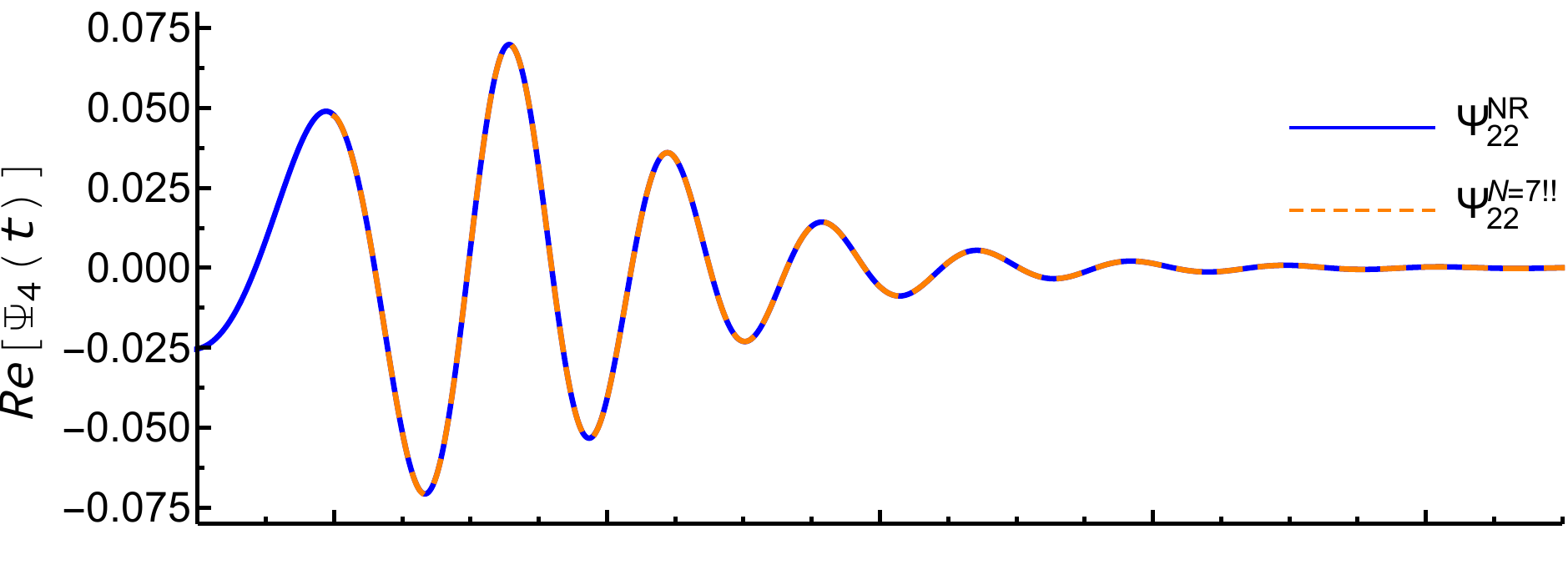}
\includegraphics[width=\linewidth,clip]{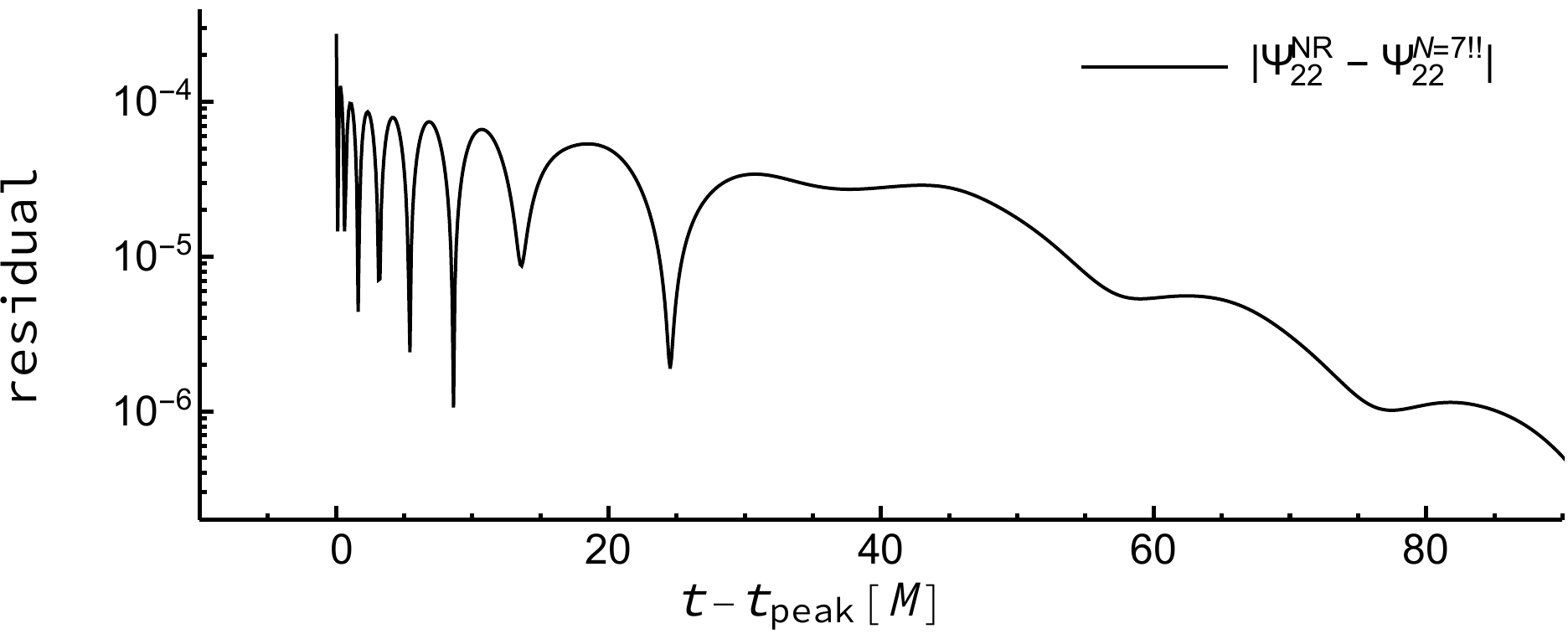}
 \caption{\label{fig:Psi22_o7C223242} Comparison of the $C_{22}$ waveform and fit using case $\{22,32,42\}$.  The upper panel shows the $(2,2)$ mode of $\rm{Re}[\Psi_4]=-\ddot{h}_+$ from the numerical-relativity waveform and the $N=7$ version of its fit with $t_i=t_{\rm peak}$.  The lower panel shows the magnitude of the difference of the two complex signals.}
\end{figure}

\begin{figure}
\includegraphics[width=\linewidth,clip]{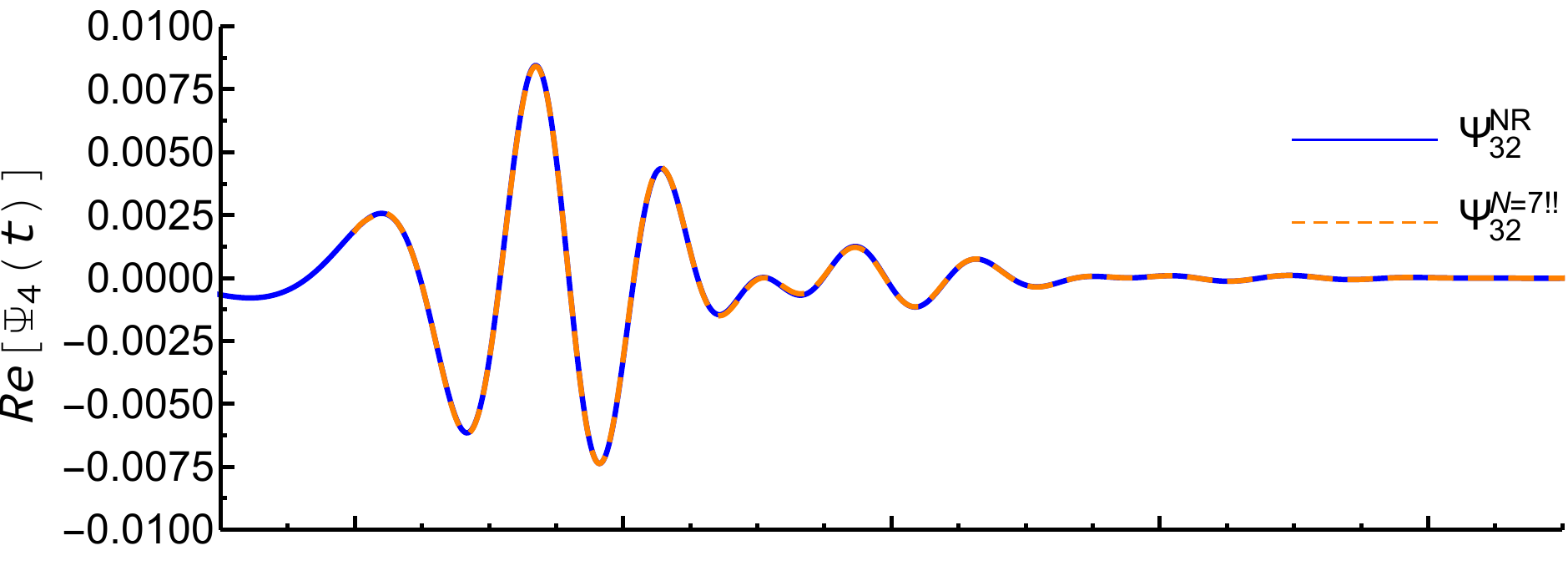}
\includegraphics[width=\linewidth,clip]{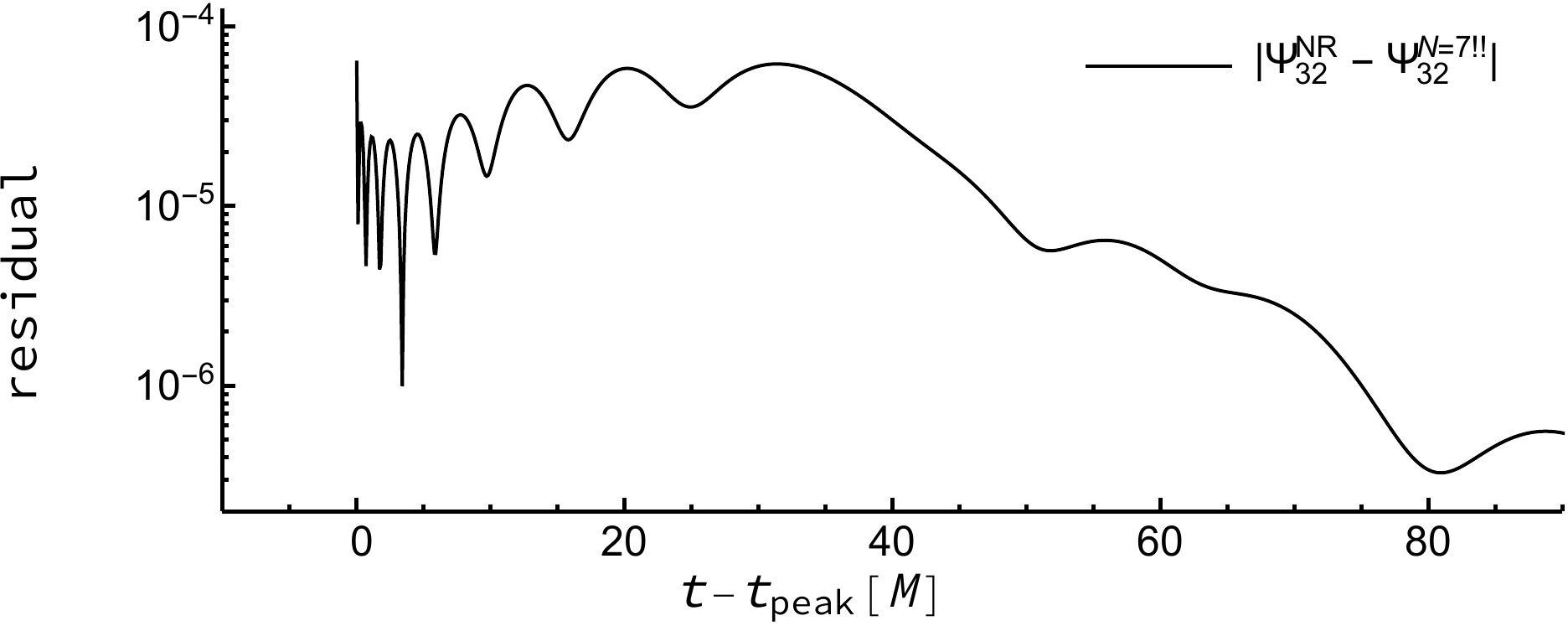}
 \caption{\label{fig:Psi32_o7C223242} Comparison of the $C_{32}$ waveform and fit using case $\{22,32,42\}$.  The upper panel shows the $(3,2)$ mode of $\rm{Re}[\Psi_4]=-\ddot{h}_+$ from the numerical-relativity waveform and the $N=7$ version of its fit with $t_i=t_{\rm peak}$.  The lower panel shows the magnitude of the difference of the two complex signals.}
\end{figure}

\begin{figure}
\includegraphics[width=\linewidth,clip]{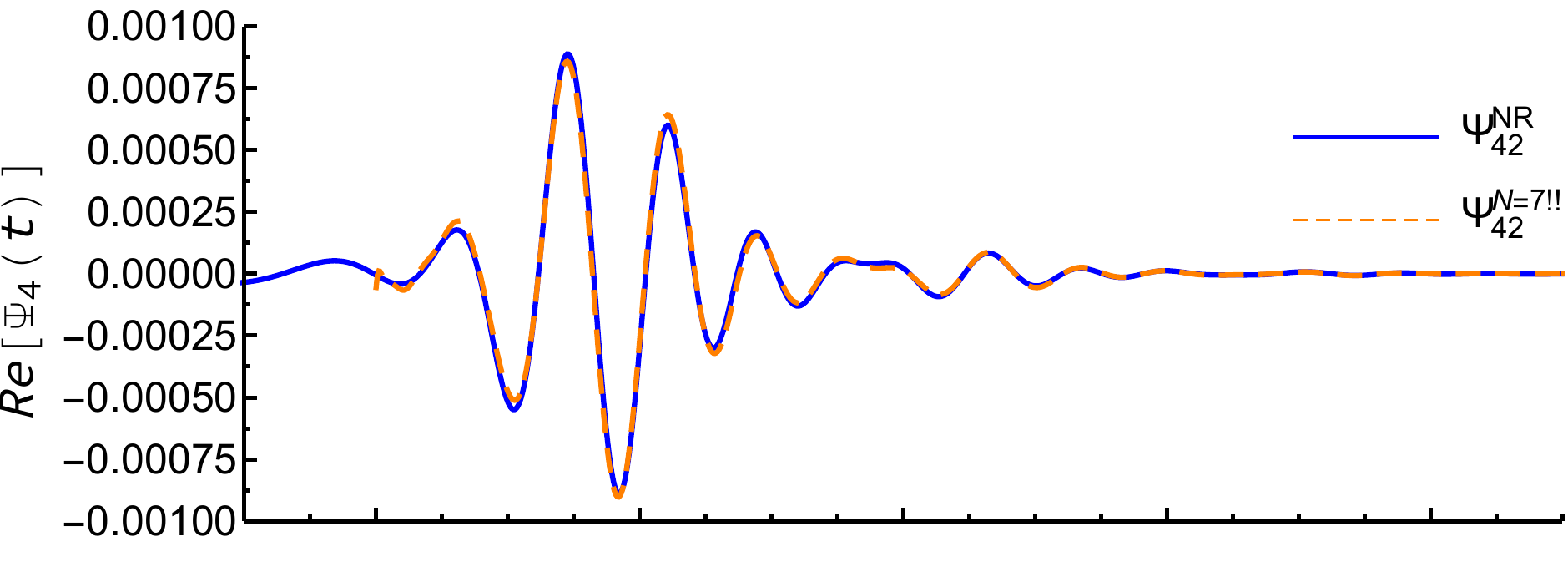}
\includegraphics[width=\linewidth,clip]{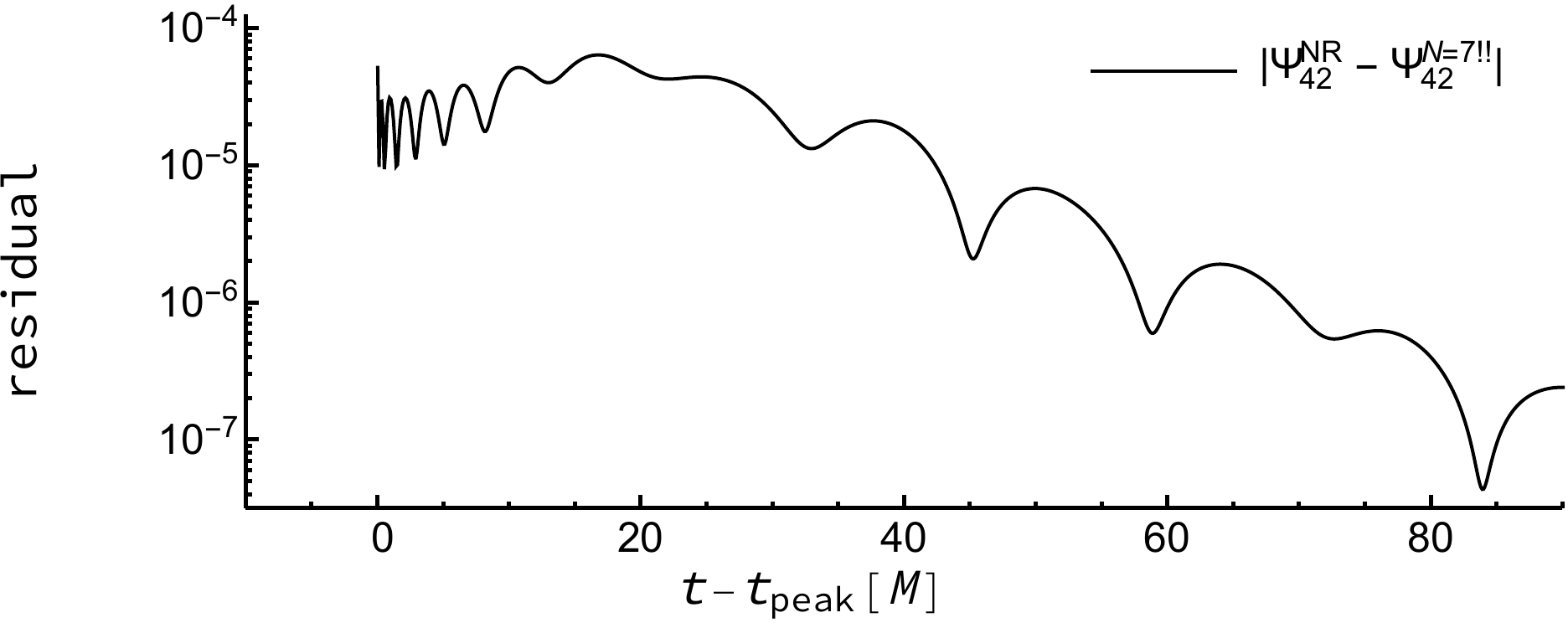}
 \caption{\label{fig:Psi42_o7C223242} Comparison of the $C_{42}$ waveform and fit using case $\{22,32,42\}$.  The upper panel shows the $(4,2)$ mode of $\rm{Re}[\Psi_4]=-\ddot{h}_+$ from the numerical-relativity waveform and the $N=7$ version of its fit with $t_i=t_{\rm peak}$.  The lower panel shows the magnitude of the difference of the two complex signals.}
\end{figure}

In Figs.~\ref{fig:Psi22_o7C223242}--\ref{fig:Psi42_o7C223242} we compare the waveform $\Psi_4$ with the fit using $N=7!!$.  The upper plot in each of these figures compares the real part of the waveform $\Psi_4$ with the fit, while the lower plot shows the magnitude of the difference of the complex data and the complex fit.  In each case, the values for the $C^+_{\ell2n}$ fit coefficients are taken at $t_i=t_{\rm peak}$.  Figure~\ref{fig:Psi22_o7C223242} shows this comparison for $\ell=2$.  That is for the $\Psi_{22}$ data and fit.
Figure~\ref{fig:Psi32_o7C223242} shows the comparison for the $\ell=3$, $\Psi_{32}$ data and fit.
Figure~\ref{fig:Psi42_o7C223242} shows the comparison for the $\ell=4$, $\Psi_{42}$ data and fit.

Finally, Tables~\ref{table:h7223242}--\ref{table:h7topsi7223242} show the amplitude and phase values for each fit coefficient obtained by fitting $C_{22}$, $C_{32}$, and $C_{42}$ from either $h$ or $\Psi_4$ with the $\omega^+_{22n}$, $\omega^+_{32n}$, and $\omega^+_{42n}$ modes with $n=1,0,\ldots,7$.  The values are for $t_i=t_{\rm peak}$.
\begin{table}
\begin{tabular}{c|d{6}d{6}|d{6}d{7}}
 Mode & \multicolumn{1}{c}{Amplitude} & \multicolumn{1}{c}{Phase/$\pi$} & \multicolumn{1}{c}{$\sigma$(Amp)} & \multicolumn{1}{c}{$\sigma$(Phase)/$\pi$} \\
 \hline\hline
 $C^+_{220}$ & 0.971 & 0.474 & 0.00010 & 0.000033 \\
 $C^+_{221}$ & 4.16 & -0.202 & 0.0033 & 0.00026 \\
 $C^+_{222}$ & 11.6 & 0.945 & 0.037 & 0.0010 \\
 $C^+_{223}$ & 25.8 & -0.0328 & 0.20 & 0.0025 \\
 $C^+_{224}$ & 41.4 & 0.915 & 0.55 & 0.0042 \\
 $C^+_{225}$ & 40.9 & -0.158 & 0.76 & 0.0059 \\
 $C^+_{226}$ & 21.9 & 0.773 & 0.52 & 0.0076 \\
 $C^+_{227}$ & 4.93 & -0.269 & 0.14 & 0.0091 \\
 \hline
 $C^+_{320}$ & 0.0319 & -0.263 & 0.00010 & 0.0010 \\
 $C^+_{321}$ & 0.147 & -0.742 & 0.0033 & 0.0072 \\
 $C^+_{322}$ & 1.53 & 0.798 & 0.036 & 0.0075 \\
 $C^+_{323}$ & 8.22 & -0.0707 & 0.18 & 0.0068 \\
 $C^+_{324}$ & 20.5 & 0.978 & 0.46 & 0.0071 \\
 $C^+_{325}$ & 27.1 & -0.000977 & 0.65 & 0.0076 \\
 $C^+_{326}$ & 18.4 & -0.993 & 0.47 & 0.0082 \\
 $C^+_{327}$ & 5.08 & 0.00955 & 0.14& 0.0088 \\
 \hline
 $C^+_{420}$ & 0.00265 & -0.650 & 0.00010 & 0.013 \\
 $C^+_{421}$ & 0.0753 & 0.702 & 0.0034 & 0.014 \\
 $C^+_{422}$ & 0.709 & -0.176 & 0.036 & 0.016 \\
 $C^+_{423}$ & 3.01 & 0.866 & 0.18 & 0.019 \\
 $C^+_{424}$ & 6.77 & -0.128 & 0.44 & 0.021 \\
 $C^+_{425}$ & 8.45 & 0.860 & 0.61 & 0.023 \\
 $C^+_{426}$ & 5.54 & -0.162 & 0.43 & 0.025 \\
 $C^+_{427}$ & 1.50 & 0.810 & 0.12 & 0.026 \\
\end{tabular}
\caption{\label{table:h7223242} The magnitude and phase of the QNM amplitudes from fitting $h$ with fitting case $\{22,32,42\}$ with $N=7$ at $t_i=t_{\rm peak}$ and using SVD tolerance $\tau=0$.}
\end{table}
\begin{table}
\begin{tabular}{c|d{6}d{5}|d{7}d{7}}
 Mode & \multicolumn{1}{c}{Amplitude} & \multicolumn{1}{c}{Phase/$\pi$} & \multicolumn{1}{c}{$\sigma$(Amp)} & \multicolumn{1}{c}{$\sigma$(Phase)/$\pi$} \\
 \hline\hline
 $C^+_{220}$ & 0.306 & 0.378 & 0.000020 & 0.000020 \\
 $C^+_{221}$ & 1.46 & -0.480 & 0.00065 & 0.00014 \\
 $C^+_{222}$ & 4.72 & 0.535 & 0.0073 & 0.00049 \\
 $C^+_{223}$ & 11.9 & -0.513 & 0.039 & 0.0010 \\
 $C^+_{224}$ & 20.9 & 0.408 & 0.11 & 0.0016 \\
 $C^+_{225}$ & 21.7 & -0.672 & 0.15 & 0.0022 \\
 $C^+_{226}$ & 11.9 & 0.261 & 0.10 & 0.0027 \\
 $C^+_{227}$ & 2.70 & -0.776 & 0.028 & 0.0033 \\
 \hline
 $C^+_{320}$ & 0.0237 & -0.332 & 0.000020 & 0.00027 \\
 $C^+_{321}$ & 0.187 & 0.878 & 0.00065 & 0.0011 \\
 $C^+_{322}$ & 0.700 & 0.0147 & 0.0070 & 0.0032 \\
 $C^+_{323}$ & 1.54 & -0.878 & 0.034 & 0.0071 \\
 $C^+_{324}$ & 2.17 & 0.235 & 0.089 & 0.013 \\
 $C^+_{325}$ & 2.08 & -0.638 & 0.13 & 0.019 \\
 $C^+_{326}$ & 1.30 & 0.477 & 0.093 & 0.023 \\
 $C^+_{327}$ & 0.375 & -0.442 & 0.027 & 0.023 \\
 \hline
 $C^+_{420}$ & 0.00229 & -0.759 & 0.000020 & 0.0028 \\
 $C^+_{421}$ & 0.0419 & 0.595 & 0.00067 & 0.0051 \\
 $C^+_{422}$ & 0.346 & -0.232 & 0.0071 & 0.0065 \\
 $C^+_{423}$ & 1.39 & 0.851 & 0.034 & 0.0079 \\
 $C^+_{424}$ & 2.99 & -0.107 & 0.087 & 0.0092 \\
 $C^+_{425}$ & 3.55 & 0.913 & 0.12 & 0.011 \\
 $C^+_{426}$ & 2.21 & -0.0786 & 0.084 & 0.012 \\
 $C^+_{427}$ & 0.558 & 0.923 & 0.024 & 0.014 \\
\end{tabular}
\caption{\label{table:Psi7223242} The magnitude and phase of the QNM amplitudes from fitting $\Psi_4$ with fitting case $\{22,32,42\}$ with $N=7$ at $t_i=t_{\rm peak}$ and using SVD tolerance $\tau=0$.}
\end{table}
\begin{table}
\begin{tabular}{c|d{6}d{5}}
 Mode & \multicolumn{1}{c}{Amplitude} & \multicolumn{1}{c}{Phase/$\pi$} \\
 \hline\hline
 $C^+_{220}$ & 0.307 & 0.377 \\
 $C^+_{221}$ & 1.51 & -0.484 \\
 $C^+_{222}$ & 5.35 & 0.502 \\
 $C^+_{223}$ & 16.0 & -0.605 \\
 $C^+_{224}$ & 34.9 & 0.248 \\
 $C^+_{225}$ & 46.4 & -0.887 \\
 $C^+_{226}$ & 33.5 & 0.00429 \\
 $C^+_{227}$ & 10.1 & 0.930 \\
 \hline
 $C^+_{320}$ & 0.0204 & -0.334 \\
 $C^+_{321}$ & 0.101 & -0.951 \\
 $C^+_{322}$ & 1.21 & 0.461 \\
 $C^+_{323}$ & 7.93 & -0.521 \\
 $C^+_{324}$ & 24.8 & 0.434 \\
 $C^+_{325}$ & 41.9 & -0.623 \\
 $C^+_{326}$ & 36.4 & 0.323 \\
 $C^+_{327}$ & 12.7 & -0.723 \\
 \hline
 $C^+_{420}$ & 0.00277 & -0.707 \\
 $C^+_{421}$ & 0.0827 & 0.533 \\
 $C^+_{422}$ & 0.855 & -0.451 \\
 $C^+_{423}$ & 4.14 & 0.493 \\
 $C^+_{424}$ & 11.0 & -0.590 \\
 $C^+_{425}$ & 16.4 & 0.321 \\
 $C^+_{426}$ & 13.1 & -0.766 \\
 $C^+_{427}$ & 4.29 & 0.151 \\
\end{tabular}
 \caption{\label{table:h7topsi7223242} The magnitude and phase of the QNM amplitudes for $\Psi_4$ from fitting $h$ with fitting case $\{22,32,42\}$ with $N=7$ at $t_i=t_{\rm peak}$ and using SVD tolerance $\tau=0$.  The coefficients used to produce Table~\ref{table:h7223242} were converted to coefficients for $\Psi_4$ using Eq.~(\ref{eqn:htoPsi4}).  The results in this table can be directly compared to those in Table~\ref{table:Psi7223242}.}
\end{table}

\subsection{Nonlinear fitting of model parameters}

So far, we have considered only the linear problem of fitting for the complex expansion coefficients $C^\pm_{\ell{m}n}$ of the QNMs, assuming the QNMs correspond to a rotating black hole with known mass and angular momentum.  In the preceding sections, we fixed the remnant parameters to $\mathcal{R}=\mathcal{R}_{\rm NR}$ with the values obtained from direct measurements of the ADM mass and angular momentum of the remnant black hole resulting from the simulated collision of a black-hole binary.  Now we want to consider determining the value of these model parameters by minimizing the mismatch ${\mathcal M}$.

As mention in Sec.~\ref{sec:implementation}, the remnant parameters to which we can fit are   
\begin{enumerate}
	\item the dimensionless ratio of the mass of the final remnant black hole $M_f$ to the mass scale of the numerical simulation $M$,
	\begin{equation}\label{eqn:massratio}
		\delta\equiv\frac{M_f}{M},
	\end{equation}
	\item the dimensionless ratio of the magnitude of the remnant black hole's angular momentum $J_f$ to the square of the remnant mass,
	\begin{equation}
		\chi_f\equiv\frac{J_f}{M_f^2}=\frac{a}{M_f},
	\end{equation}
	\item and the inclination angle $\bar\beta$ of the angular momentum vector relative to the $z$-axis of the simulation coordinate system (see Fig.~\ref{fig:coordrot}).
\end{enumerate}
A fourth remnant parameter is the the rotation angle $\bar\alpha$ of the angular momentum vector relative to the $z$-axis of the simulation coordinate system (see Fig.~\ref{fig:coordrot}).  However, as mentioned previously, this parameter cannot be determined as it corresponds to a constant phase change for each of the complex expansion coefficients $C^\pm_{\ell{m}n}$.  So, we define the set of model parameters as the three-dimensional set
\begin{equation}
	\mathcal{P}_m = \left\{\delta,\chi_f,\bar\beta\right\}.
\end{equation}
Given a guess for the model parameters $\mathcal{P}_m$, we can set the remnant parameters as $\mathcal{R}=\left\{\delta,\chi_f,\bar\beta,0\right\}$ and compute $\rho_{\rm max}$ via Eq.~(\ref{eqn:rho2max}), or the mismatch $\mathcal{M}=1-\rho_{\rm max}$, as described in Sec.~\ref{sec:methods}.

\begin{figure}
\includegraphics[width=\linewidth,clip]{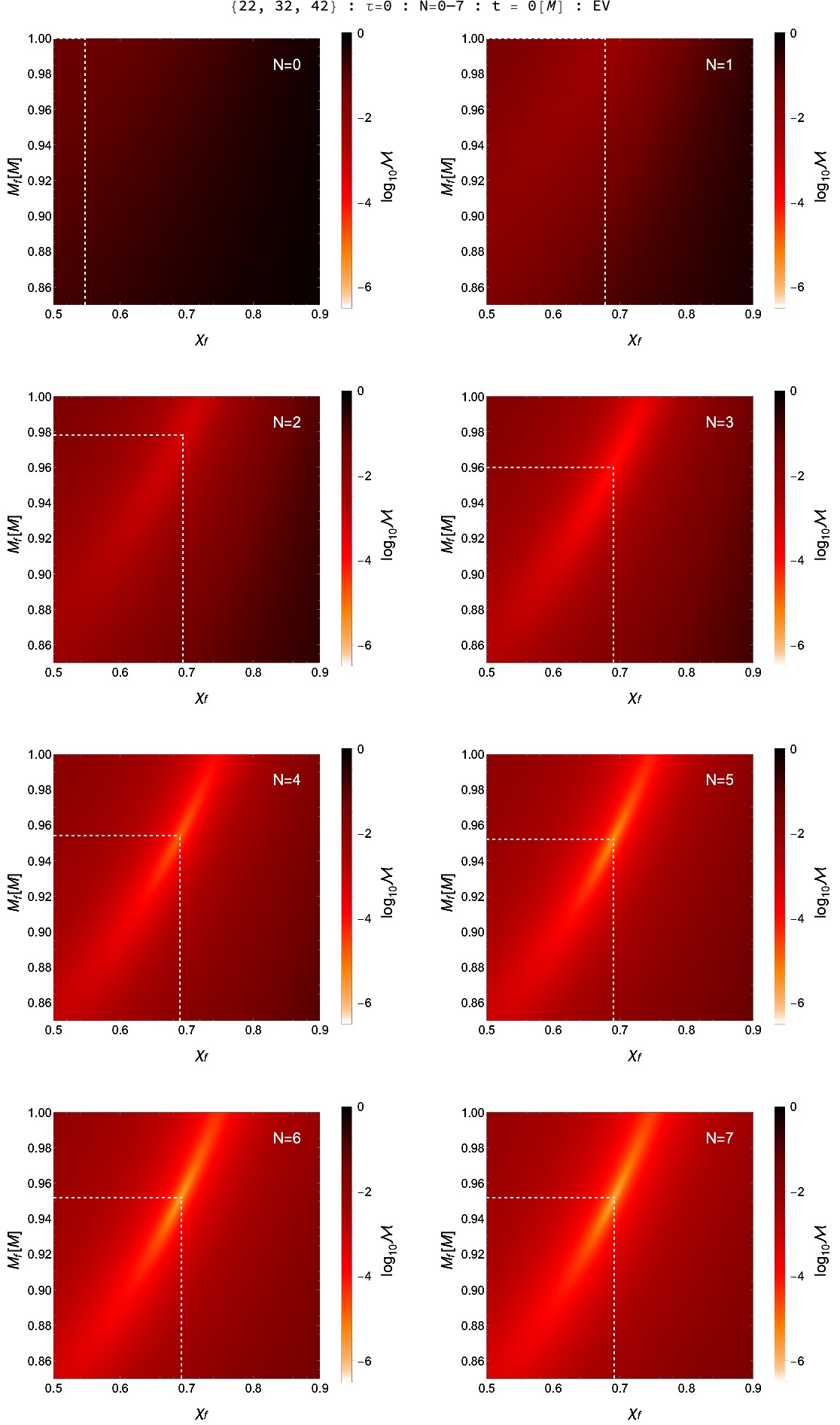}
 \caption{\label{fig:MMcp_h42_o7C223242_0}  Mismatches as a function of $\delta$ and $\chi_f$ at $t_i=t_{\rm peak}$ for fitting case $\{22,32,42\}$.  Fitting was performed using the EV method with an SVD tolerance of $\tau=0$.  The number $N$ associated with each panel denotes the maximum value of the overtone index $n$ used in the fitting mode set $\{{\rm QNM}\}$.  The dotted lines indicate the location of the minimum of the overlap in each panel.  The coordinates of the remnant black hole, as obtained directly from the simulation, are $(0.692,0.952)$.}
\end{figure}

\begin{figure}
\includegraphics[width=\linewidth,clip]{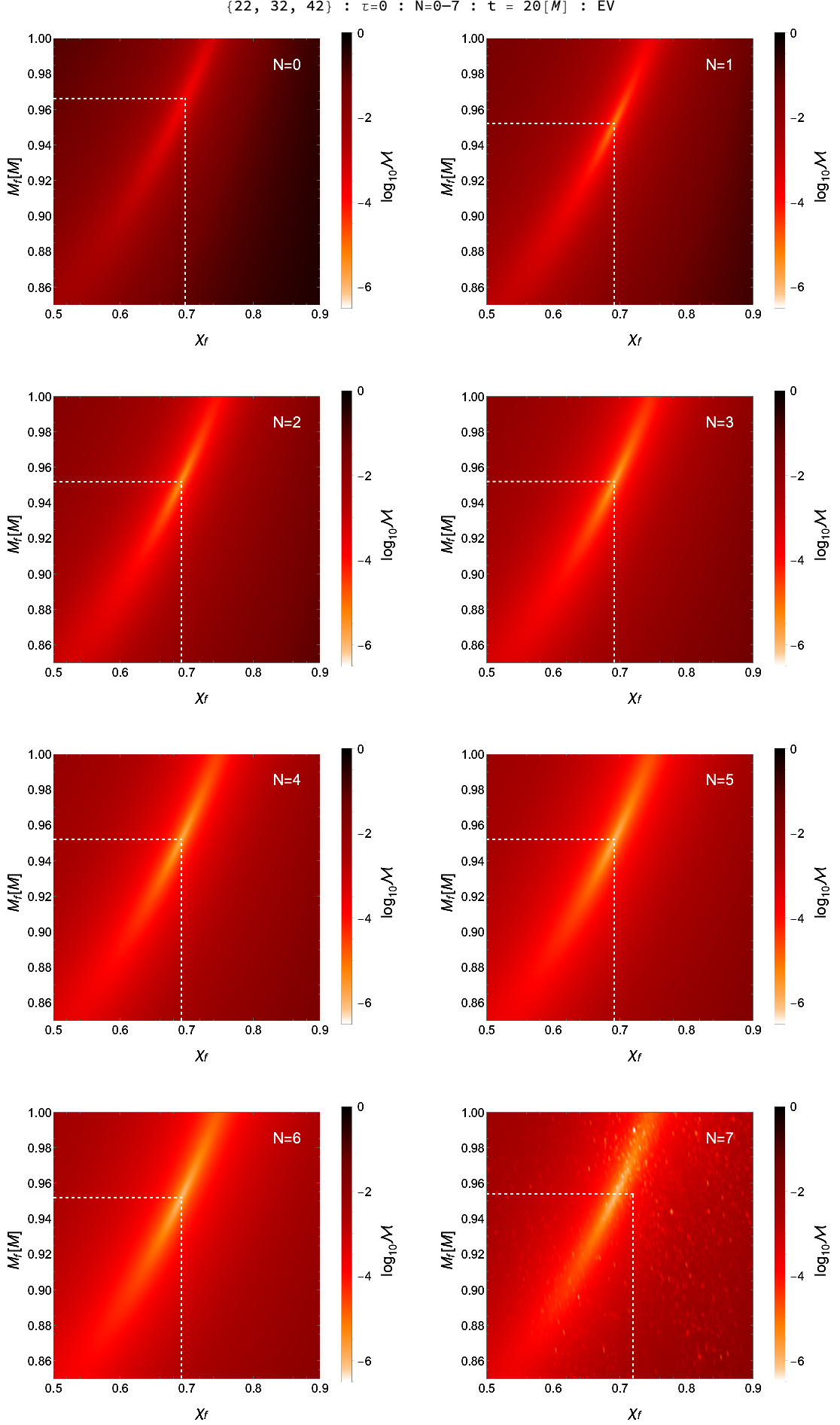}
 \caption{\label{fig:MMcp_h42_o7C223242_20}  The same as Fig.~\ref{fig:MMcp_h42_o7C223242_0}, but at $t_i-t_{\rm peak}=20M$.}
\end{figure}

\begin{figure}
\includegraphics[width=\linewidth,clip]{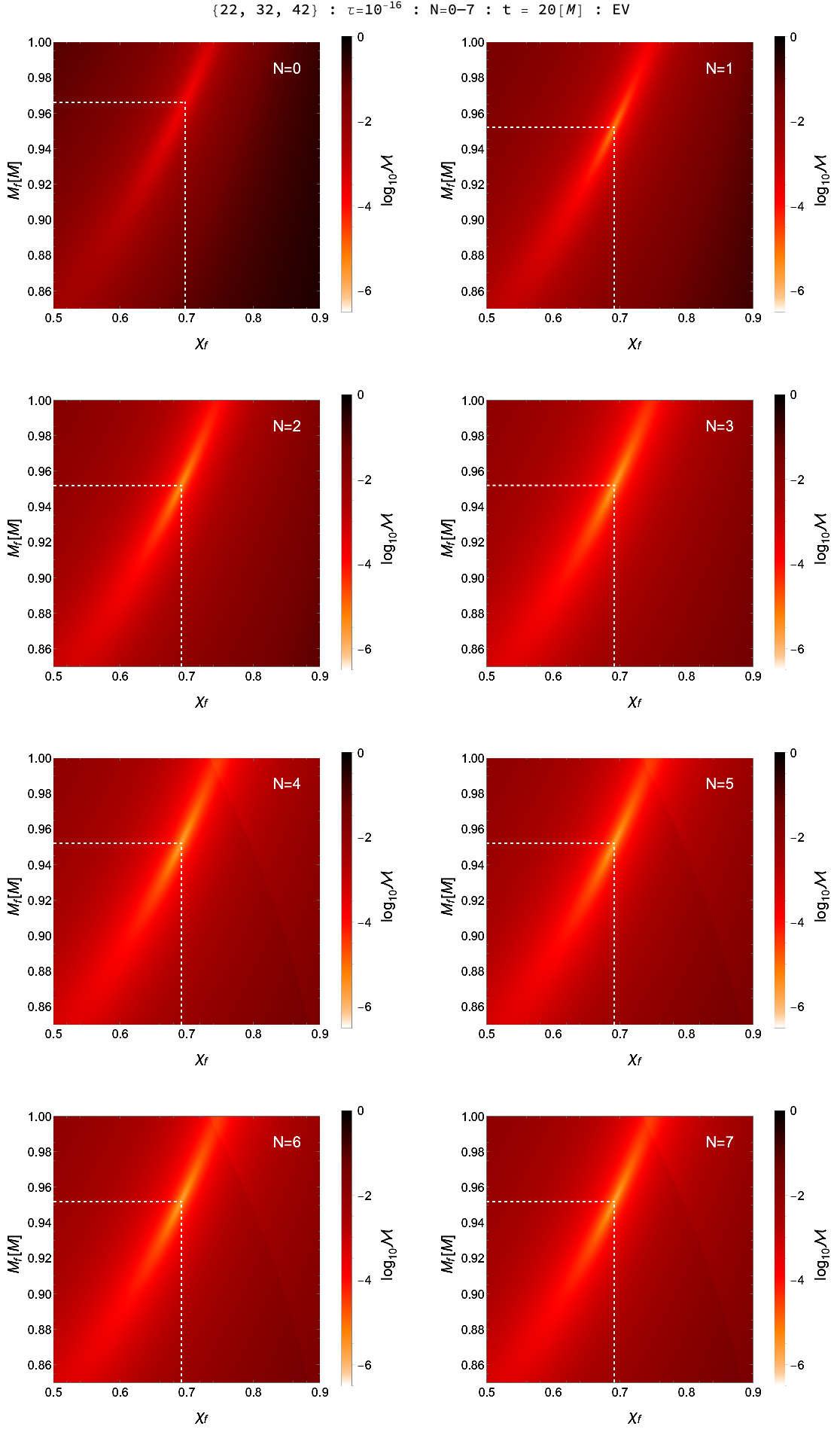}
 \caption{\label{fig:MMcp_h42_o7C223242_20_T16}  The same as Fig.~\ref{fig:MMcp_h42_o7C223242_20}, but with $\tau=10^{-16}$.}
\end{figure}

Here, we consider the reduced parameter space of $\mathcal{P}_m$ with $\bar\beta=0$, which is consistent with the data set SXS:BBH:0305.  Figure~\ref{fig:MMcp_h42_o7C223242_0} shows a set of color-density plots of the mismatch $\mathcal{M}$ as a function of the two remaining model parameters, $\delta$ and $\chi_f$ for $t_i=t_{\rm peak}$ and the fitting case $\{22,32,42\}$ computed using the EV method with overtones $n$ up to $N=1,2,\ldots,7$.
Each panel in Fig.~\ref{fig:MMcp_h42_o7C223242_0} show the effect of increasing the maximum number of overtones used in the fitting.  The most obvious feature is that the model parameters $\mathcal{P}_m$ that produce the minimum overlap $\mathcal{M}$ do not agree well with the remnant parameters $\mathcal{R}_{\rm NR}$ for small values of the maximum overtone number $N$.  This is not surprising giving Fig.~\ref{fig:h7_C223242} which shows how the mismatch decreases dramatically at $t_i=t_{\rm peak}$ as $N$ increases.  However, Fig.~\ref{fig:h7_C223242} does not illustrate the importance of including higher overtones in obtaining reasonable values for the model parameters, something that Fig.~\ref{fig:MMcp_h42_o7C223242_0} illustrates clearly.  Of course, the importance of overtones decreases as we move past $t_i=t_{\rm peak}$.  Figure~\ref{fig:MMcp_h42_o7C223242_20} shows the same information, but for $t_i-t_{\rm peak}=20M$.
Even the $N=1$ panel shows the minimum of the mismatch is located in reasonable agreement with the remnant parameters obtained directly from the simulation.  This figure also shows one of the potential problems of using high overtones.  In the $N=7$ panel, we see that there is a considerable amount of noise in the value of $\mathcal{M}$ and the location of the minimum in $\mathcal{M}$ is erroneously shifted.  The noise is due in part to roundoff error effects in computing $\mathcal{M}$, and in part to the numerical error intrinsic to the simulation data.  This noise can be suppressed to some extent by using extended numerical precision when computing $\mathcal{M}$, but it cannot eliminate all of the noise because of the intrinsic errors in the simulation data.  A better approach is to use the SVD pseudoinverse (see Sec.~\ref{sec:implementation}) and set a nonvanishing tolerance.  Figure~\ref{fig:MMcp_h42_o7C223242_20_T16} shows the same information as Fig.~\ref{fig:MMcp_h42_o7C223242_20}, but with the tolerance set to $\tau=10^{-16}$.
The noise seen in the $N=7$ panel of Fig.~\ref{fig:MMcp_h42_o7C223242_20} is effectively removed and the minimum of $\mathcal{M}$ is no longer erroneously shifted.  Unfortunately, there is a penalty for using a nonvanishing SVD tolerance that can be seen in the $N=4\text{--}7$ panels of Fig.~\ref{fig:MMcp_h42_o7C223242_20_T16}.  Clearly seen in these panels is a line of discontinuity in the value of $\mathcal{M}$.  This arises because of the discontinuous nature of the decision to effectively remove QNMs that are irrelevant to the fit.

\begin{figure}
\includegraphics[width=\linewidth,clip]{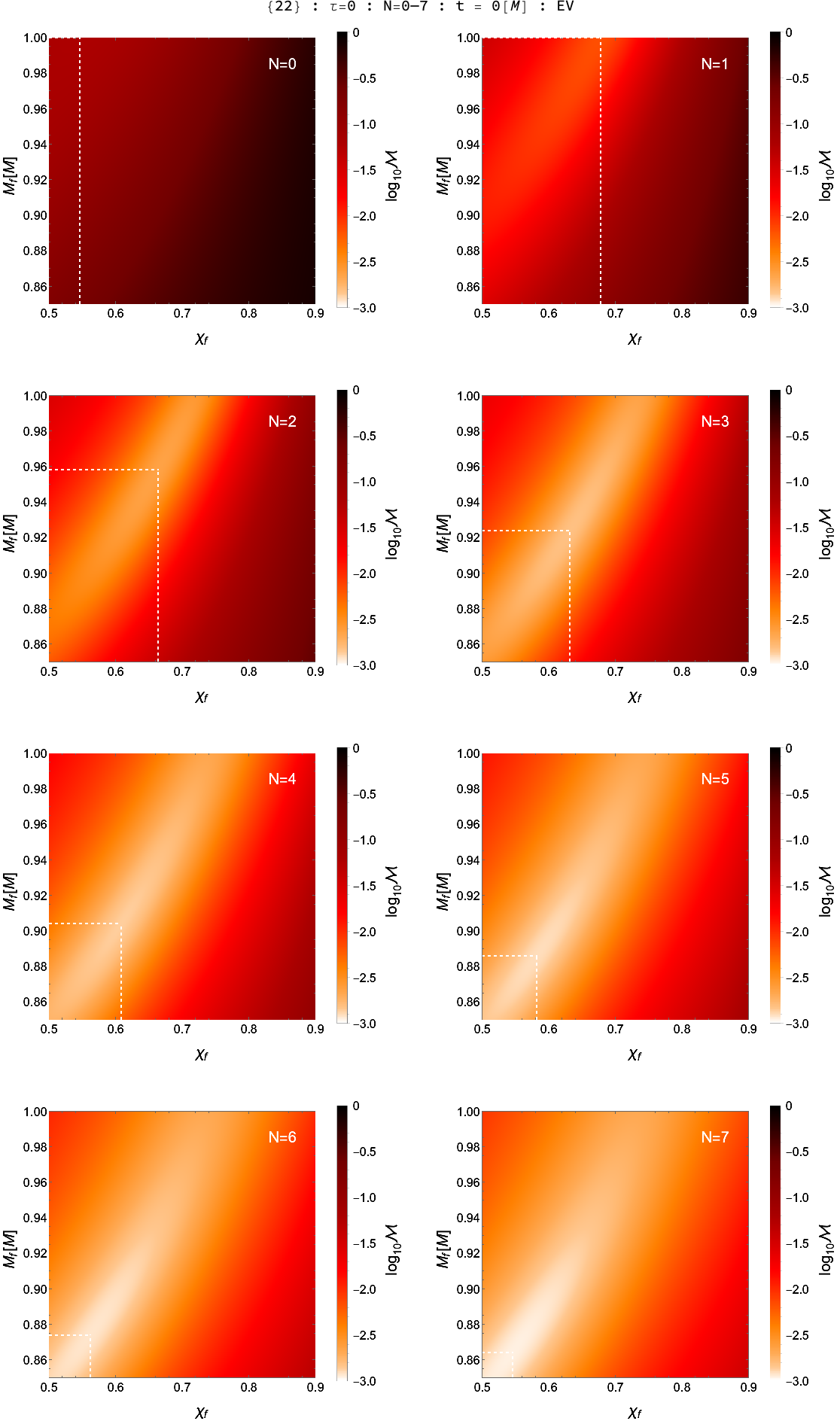}
 \caption{\label{fig:MMcp_h22_o7C22_0}  The same as Fig.~\ref{fig:MMcp_h42_o7C223242_0}, but for fitting case $\{22\}$.}
\end{figure}

The mismatch also depends strongly on the set of simulation modes $\{{\rm NR}\}$ that are used.  If we use fitting case $\{22,32\}$, the results are nearly identical, visually, to those of Fig.~\ref{fig:MMcp_h42_o7C223242_0}.  However if we reduce to fitting case $\{22\}$, as shown in Fig.~\ref{fig:MMcp_h22_o7C22_0}, we see a clear change in the mismatch.
In this case, the minima are much less localized, and the precise minima are in poor agreement with the remnant black-hole parameters $\mathcal{R}_{\rm NR}$.  It should be pointed out that this is in stark contrast to what is found in GIST using the restricted fitting model of Eq.~(\ref{eqn:GISTfitmodel}) in which the spheroidal-harmonic expansion coefficients are ignored (fitting case $\{22\mbox{-$\mathcal{A}$}\}$).  

In Figs.~\ref{fig:MMcp_h42_o7C223242_0}, \ref{fig:MMcp_h42_o7C223242_20}, \ref{fig:MMcp_h42_o7C223242_20_T16}, and \ref{fig:MMcp_h22_o7C22_0} we have computed the mismatch using Eq.~(\ref{eqn:rho2max}) with $\mathbb{B}^{-1}$ computed from $\mathbb{B}$.  Similar results were obtained using the Least-Squares Method with the mismatch computed using Eq.~(\ref{eqn:rho2explicit}) and $\mathbb{B}$.  The results will be somewhat different if the mismatch is computed using $\mathbb{B}_{\rm ml}$.
\begin{figure}
\includegraphics[width=\linewidth,clip]{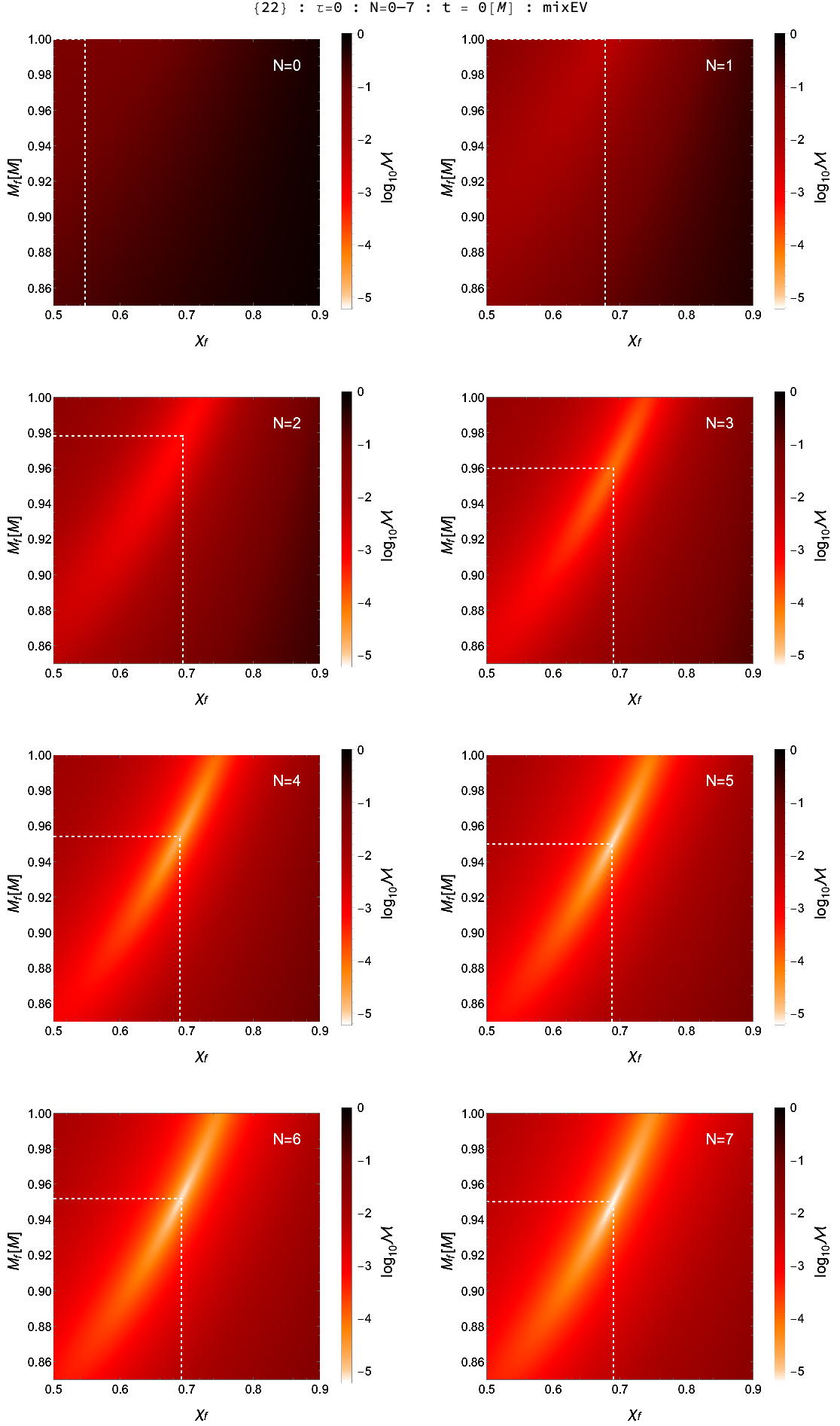}
 \caption{\label{fig:MMcp_h22_mix_o7C22_0}  The same as Fig.~\ref{fig:MMcp_h22_o7C22_0}, but with $\mathcal{M}$ computed using $\mathbb{B}_{\rm ml}$.}
\end{figure}
Figure~\ref{fig:MMcp_h22_mix_o7C22_0} shows the same set of color-density plots as shown in Fig.~\ref{fig:MMcp_h22_o7C22_0}, but with the mismatch computed using Eq.~(\ref{eqn:rho2explicit}) and $\mathbb{B}_{\rm ml}$.  In this case, the expansion coefficients $C^+_{22n}$ are unchanged.
\begin{figure}
\includegraphics[width=\linewidth,clip]{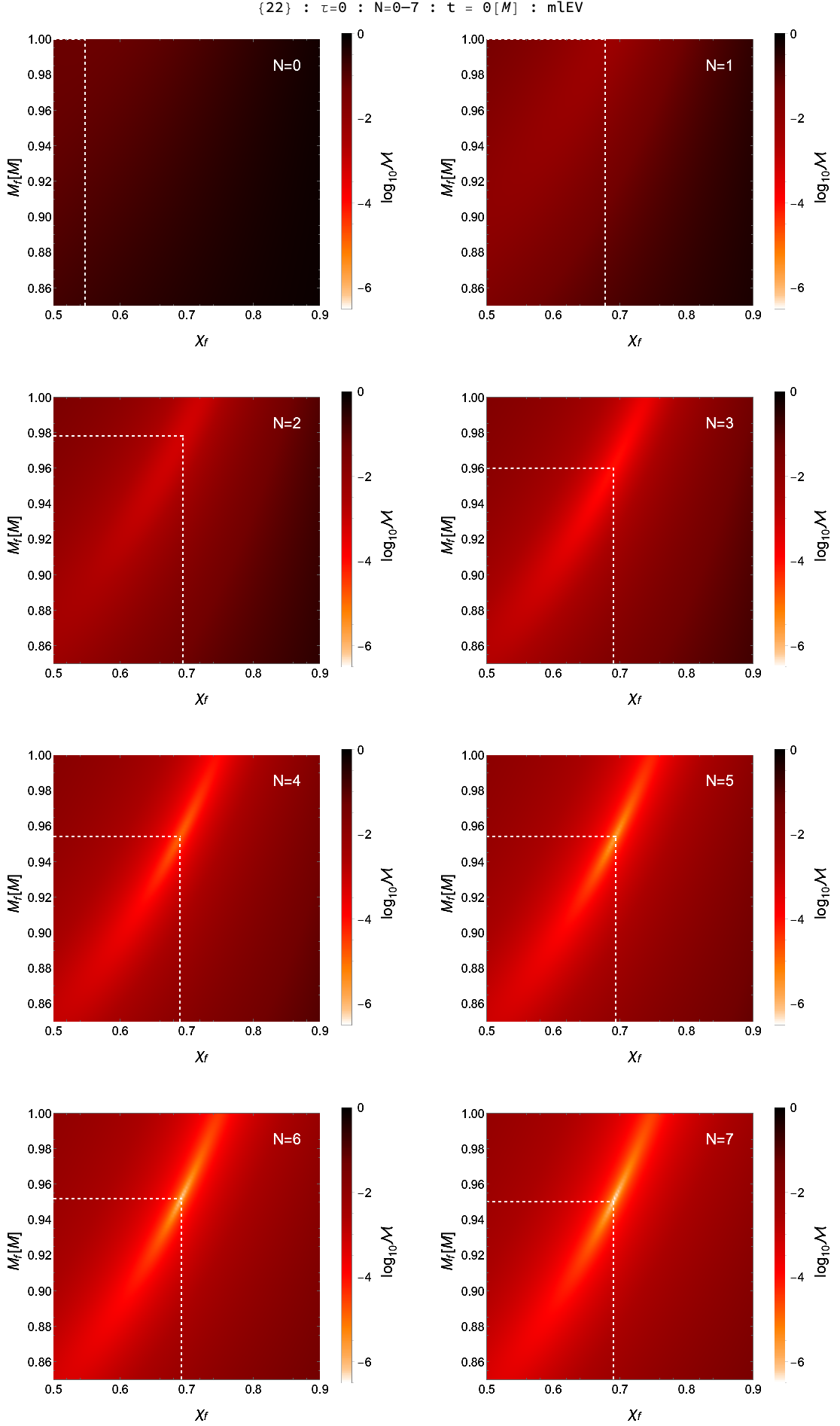}
 \caption{\label{fig:MMcp_h22_ml_o7C22_0}  The same as Fig.~\ref{fig:MMcp_h22_o7C22_0}, but fitting was performed using the mlEV method (same as GIST, or LS method with $\mathcal{M}$ computed using $\mathbb{B}_{\rm ml}$).}
\end{figure}
In contrast, Fig.~\ref{fig:MMcp_h22_ml_o7C22_0} shows the same set of color-density plots as shown in Fig.~\ref{fig:MMcp_h22_o7C22_0}, but in this case the mlEV method was used, altering both the expansion coefficients $C^+_{22n}$ and mismatch $\mathcal{M}$.  For the $\{22\}$ case, this approach yields the results found in GIST.

While the color-density plots provide very useful information, it is also very instructive to look at what we will call `fit-series' plots.  In all cases, these plots are based on the model parameters $\mathcal{P}_m$ for the minimum of the mismatch $\mathcal{M}$ for each value of $t_i-t_{\rm peak}$.
\begin{figure}
\includegraphics[width=\linewidth,clip]{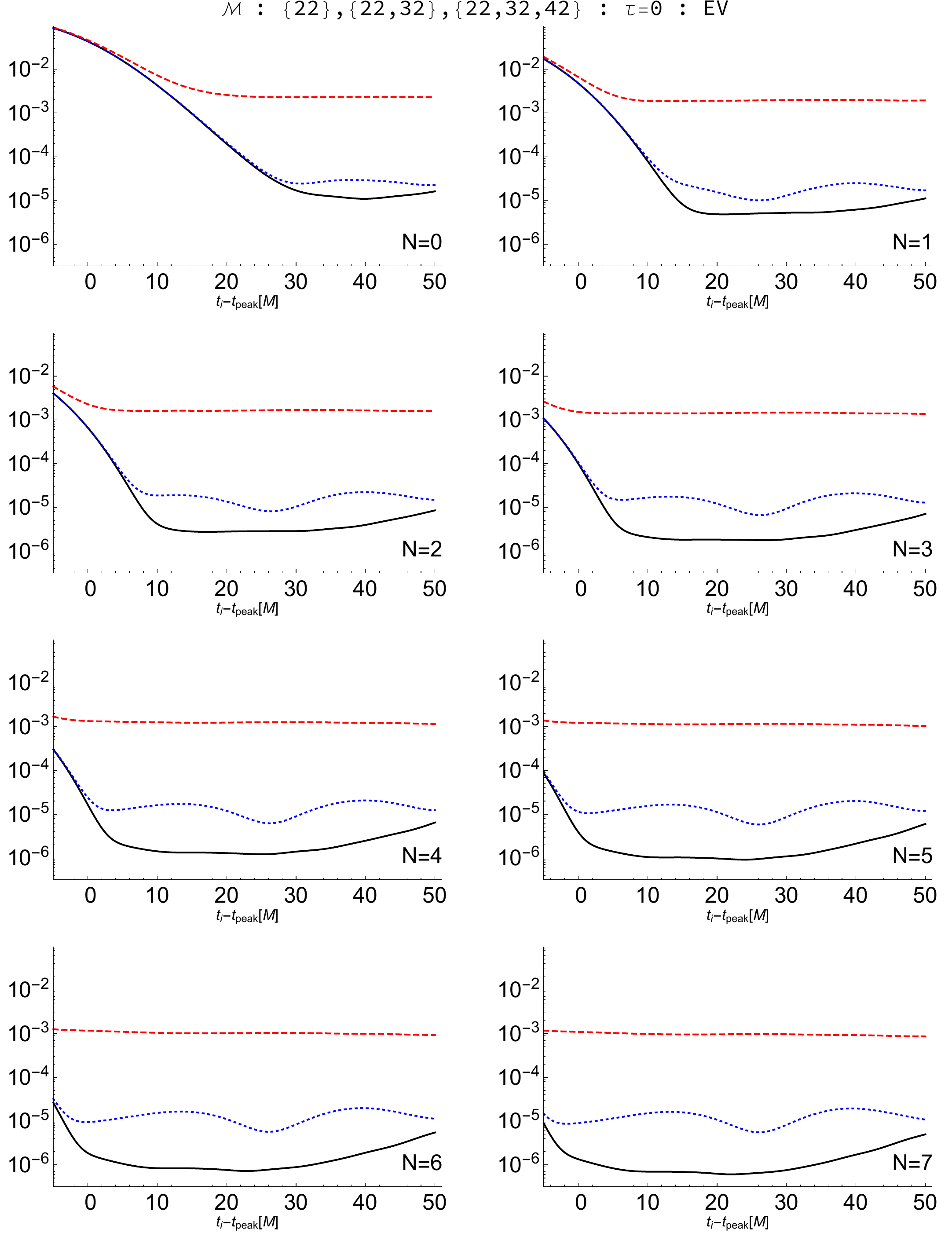}
 \caption{\label{fig:FitMM_h_CvsN}  The fit value of the mismatch $\mathcal{M}$ as a function of $t_i-t_{\rm peak}$.  Each panel compares the results when using fitting cases $\{22\}$ (dashed red line), $\{22,32\}$ (dotted blue line), and $\{22,32,42\}$ (solid black line).  The number $N$ associated with each panel denotes the maximum value of the overtone index $n$ used in each fitting case.  The overlaps are computed using the eigenvalue method with SVD tolerance $\tau=0$.}
\end{figure}
Figure~\ref{fig:FitMM_h_CvsN} compares the minimum mismatch $\mathcal{M}$ for the three fitting cases of $\{22\}$, $\{22,32\}$, and $\{22,32,42\}$.  Each panel in the figure shows a different value for $N$, the maximum overtone $n$ used in each fitting case.  As seen before, including the $C_{32}$ and $C_{42}$ data into the waveform being fit significantly reduces the mismatch compared to just using the $C_{22}$ data.  It is also clear that as more overtones are included in the set of fit QNMs, the range of values of $t_i-t_{\rm peak}$ over which the mismatch is very small extends closer to and then beyond $t_i=t_{\rm peak}$.

\begin{figure}
\includegraphics[width=\linewidth,clip]{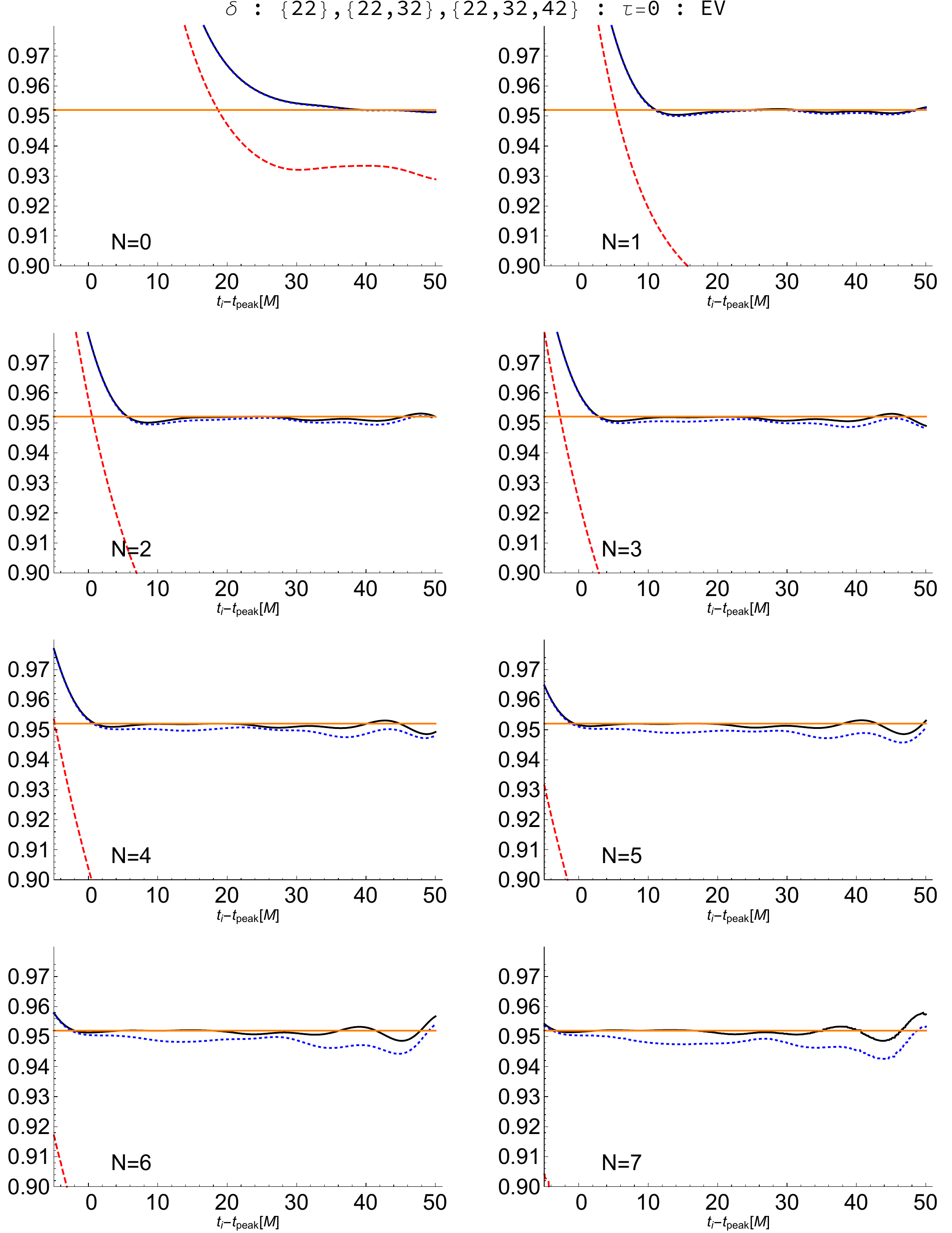}
 \caption{\label{fig:Fitdelta_h_CvsN}  The fit value of the mass ratio $\delta$ as a function of $t_i-t_{\rm peak}$.  Each panel compares the results when using fitting cases $\{22\}$ (dashed red line), $\{22,32\}$ (dotted blue line), and $\{22,32,42\}$ (solid black line).  The mass of the remnant black hole obtained directly from the numerical simulation is shown as a horizontal line (solid orange).  The overlaps are computed using the eigenvalue method with SVD tolerance $\tau=0$.}
\end{figure}

\begin{figure}
\includegraphics[width=\linewidth,clip]{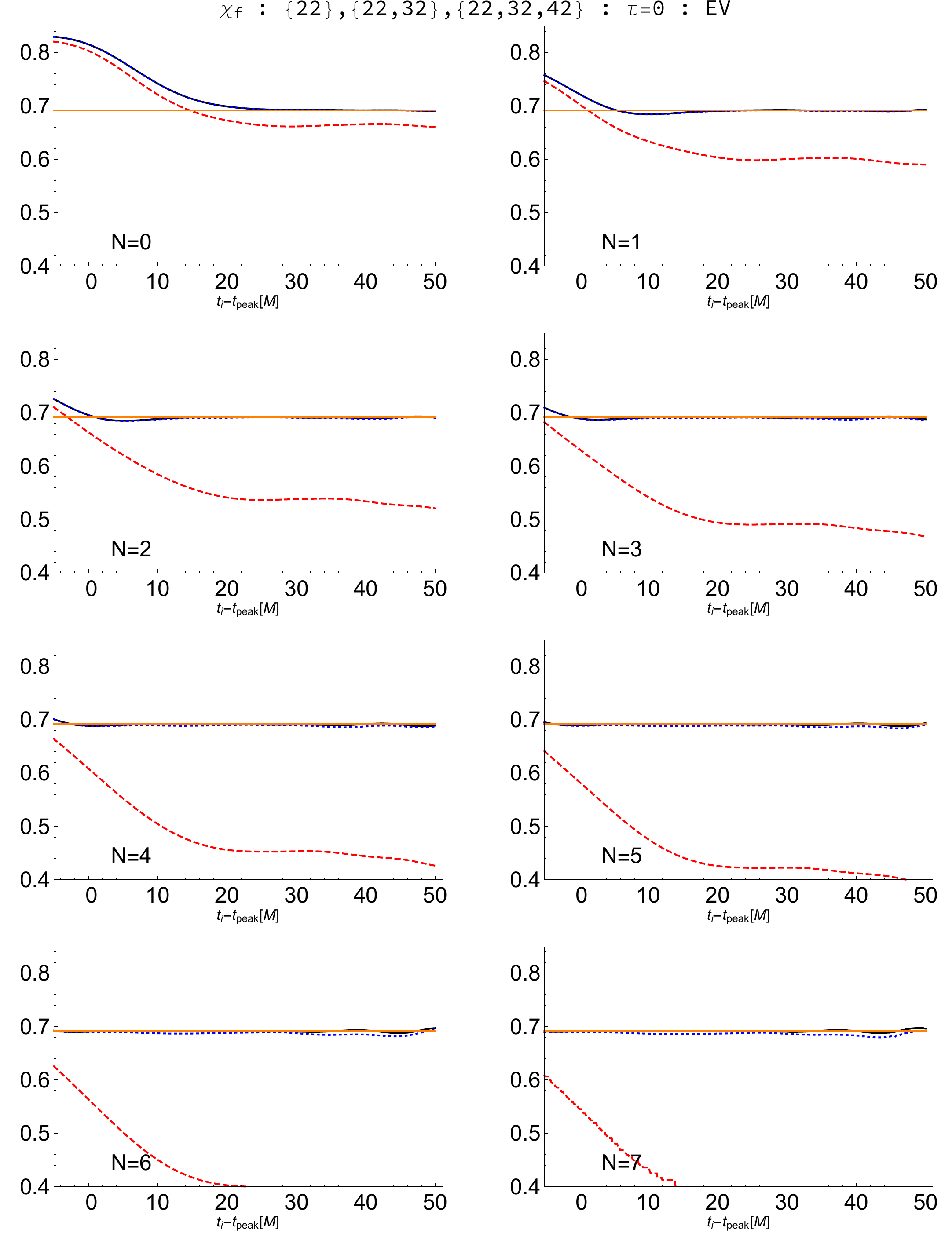}
 \caption{\label{fig:Fitchi_h_CvsN}  The fit value of the dimensionless angular momentum $\chi$ as a function of $t_i-t_{\rm peak}$.  Each panel compares the results when using fitting cases $\{22\}$ (dashed red line), $\{22,32\}$ (dotted blue line), and $\{22,32,42\}$ (solid black line).  The number $N$ associated with each panel denotes the maximum value of the overtone index $n$ used in each fitting case.  The dimensionless angular momentum of the remnant black hole obtained directly from the numerical simulation is shown as a horizontal line (solid orange).  The overlaps are computed using the eigenvalue method with SVD tolerance $\tau=0$.}
\end{figure}

Figures~\ref{fig:Fitdelta_h_CvsN} and \ref{fig:Fitchi_h_CvsN} show the fit values for $\delta$ and $\chi_f$ for the fits that produce the minimum mismatches in Fig.~\ref{fig:FitMM_h_CvsN}.
The panels in each plot include a reference line showing the value of either $\delta$ or $\chi_f$ from $\mathcal{R}_{\rm NR}$.  First, we note that the fit values for $\delta$ and $\chi_f$ obtained from the $\{22\}$ fitting case are only reasonably close to the expected values for small values of $N$ and when the fitting starts relatively late in the ringdown (larger values of $t_i-t_{\rm peak}$).  This is more significant for $\delta$ than for $\chi_f$.  As $N$ increases, the agreement with the expected value gets worse.  For both the $\{22,32\}$ and $\{22,32,42\}$ fitting cases, the agreement with the expected value is quite good.  As expected, good agreement extends to earlier times in the ringdown (smaller values of $t_i-t_{\rm peak}$) as $N$ increases indicating more overtones are being used.

For small values of $N$, the $\{22,32\}$ and $\{22,32,42\}$ fitting cases are in good visual agreement in both figures, but as $N$ increases, the $\{22,32,42\}$ fitting case gives results that are clearly in better agreement with the expected values.  This is most easily seen in Fig.~\ref{fig:Fitdelta_h_CvsN} for $\delta$.  We also see that agreement with the expected value gets worse at large values of $t_i-t_{\rm peak}$ as $N$ increases.  The reason for this is the short decay time of the higher overtones.  These modes should not contribute significantly at late times in the ring down.  However, these modes may be given undue weight in the fitting because $\tau=0$, causing increased error in the fit.

\begin{figure}
\includegraphics[width=\linewidth,clip]{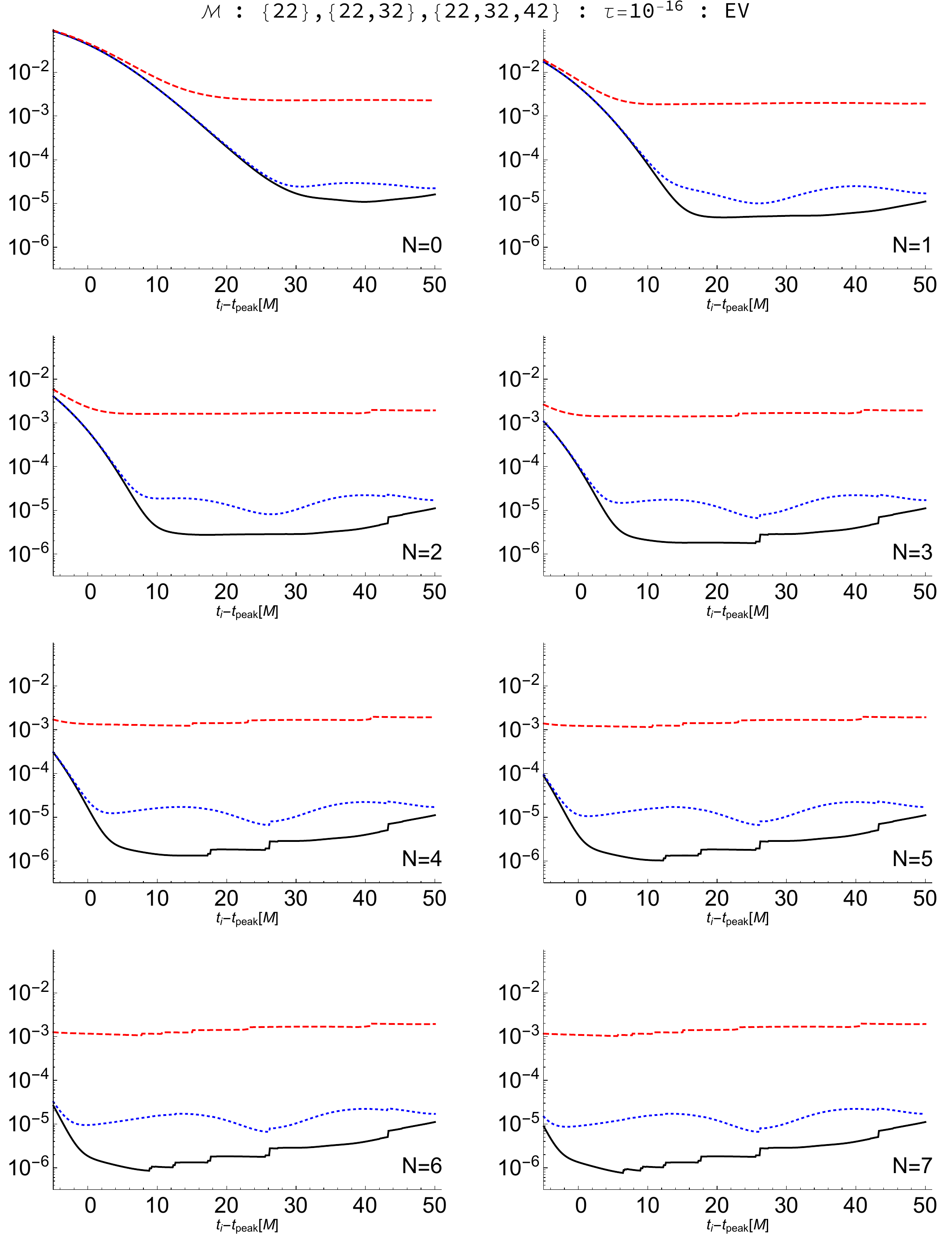}
 \caption{\label{fig:FitMM_h_CvsN_Tol16}  Same as Fig.~\ref{fig:FitMM_h_CvsN} but with SVD tolerance $\tau=10^{-16
 }$.}
\end{figure}

\begin{figure}
\includegraphics[width=\linewidth,clip]{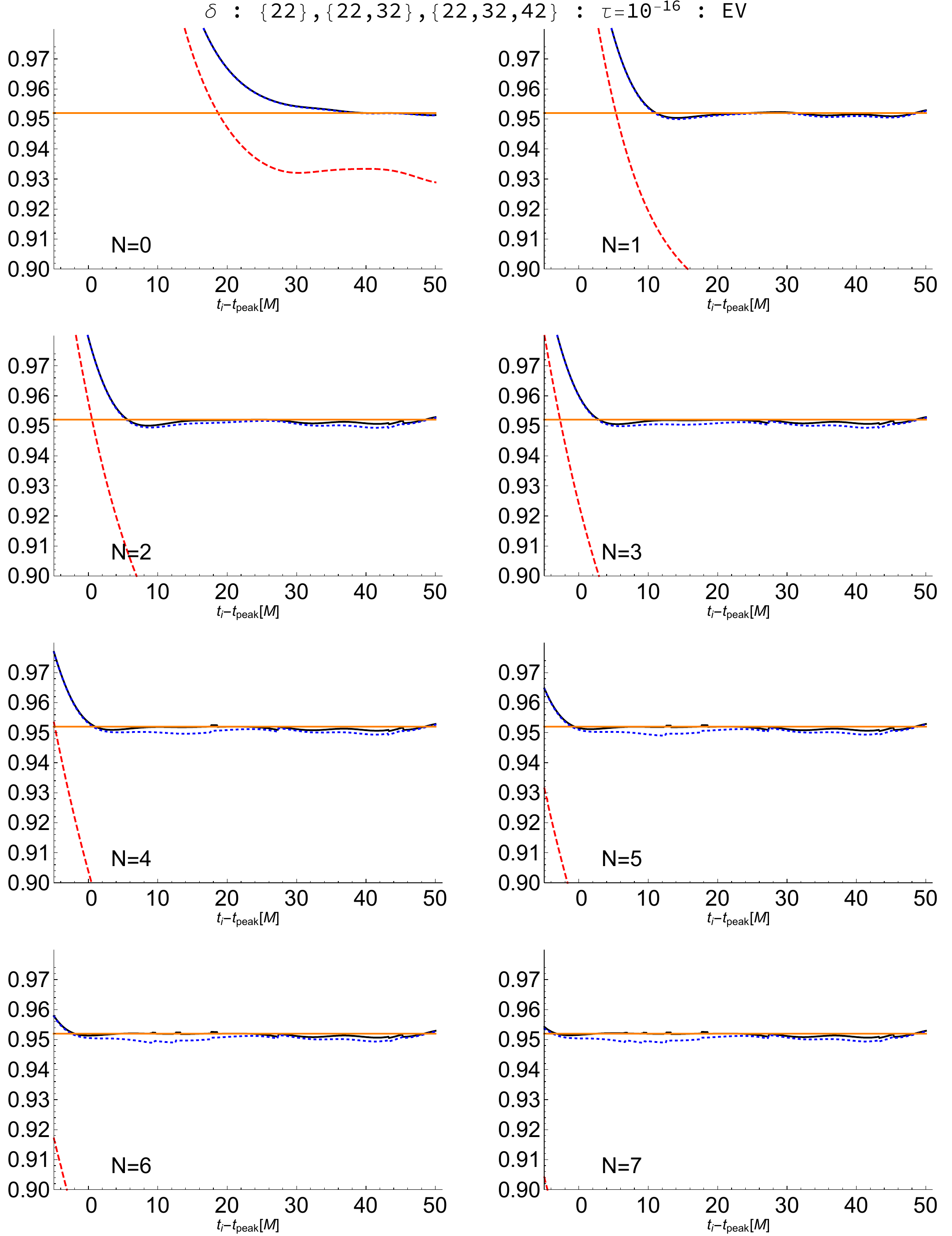}
 \caption{\label{fig:Fitdelta_h_CvsN_Tol16}  Same as Fig.~\ref{fig:Fitdelta_h_CvsN} but with SVD tolerance $\tau=10^{-16
 }$.}
\end{figure}

\begin{figure}
\includegraphics[width=\linewidth,clip]{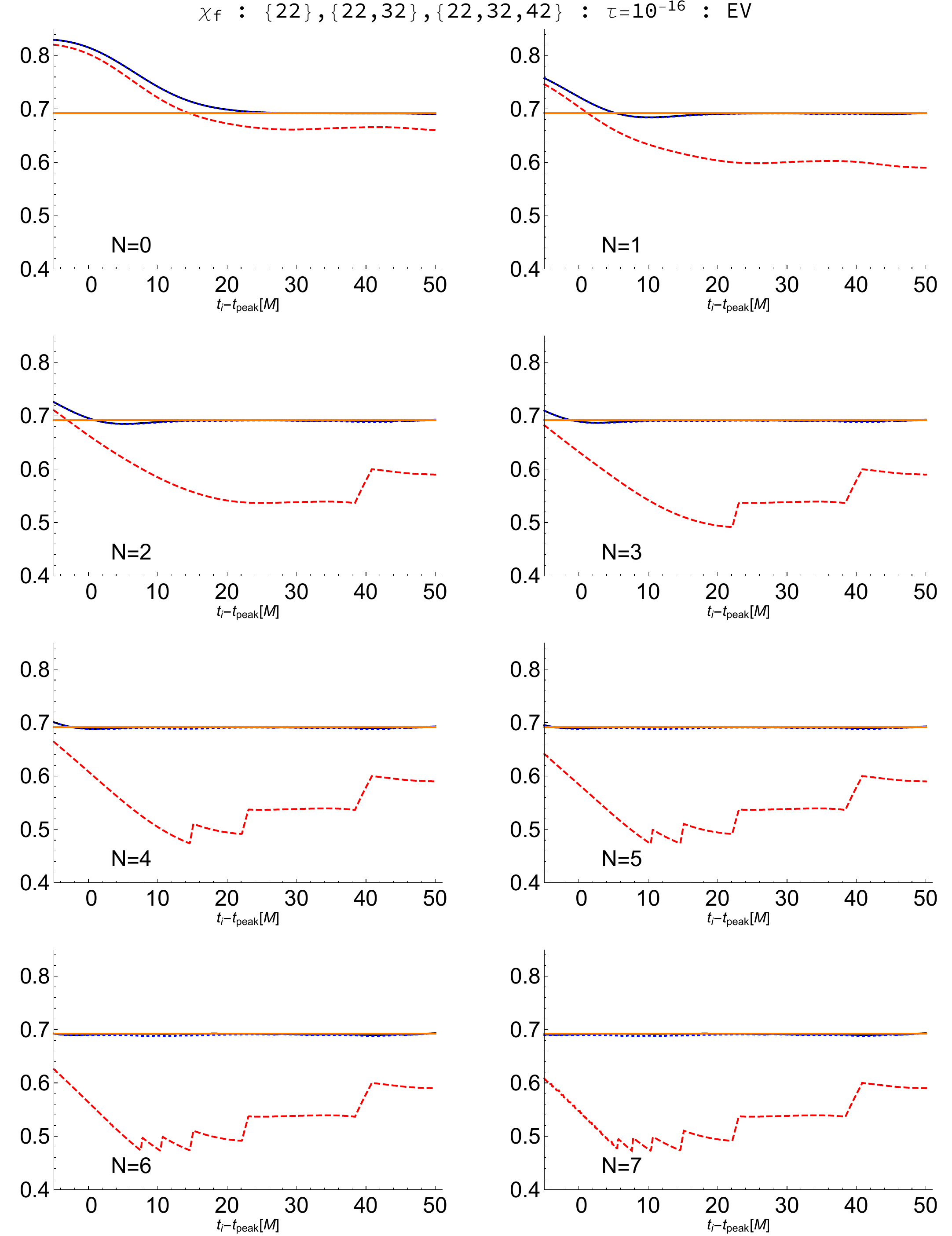}
 \caption{\label{fig:Fitchi_h_CvsN_Tol16}  Same as Fig.~\ref{fig:Fitchi_h_CvsN} but with SVD tolerance $\tau=10^{-16
 }$.}
\end{figure}

The cure for this problem it to use a nonvanishing SVD tolerance during the fitting process.  Figures~\ref{fig:FitMM_h_CvsN_Tol16}, \ref{fig:Fitdelta_h_CvsN_Tol16}, and \ref{fig:Fitchi_h_CvsN_Tol16} show the results of fitting the same cases as in Figs.~\ref{fig:FitMM_h_CvsN}, \ref{fig:Fitdelta_h_CvsN} and \ref{fig:Fitchi_h_CvsN} but with the SVD tolerance set to $\tau=10^{-16}$.
The results in all cases are as good or better than when the SVD tolerance is set to $\tau=0$.  In particular, good agreement is seen for the fit values of $\delta$ and $\chi_f$ for all values of $t_i-t_{\rm peak}$ for large values of $N$.  The expected cost of these improved results is the introduction of discontinuities in the quantities as functions of $t_i-t_{\rm peak}$.  

\begin{figure}
\includegraphics[width=\linewidth,clip]{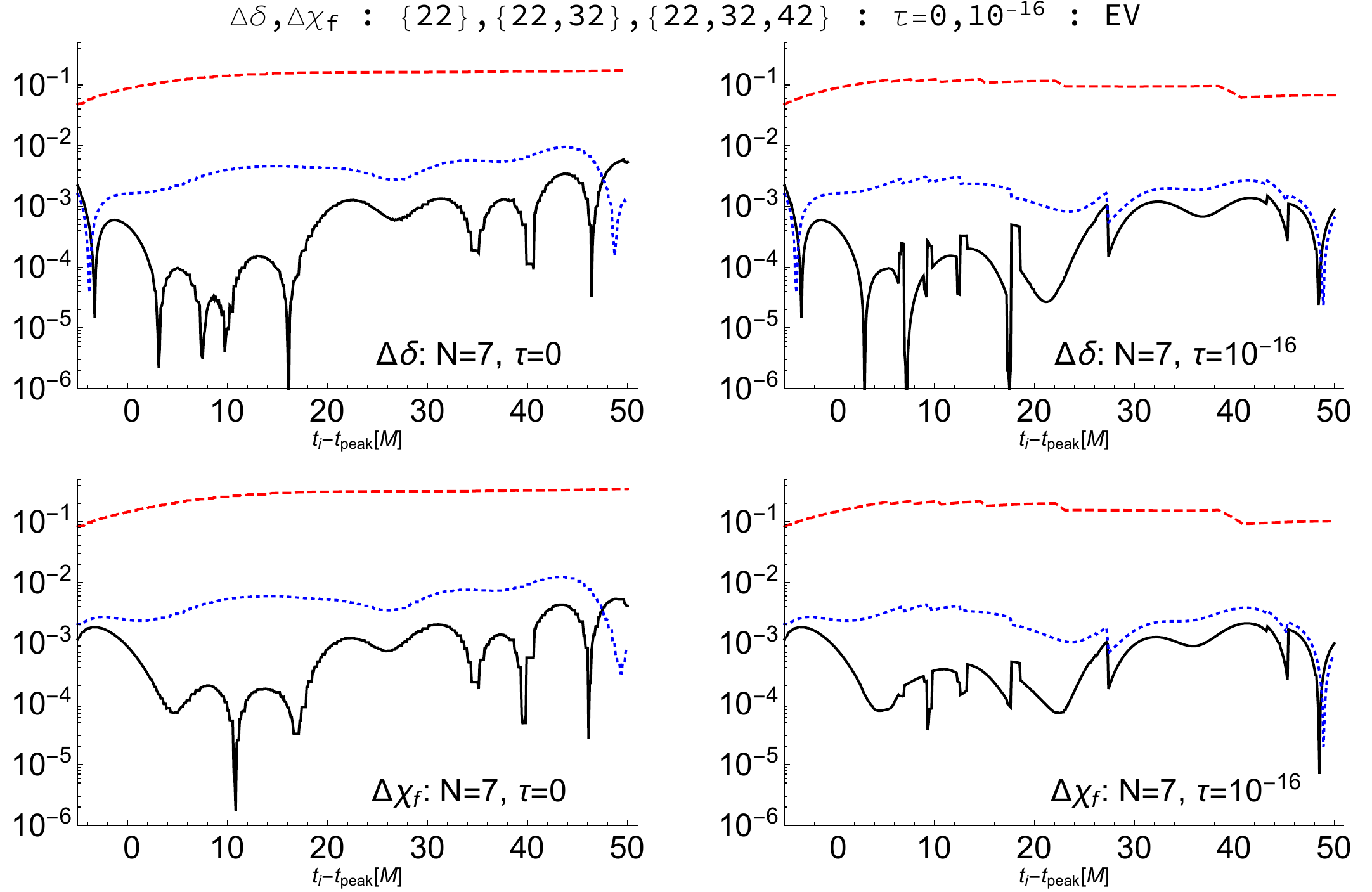}
 \caption{\label{fig:FitDeltas_h_CvsN_Cmp}  The deviation of the fit values of the mass ratio $\delta$ and $\chi_f$ from $\mathcal{R}_{\rm NR}$ plotted as a function of $t_i-t_{\rm peak}$ for $N=7$.  The upper two panels display $\Delta\delta\equiv|0.952033-\delta|$ for the cases of $\tau=0$ and $\tau=10^{-16}$.  The lower two panels display $\Delta\chi_f\equiv|0.692085-\chi_f|$ for the cases of $\tau=0$ and $\tau=10^{-16}$.  Each panel compares the results when using fitting cases $\{22\}$ (dashed red line), $\{22,32\}$ (dotted blue line), and $\{22,32,42\}$ (solid black line).  The overlaps are computed using the eigenvalue method.}
\end{figure}

To more clearly see the quality of the fits and the effects of using a nonvanishing SVD tolerance, we include an abbreviated plot showing the differences, $\Delta\delta$ and $\Delta\chi_f$, of $\delta$ and $\chi_f$ with their expected values.  Figure~\ref{fig:FitDeltas_h_CvsN_Cmp} includes panels showing both $\Delta\delta$ and $\Delta\chi_f$
for $N=7$ only, but with the SVD tolerance set to both $\tau=0$ and $\tau=10^{-16}$.

Of course, the nonlinear fitting of model parameters depends critically on how the mismatch is computed.  All of the examples so far have assumed that the mismatch was computed using the EV method with Eq.~(\ref{eqn:rho2max}) computed using the full $\mathbb{B}$.
\begin{figure}
\includegraphics[width=\linewidth,clip]{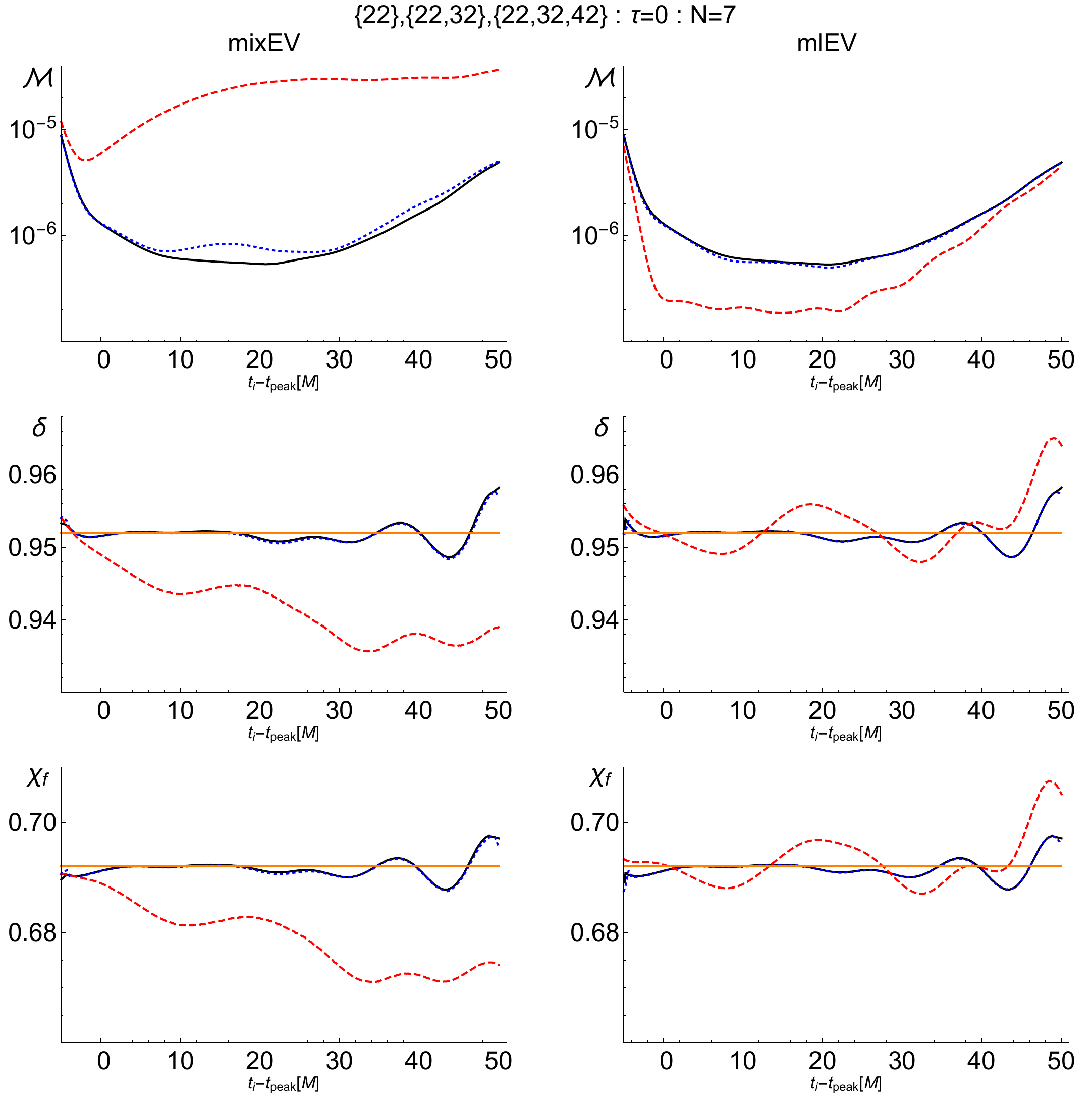}
 \caption{\label{fig:FitAll_h_N7_mixml}  The fit of the mismatch $\mathcal{M}$ (top row), mass ratio $\delta$ (middle row), and dimensionless angular momentum $\chi$ (bottom row) as functions of $t_i-t_{\rm peak}$.  Each panel compares the results when using fitting cases $\{22\}$ (dashed red line), $\{22,32\}$ (dotted blue line), and $\{22,32,42\}$ (solid black line).  For brevity, we include only examples where all eight overtones (N=7) are used for each mode, and SVD tolerance $\tau=0$ in each case.  The left column shows the results for the mismatch computed using the EV method, but with the overlap recomputed using Eq.~(\ref{eqn:rho2explicit}) and $\mathbb{B}_{\rm ml}$.  The right column shows the results for the mismatch computed using the mlEV method.}
\end{figure}
Figure~\ref{fig:FitAll_h_N7_mixml} illustrates how the fitting changes if we compute the overlap using $\mathbb{B}_{\rm ml}$ instead of $\mathbb{B}$.  For brevity, the figure displays only the $N=7$ examples where all eight overtones are used for each mode.  The left-hand column shows results obtained by computing the mismatch using the EV method, but with the overlap recomputed using Eq.~(\ref{eqn:rho2explicit}) and $\mathbb{B}_{\rm ml}$.  The most significant difference from the results plotted in Figs.~\ref{fig:FitMM_h_CvsN}, \ref{fig:Fitdelta_h_CvsN}, and \ref{fig:Fitchi_h_CvsN} is seen in the $\{22\}$ cases.  While still not yielding very good fits, the fit values for $\delta$ and $\chi_f$ are significantly improved.  The right-hand column shows the results obtained by computing the mismatch using the mlEV method.  For this case, the most striking observation is that the smallest mismatches occur for the $\{22\}$ case, even though this case does not yield the best fit values for $\delta$ and $\chi_f$.  While using $\mathbb{B}_{\rm ml}$ to compute the mismatch has a significant effect on the $\{22\}$ case, the changes are relatively insignificant for the $\{22,32\}$ and $\{22,32,42\}$ cases.
 
\subsection{Fidelity of the amplitudes and the onset of ringdown}\label{sec:amplitudes}

So far, we have focused on how the mismatch $\mathcal{M}$ and model parameters $\mathcal{P}_m$ behave as we vary the set of $\{\rm NR\}$ of simulation modes being fit to, and the set $\{\rm QNM\}$ of QNMs being used to perform the fit.  By using enough simulation modes and QNMs, it is possible to obtain good fits with small mismatches over a wide range in values of start times $t_i$.  In fact, with enough QNM overtones it is possible to obtain good fits with $t_i<t_{\rm peak}$.  Does this mean that the merged black hole has settled down sufficiently that the system is already nearly linear at $t_{\rm peak}$?  As the perturbed black hole settles down through the emission of gravitational waves, there will be a point at which the spacetime becomes well approximated by linear perturbations of the Kerr geometry.  The earliest time at which this occurs is usually referred to as the onset of ringdown.

As well summarized by GIST, there have been many attempts\cite{BCP-2007,kamaretsos-etal-2012a,londonShoemakerhealy-2014,thranelaskyleven-2017,carullo-etal-2018,bhagwat-etal-2018,baibhav-etal-2018,baibhavberti-2019} to determine the time of the onset of ringdown by studying waveforms from numerical relativity simulations, with considerable disagreement among the results.  This is a very interesting topic, but we will only delve into this topic indirectly.  Here, we will consider the fidelity of the QNM amplitudes obtained by fitting the waveform.  That is, how robustly are the amplitudes determined.

We will consider the amplitudes from two perspectives.  First is the perspective of absolute amplitudes, and second is the perspective of relative amplitudes, both as functions of time.  When we fit a ringdown signal starting at some time $t_i$ in the signal, the amplitudes that are obtained are always scaled to be the amplitudes at some fiducial time in the waveform.  In the case of this work, that fiducial time is chosen to be $t_{\rm peak}$.  Note, this does not mean that we are assuming that $t_{\rm peak}$ represents the onset of ringdown.  If we plot the amplitude of $C^\pm_{\ell{m}n}$ as a function of $t_i$, but keep the amplitude scaled to its value at $t_{\rm peak}$, then a mode which is fit robustly will have a constant value.  This is the perspective of absolute amplitudes.
\begin{figure}
\includegraphics[width=\linewidth,clip]{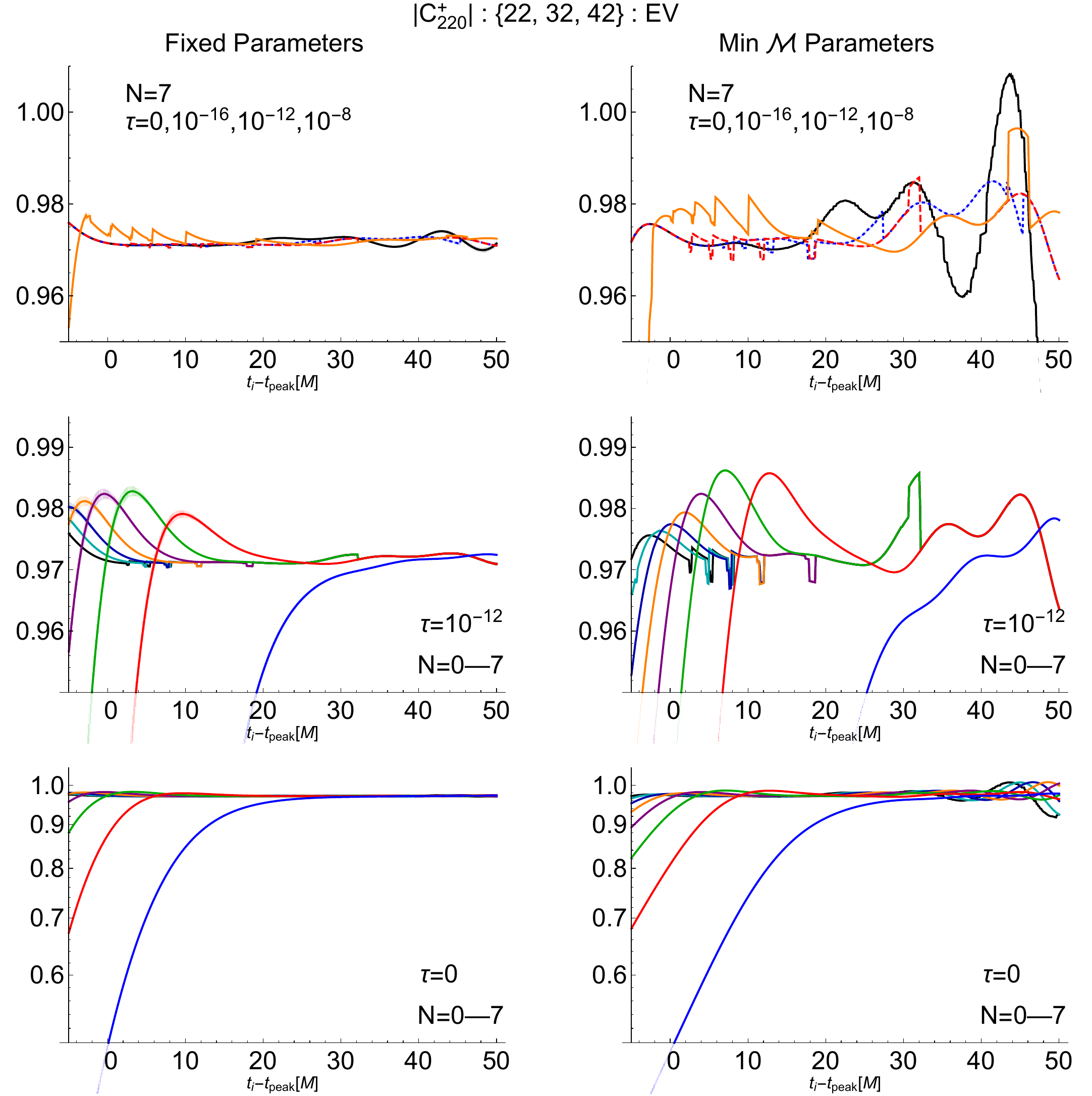}
 \caption{\label{fig:AbsAmp220_h_o7223242}  Absolute amplitude plots of $|C^+_{220}|$ obtained using fitting case $\{22,32,42\}$.  Amplitudes displayed in the left column are from fits where the remnant parameters were fixed at $\mathcal{R}=\mathcal{R}_{\rm NR}$, while those in the right column had $\mathcal{R}=\mathcal{P}_m(t_i)$ chosen to minimize the mismatch $\mathcal{M}$ at each time $t_i$.  Results in the top row are for $N=7$ but with the SVD tolerance varied.  The black line represents $\tau=0$, the dotted blue line represents $\tau=10^{-16}$, the dashed red line represents $\tau=10^{-12}$, and the orange line represents $\tau=10^{-8}$.  The lower two rows compare the amplitude as $N$ is varied, but the SVD tolerance if fixed.  The middle row uses a linear scale with $\tau=10^{-12}$, while the bottom row uses a log scale with $\tau=0$.  In these panels, the rightmost line(blue) is $N=0$ and the black line is $N=7$.  Error bars from the fit at each $t_i$ are included in all curves, but are only discernible on some curves in the middle-left panel.}
\end{figure}
Figure~\ref{fig:AbsAmp220_h_o7223242} shows several absolute amplitude plots of $|C^+_{220}|$ obtained from fitting case $\{22,32,42\}$.  The panels in the left column represent fits performed with the remnant parameters $\mathcal{R}=\mathcal{R}_{\rm NR}$, while the panels in the right column represent fits performed with the remnant parameters $\mathcal{R}=\mathcal{P}_{m}$ chosen to minimize the mismatch $\mathcal{M}$ at each $t_i$.  In the top row, the two panels display results for the $N=7$ case where a total of $24$ QNMs are used ($n=0,\ldots,7$ for $C^+_{22n}$, $C^+_{32n}$, $C^+_{42n}$) and the fits are performed with four different values of the SVD tolerance ($\tau=0$, $10^{-16}$, $10^{-12}$, and $10^{-8}$).  Most obvious in these panels is that using $\tau=10^{-8}$ seems to remove too many modes for $t_i$ approaching $t_{\rm peak}$
.  Perhaps not surprising, the value of $|C^+_{220}|$ is most consistent when the remnant parameters are held fixed, and variation in the amplitude is most pronounced for $t_i-t_{\rm peak}>30M$.  The bottom two rows in the figure compare results for different values of $N$, each at a specified value of $\tau$.  The middle row of panels plots the case for $\tau=10^{-12}$.  We see the general trend that the amplitude becomes more robust as $N$ increases.  The bottom row of panels presents similar plots to the middle row, but with $\tau=0$ and using a log scale for the amplitude to emphasize the exponential behavior of the amplitudes for early $t_i$.  Again we see the general trend that the amplitude becomes more robust as $N$ increases.

An alternative perspective is gained by plotting, not the magnitude of a given mode at a fiducial time, but its exponentially decaying amplitude as a function of time.  Let us define a mode's relative amplitude as
\begin{align}
	\mathcal{R}_{\acute\ell\acute{m}}C^\pm_{\ell{m}n}(t)&\equiv
	|D^\ell_{\acute{m}m}|\!\times\!|\YSHn{\acute\ell\ell(\pm m)n}|\!\times\!|C^\pm_{\ell{m}n}(t_i)| \\
	&\mbox{}\qquad\qquad\qquad\qquad \times\! e^{{\rm Im}(\omega_{\ell(\pm m)n})(t-t_{\rm peak})}, \nonumber
\end{align}
which follows directly from Eq.~(\ref{eqn:Clm_expansion}).  Note that $|C^\pm_{\ell{m}n}(t_i)|$ is the mode amplitude determined from a fit starting at $t_i$, and is always evaluated at the fiducial time (in this case $t_{\rm peak}$), but is then multiplied by the appropriate exponential damping behavior.  This gives the dominant scaling in a mode's relative amplitude.  But, the relative amplitude also depends on which simulation mode $C_{\acute\ell\acute{m}}$ is being considered.  This comes in primarily through the spheroidal-harmonic expansion coefficient $\YSHn{\acute\ell\ell{m}n}$ but also through the Wigner rotation matrix $D^\ell_{\acute{m}m}$, both of which depend on the remnant parameters $\mathcal{R}$ used to compute $|C^\pm_{\ell{m}n}(t_i)|$.  Let us consider a specific example to see how the relative amplitude plots behave. 
\begin{figure}
\includegraphics[width=\linewidth,clip]{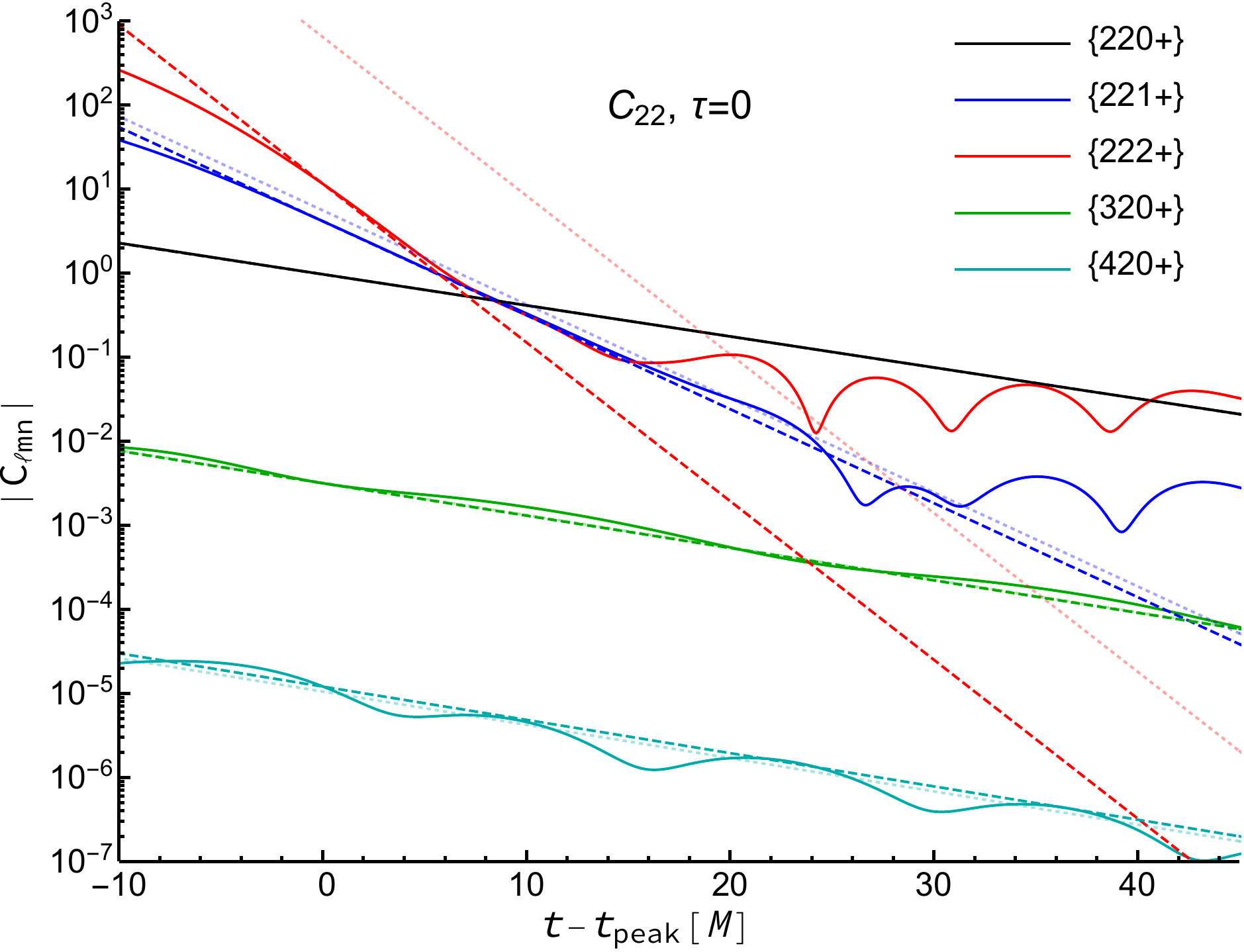}
 \caption{\label{fig:RelAmp22a_h_o7223242}  Relative amplitude plots of a subset $\{{\rm QNM_{rob}}\}$ of the QNMs for fitting case $\{22,32,42\}$ with $N=7$ and SVD tolerance $\tau=0$ as they contribute to $C_{22}$.  The solid lines are continuously fit, while the dashed lines are discretely fit with $t_i=t_{\rm peak}$ and the dotted lines are discretely fit with $t_i-t_{\rm peak}=20M$.}
\end{figure}
Figure~\ref{fig:RelAmp22a_h_o7223242} presents a relative amplitude plot of a set of QNMs as they contribute to $C_{22}$.  Fitting was performed with case $\{22,32,42\}$, the remnant parameters were fixed to $\mathcal{R}_{\rm NR}$, and the SVD tolerance was $\tau=0$.  The data in the figure was for $N=7$, but we only plot a subset of the modes for clarity.  We will define the subset as $\{{\rm QNM_{rob}}\}\equiv\{220+,221+,222+,320+,420+\}$.  First consider the solid lines.  These correspond to choosing $t_i=t$.  That is, at each time $t$, the mode amplitudes were obtained from the fit with $t_i=t$.  We will refer to an amplitude curve obtained in this way as ``continuously fit.''  The dashed and dotted lines, on the other hand, are ``discretely fit.''  For the dashed lines, the amplitudes $|C^\pm_{\ell{m}n}(t_i)|$ are obtained from the fit at $t_i=t_{\rm peak}$, and for the dotted lines from the fit at $t_i-t_{\rm peak}=20M$.

On a log-linear plot, discretely fit relative amplitude lines will have a constant slope determined by ${\rm Im}(\omega_{\ell{m}n})$.  If a given mode is robustly fit, then its continuously fit relative amplitude curve will also have constant slope and will coincide with any discretely fit counterpart with $t_i=t$ in the range where the amplitude is robust.  In Fig.~\ref{fig:RelAmp22a_h_o7223242}, we see that the $\{220+\}$ QNM is robust over the entire range of the plot.  The $\{320+\}$ and $\{420+\}$ QNMs are also reasonably robust over the entire range of the plot, but we do see some oscillation in the continuously fit curves.  All three of the $n=0$ modes have roughly the same slope, but the amplitudes are separated by a little more than $2$ orders of magnitude.  If we consider the $\{221+\}$ QNM, we see that it has a steeper slope.  The mode is reasonably robust for $-5M \lesssim t_i-t_{\rm peak} \lesssim 15M$.  For $t_i-t_{\rm peak} \gtrsim 20M$ the mode is clearly not robust.  For the $\{222+\}$ QNM, the robust range is even smaller and the deviations at large $t_i-t_{\rm peak}$ even greater.

Relative amplitude plots like Fig.~\ref{fig:RelAmp22a_h_o7223242} convey a great deal if useful information.  In addition to showing clearly where given QNMs are being robustly fit, they also show the relative importance of each mode at different times.  For late times, we see that the $\{220+\}$ mode is dominant.  If we focus on the dashed lines, discretely fit to $t_i=t_{\rm peak}$, we see how the overtones dominate at early times.  
\begin{figure}
\includegraphics[width=\linewidth,clip]{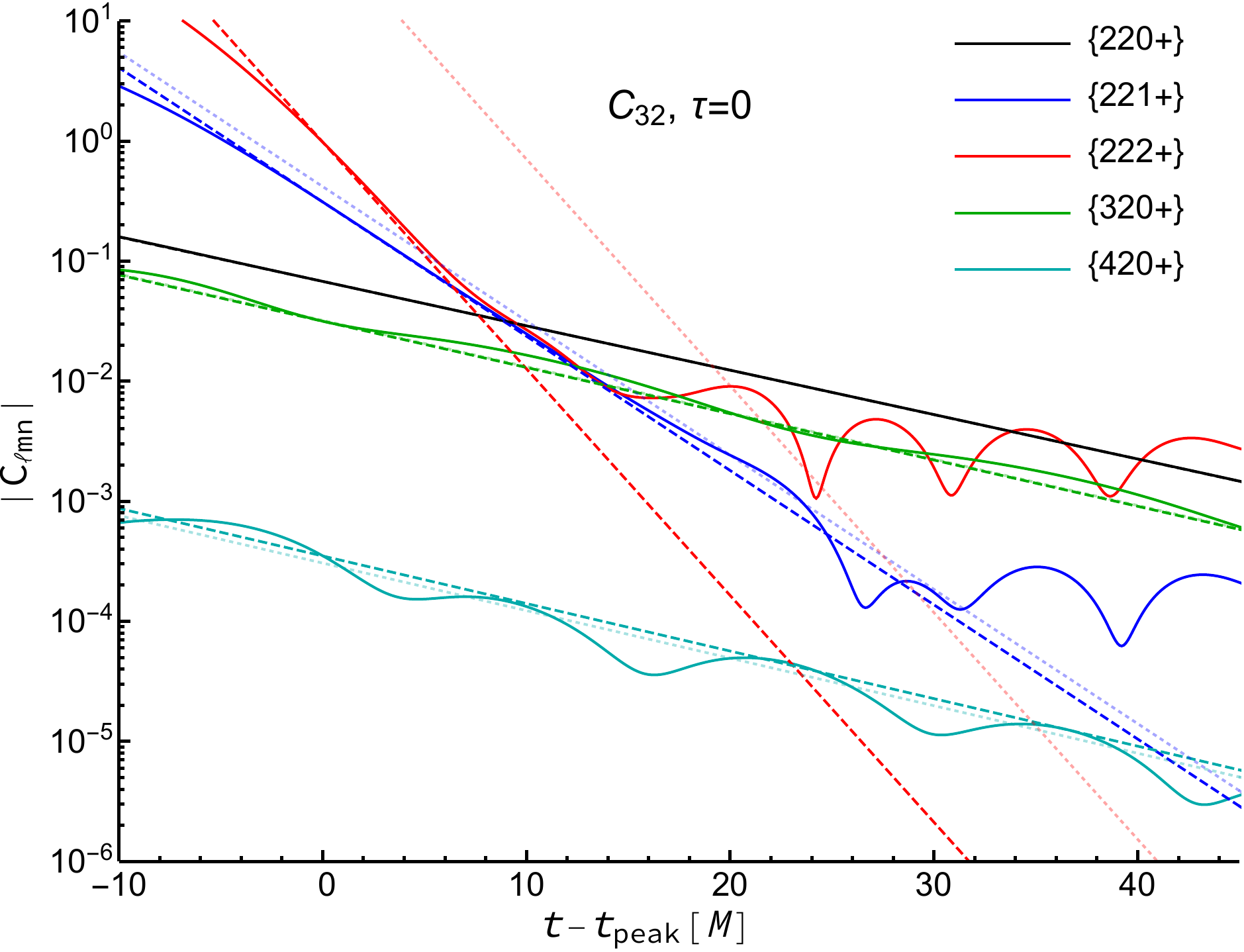}
 \caption{\label{fig:RelAmp32a_h_o7223242}  Relative amplitude plots of a subset $\{{\rm QNM_{rob}}\}$ of the QNMs for fitting case $\{22,32,42\}$ with $N=7$ and SVD tolerance $\tau=0$ as they contribute to $C_{32}$.  See Fig.~\ref{fig:RelAmp22a_h_o7223242} for additional detail.}
\end{figure}
\begin{figure}
\includegraphics[width=\linewidth,clip]{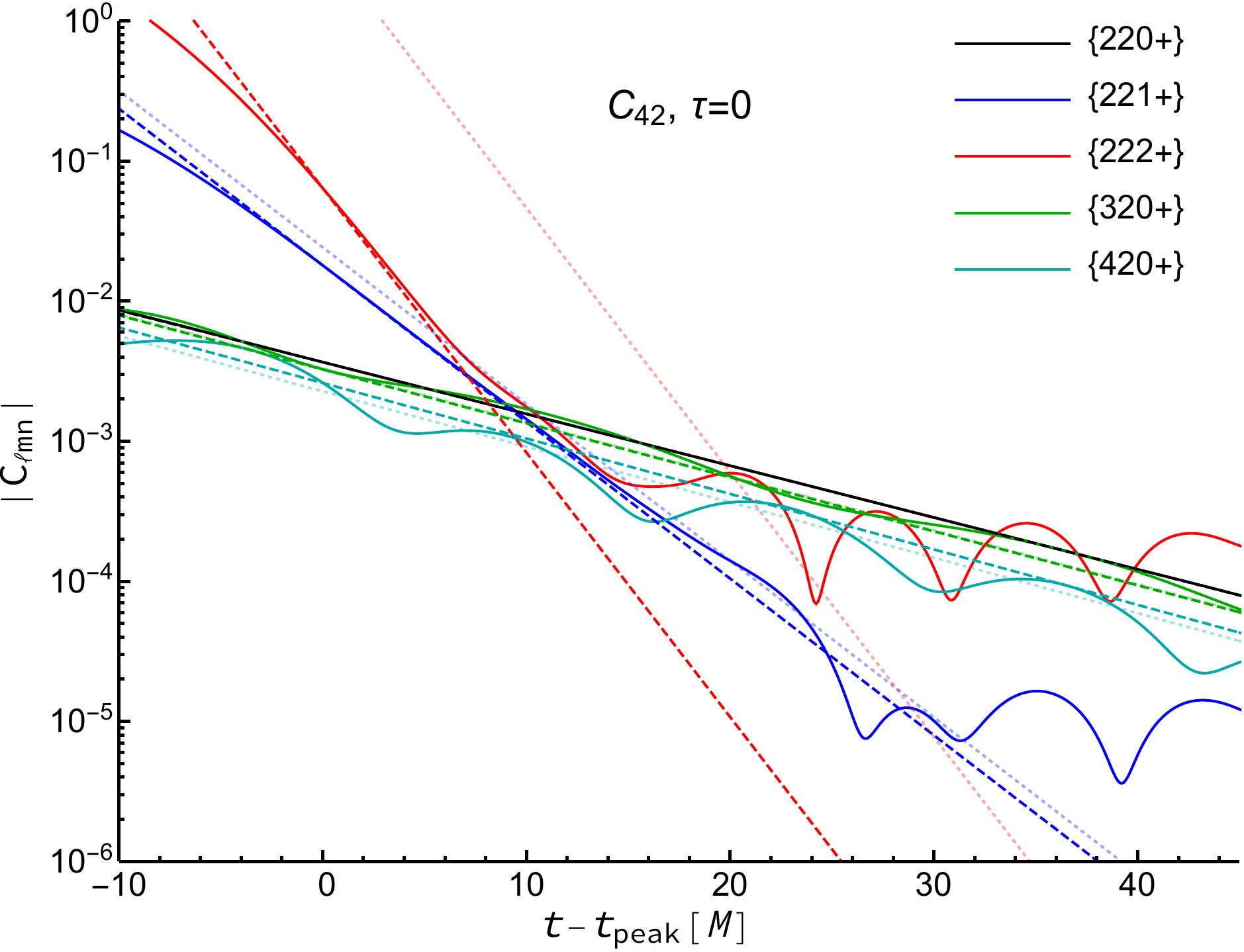}
 \caption{\label{fig:RelAmp42a_h_o7223242}  Relative amplitude plots of a subset $\{{\rm QNM_{rob}}\}$ of the QNMs for fitting case $\{22,32,42\}$ with $N=7$ and SVD tolerance $\tau=0$ as they contribute to $C_{42}$.  See Fig.~\ref{fig:RelAmp22a_h_o7223242} for additional detail.}
\end{figure}
Figures~\ref{fig:RelAmp32a_h_o7223242} and \ref{fig:RelAmp42a_h_o7223242} show similar plots but scaled for the $C_{32}$ and $C_{42}$ simulation modes.  In Fig.~\ref{fig:RelAmp32a_h_o7223242}, we see that the relative amplitudes of the $\{220+\}$ and $\{320+\}$ QNMs for $C_{32}$ only differ in magnitude by about a factor of 2, while in Fig.~\ref{fig:RelAmp42a_h_o7223242}, we see that the relative amplitudes of the $\{220+\}$, $\{320+\}$, and $\{420+\}$ QNMs for $C_{42}$ are roughly comparable in magnitude.

One of the important problems with QNM fitting is also clearly demonstrated in Figs.~\ref{fig:RelAmp22a_h_o7223242} \ref{fig:RelAmp32a_h_o7223242}, and \ref{fig:RelAmp42a_h_o7223242}.  If we look at the continuously fit relative amplitude curves at late times, we see that the higher overtones have amplitudes which are too large.  These modes should have small relative amplitudes, but the fitting process has given them too much weight.  We have seen that this can be remedied by using a nonvanishing SVD tolerance.
\begin{figure}
\includegraphics[width=\linewidth,clip]{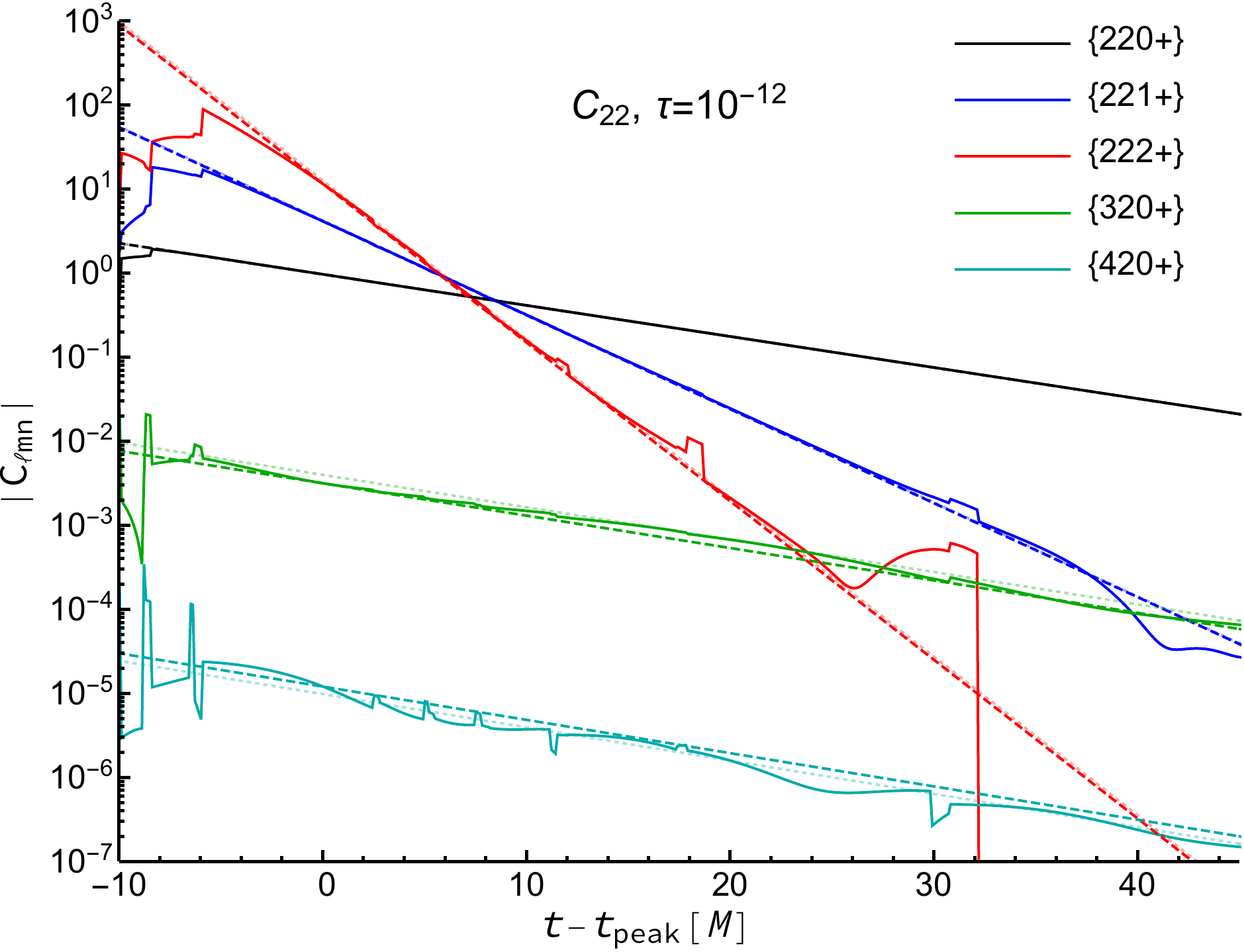}
 \caption{\label{fig:RelAmp22a_h_o7223242_Tol12}  Relative amplitude plots of a subset $\{{\rm QNM_{rob}}\}$ of the QNMs for fitting case $\{22,32,42\}$ with $N=7$ and SVD tolerance $\tau=10^{-12}$ as they contribute to $C_{22}$.  See Fig.~\ref{fig:RelAmp22a_h_o7223242} for additional detail.}
\end{figure}
Figure~\ref{fig:RelAmp22a_h_o7223242_Tol12} reproduces Fig.~\ref{fig:RelAmp22a_h_o7223242} but with the SVD tolerance set to $\tau=10^{-12}$.  The improvement in the continuously fit relative amplitude curves is dramatic, with all of the plotted modes showing reasonably robust behavior for $-6M \lesssim t_i-t_{\rm peak} \lesssim 25M$.  Of course there are some obvious discontinuities in the continuously fit curves due to the discontinuous nature of the decision to effectively remove modes that are deemed irrelevant to the fit.  Figure~9 in GIST is similar to Fig.~\ref{fig:RelAmp22a_h_o7223242_Tol12} in that it also displays continuously fit curves.  However, the continuously fit curves in GIST were constructed by manually removing individual modes whose fit was beginning to show incorrect behavior.  Such a process is likely to be more complicated in multimode fitting, and may be susceptible to some form of bias.  The use of SVD and the pseudoinverse will likely prove to be a better general approach.

\begin{figure}
\includegraphics[width=\linewidth,clip]{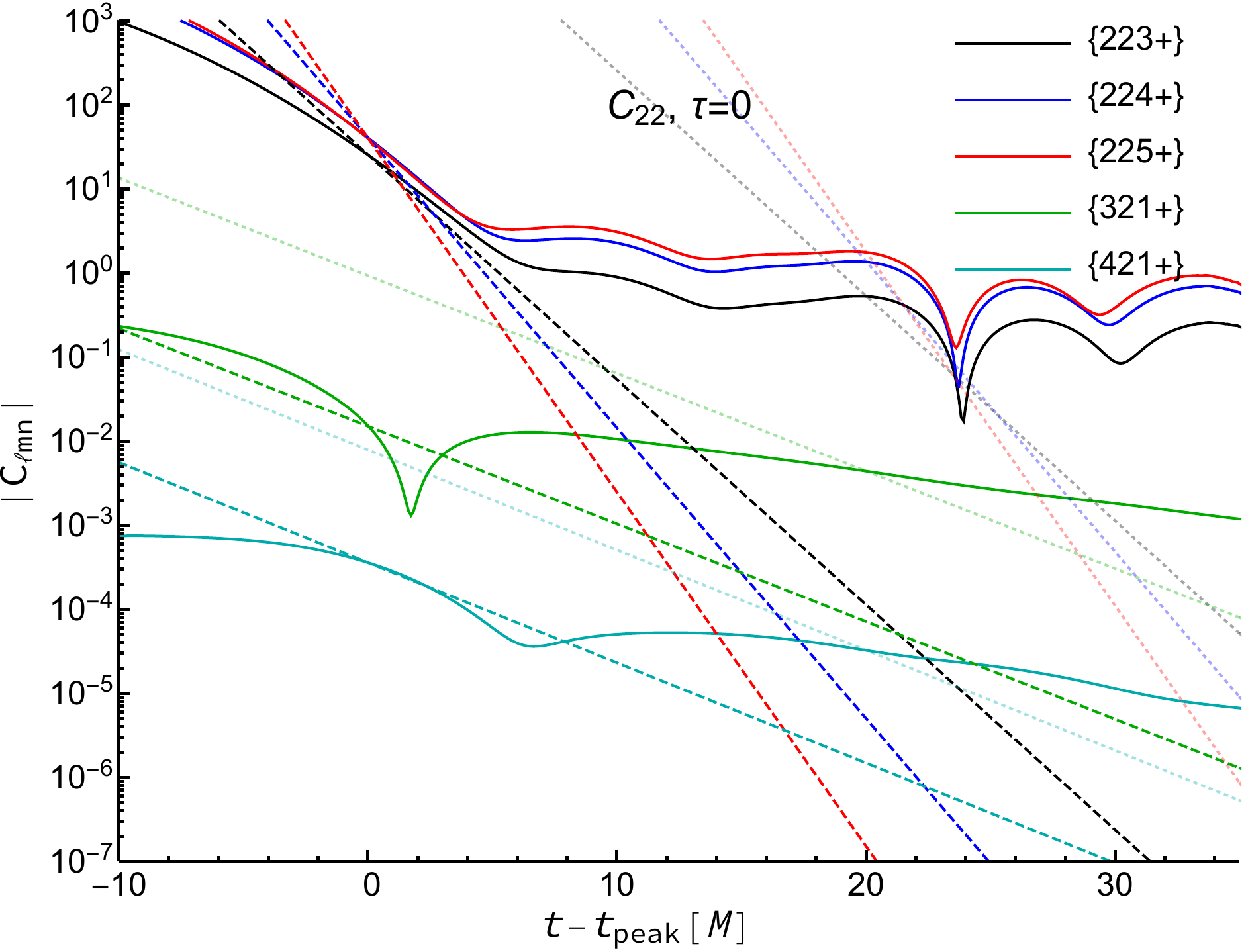}
 \caption{\label{fig:RelAmp22b_h_o7223242}  Relative amplitude plots of a subset $\{223+,224+,225+,321+,421+\}$ of the QNMs for fitting case $\{22,32,42\}$ with $N=7$ and SVD tolerance $\tau=0$ as they contribute to $C_{22}$.  See Fig.~\ref{fig:RelAmp22a_h_o7223242} for additional detail.}
\end{figure}

\begin{figure}
\includegraphics[width=\linewidth,clip]{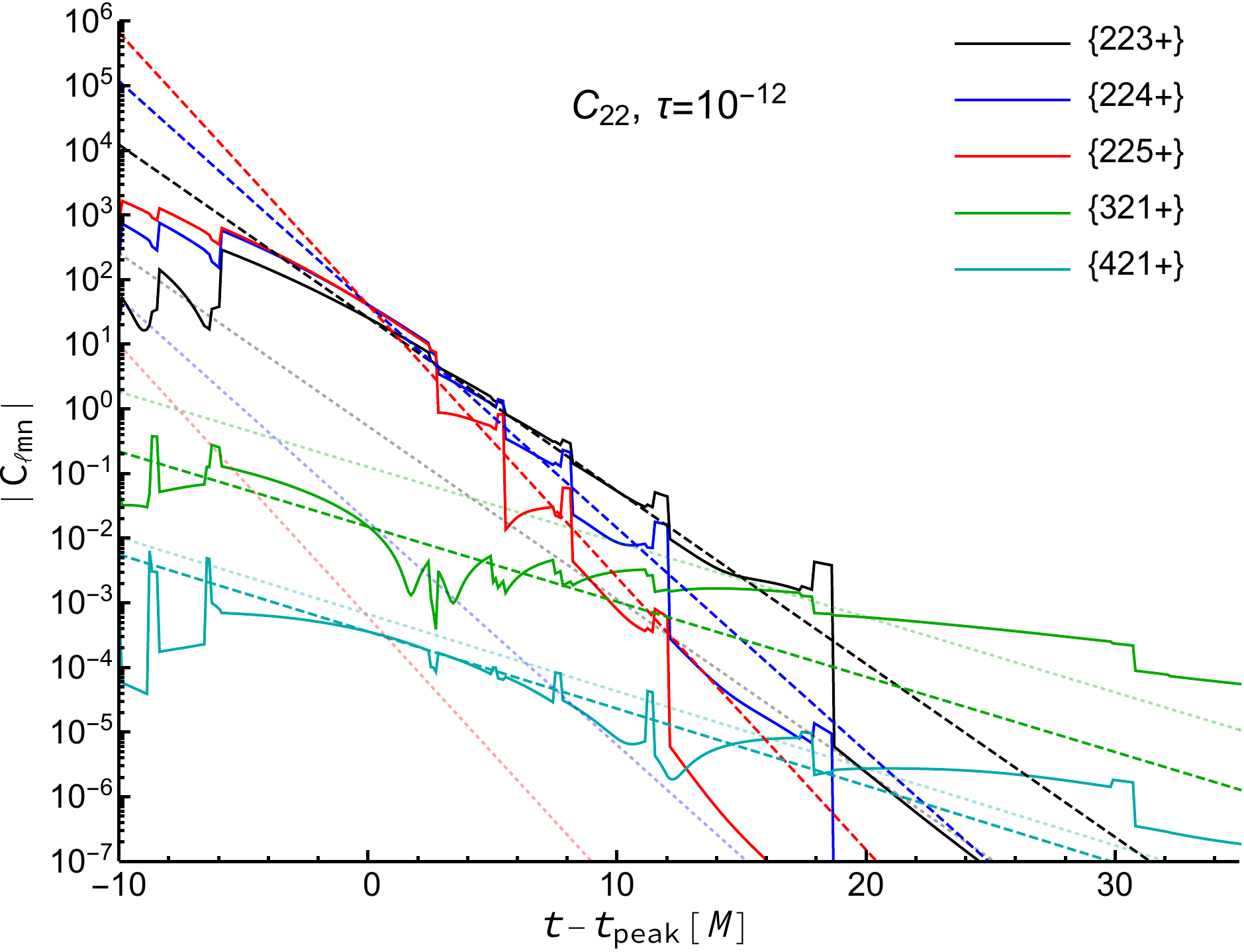}
 \caption{\label{fig:RelAmp22b_h_o7223242_Tol12}  Relative amplitude plots of a subset $\{223+,224+,225+,321+,421+\}$ of the QNMs for fitting case $\{22,32,42\}$ with $N=7$ and SVD tolerance $\tau=10^{-12}$ as they contribute to $C_{22}$.  See Fig.~\ref{fig:RelAmp22a_h_o7223242} for additional detail.}
\end{figure}

If we consider additional QNMs in the fit, we find that they are not fit robustly.  Figures~\ref{fig:RelAmp22b_h_o7223242} and \ref{fig:RelAmp22b_h_o7223242_Tol12} present relative amplitude plots for the $\{223+\}$, $\{224+\}$, $\{225+\}$, $\{321+\}$, and $\{421+\}$ QNMs as they contribute to $C_{22}$.
In Fig.~\ref{fig:RelAmp22b_h_o7223242}, the SVD tolerance is $\tau=0$ and none of the modes show any semblance of robustness.  In Fig.~\ref{fig:RelAmp22b_h_o7223242_Tol12}, the SVD tolerance is $\tau=10^{-12}$ and it appears that the $\{223+\}$ mode is marginally robust over the range $-3M \lesssim t_i-t_{\rm peak} \lesssim 15M$.  But it is clear that only a small number of the modes used in the $\{22,32,42\}$ fitting case are robust in the sense that they are consistently fit across a wide range of fitting times $t_i$.

Since only the five QNMs in the set $\{{\rm QNM_{rob}}\}$ are robustly fit, let us consider using this restricted set of QNMs to fit the simulation modes $\{{\rm NR}\}=\{22,32,42\}$.  This fitting case is denoted $\{22,32,42*\}$.
\begin{figure}
\includegraphics[width=\linewidth,clip]{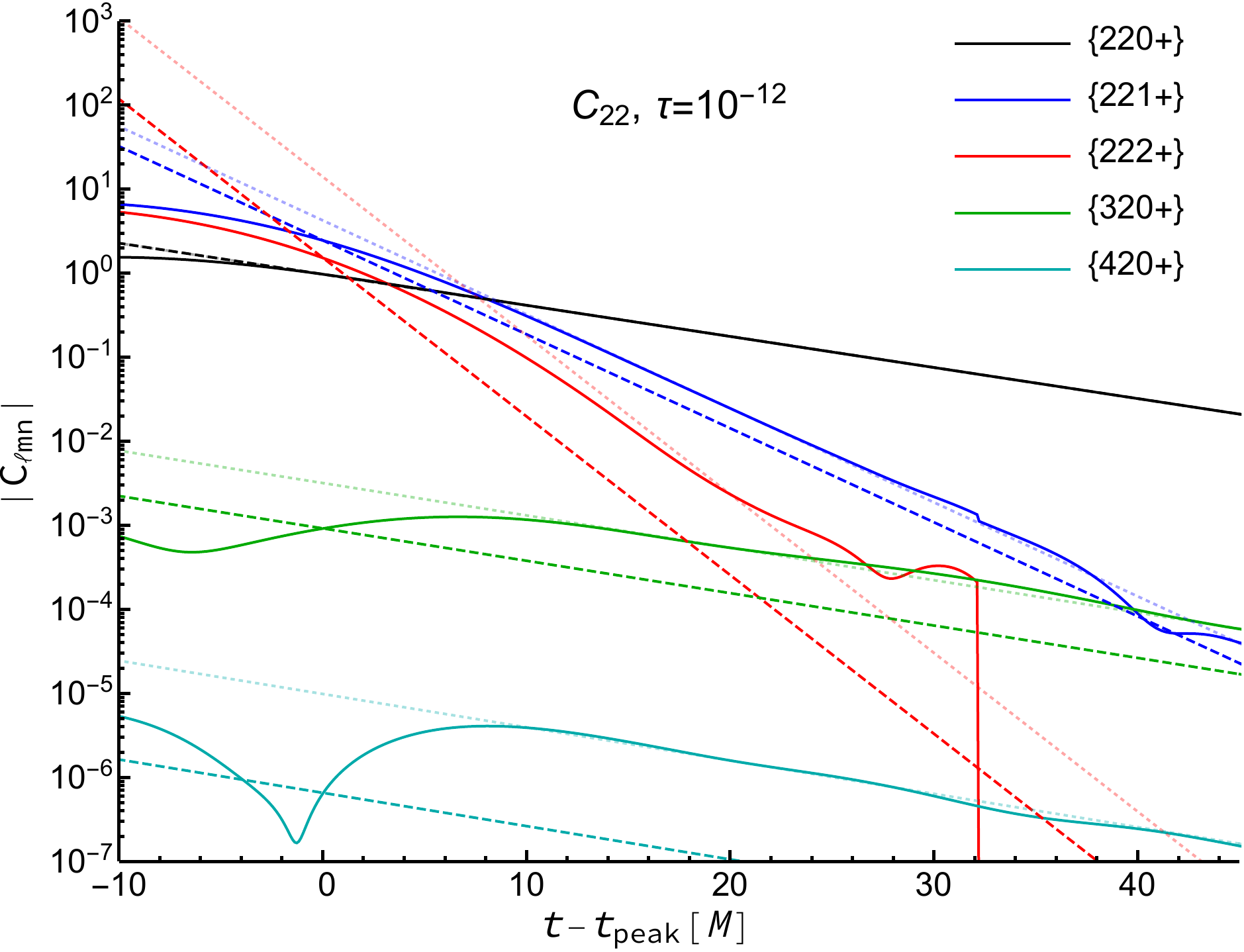}
 \caption{\label{fig:RelAmp22c_h_o7223242_Tol12}  Relative amplitude plots for fitting case $\{22,32,42*\}$ with SVD tolerance $\tau=10^{-12}$ as they contribute to $C_{22}$.  See Fig.~\ref{fig:RelAmp22a_h_o7223242} for additional detail.}
\end{figure}
The resulting relative amplitudes plot is displayed in Fig.~\ref{fig:RelAmp22c_h_o7223242_Tol12}.  We see that the relative amplitude line discretely fit at $t_i=t_{\rm peak}$ is not consistent with the previously robust amplitudes except for the $\{220+\}$ QNM.  However, when fit at $t_i-t_{\rm peak}=20M$, the agreement is reasonably good, but the continuously fit curves show robustness over a narrower range of $t_i$.  While the $\{220+\}$ QNM is reasonably robust until $t_i-t_{\rm peak}\lesssim-8M$, the other modes show poor robustness for $t_i-t_{\rm peak}\lesssim8M$ and the $\{222+\}$ QNM shows an even smaller range of robustness.
\begin{figure}
\includegraphics[width=\linewidth,clip]{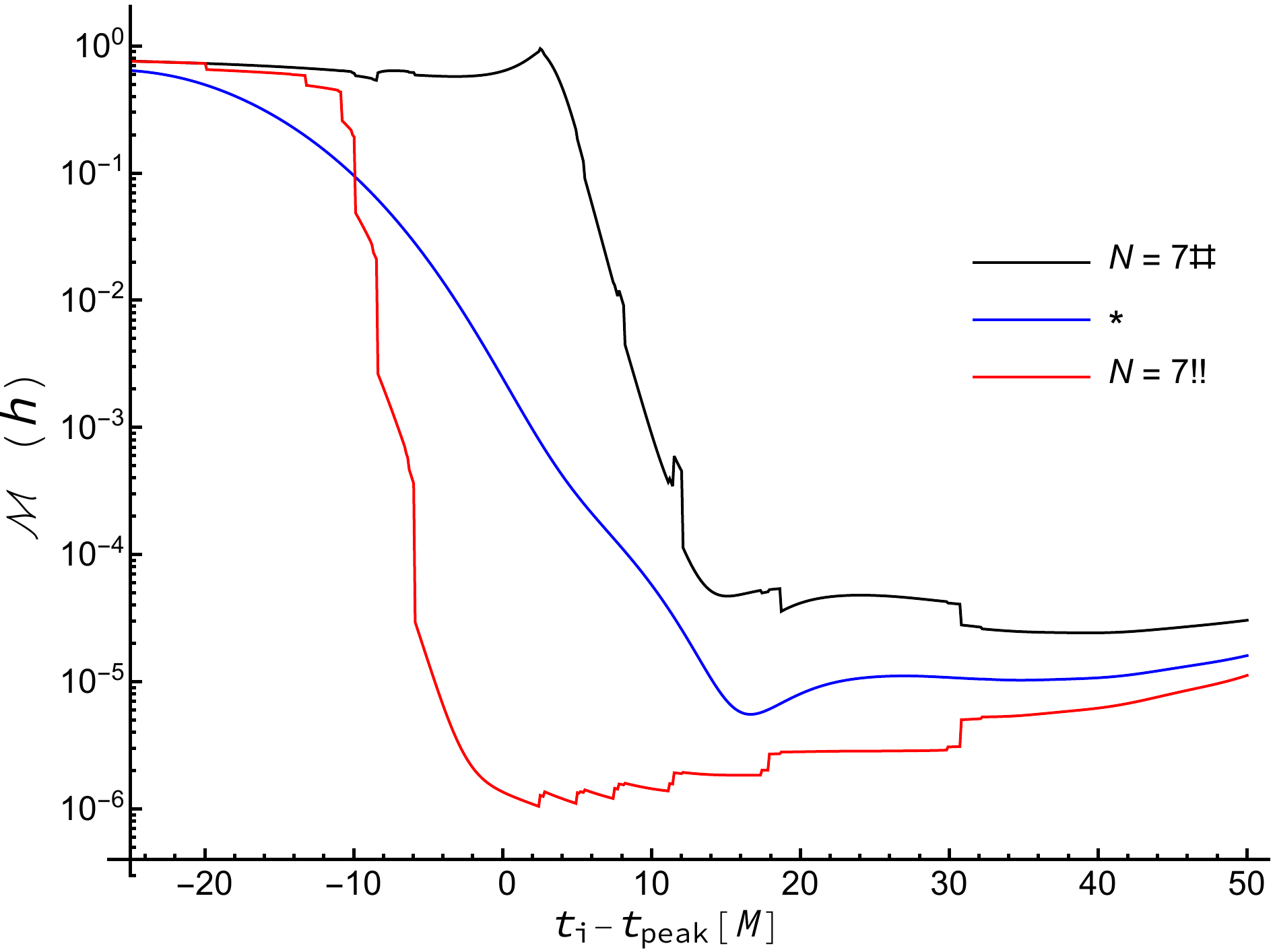}
 \caption{\label{fig:MM0305_h_o7C223242_Tol12Star}  Mismatch $\mathcal{M}$ plotted as a function of $t_i-t_{\rm peak}$ for $h$ and SVD tolerance $\tau=10^{-12}$.   The line labeled with an asterisk(*) corresponds to fitting case $\{22,32,42*\}$.  The remain two lines correspond to fitting case $\{22,32,42\}$ with $N=7$.  The line labeled $N=7!!$ was computed using all mode amplitudes.  The line labeled $N=7\#$ was computed using only the five mode amplitudes from case $\{22,32,42*\}$.}
\end{figure}
We can gain additional insights by comparing a few mismatch curves as shown in Fig.~\ref{fig:MM0305_h_o7C223242_Tol12Star}.  In this figure, the line labeled by an asterisk(*) is the mismatch for fitting case $\{22,32,42*\}$, corresponding directly with the relative amplitude plots in Fig.~\ref{fig:RelAmp22c_h_o7223242_Tol12}.  The line labeled by $N=7!!$ is the full mismatch for fitting case $\{22,32,42\}$, corresponding directly with the relative amplitude plots in Figs.~\ref{fig:RelAmp22a_h_o7223242_Tol12} and \ref{fig:RelAmp22b_h_o7223242_Tol12}.  The line labeled by $N=7\#$ is a restricted mismatch for fitting case $\{22,32,42\}$ obtained by removing the contributions from the nonrobust QNMs.  More precisely, the QNM expansion coefficients used to construct the mismatch labeled by $N=7!!$ were taken and the coefficients of the nonrobust modes were set to $0$.  Then the overlap was recomputed using Eq.~(\ref{eqn:rho2explicit}).  All fits in Fig.~\ref{fig:MM0305_h_o7C223242_Tol12Star} were performed with the EV method and $\tau=10^{-12}$.

\begin{figure}
\includegraphics[width=\linewidth,clip]{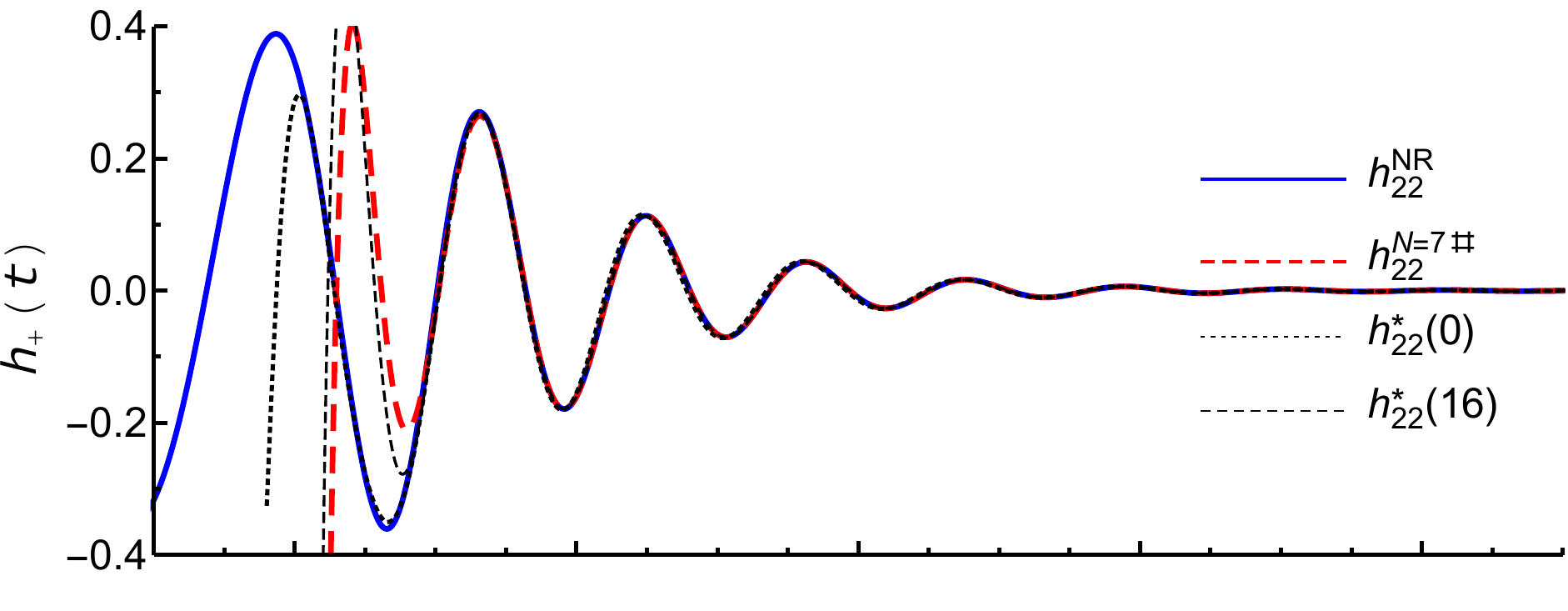}
\includegraphics[width=\linewidth,clip]{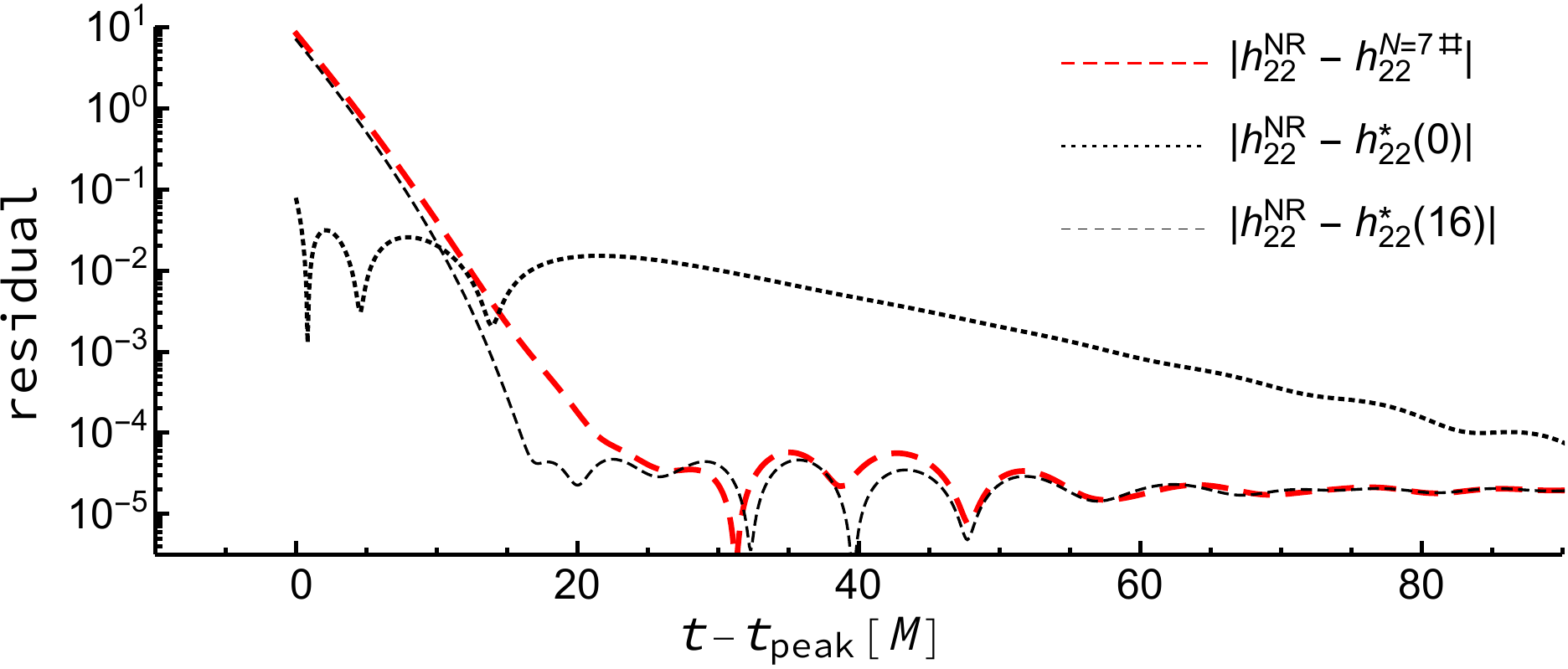}
 \caption{\label{fig:h22_o7C223242_omit} Comparison of the $C_{22}$ waveform and various fits using SVD tolerance $\tau=10^{-12}$.  The upper panel shows the $(2,2)$ mode of $h_+$ from the numerical-relativity waveform and three comparison fits.  The fit line labeled $N=7\#$ uses only the five robust modes from the $\{22,32,42\}$ fitting case with $N=7$ at $t_i=t_{\rm peak}$.  The other lines correspond to fitting case $\{22,32,42*\}$.   The line labeled $*(0)$ is for the fit at $t_i=t_{\rm peak}$, while the line labeled $*(16)$ is for the fit at $t_i-t_{\rm peak}=16M$.  The lower panel shows the magnitude of the difference of the complex signals.}
\end{figure}

Finally, Fig.~\ref{fig:h22_o7C223242_omit} shows directly the effect of using only the five robust QNMs to represent the $C_{22}$ waveform.
The upper panel in the figure shows a direct comparison of three different fits against the numerical relativity $h_+$ waveform, while the lower plot shows the magnitude of the difference between the numerical waveform and each of the fits.  The lines labeled with $N=7\#$ uses the five robust expansion coefficients from the $\{22,32,42\}$ fitting case with $N=7$ at $t_i=t_{\rm peak}$.  The other lines use the expansion coefficients from the $\{22,32,42*\}$ fitting case.  The lines labeled by $*(0)$ use the expansion coefficients from $t_i=t_{\rm peak}$, while the lines labeled by $*(16)$ use the expansion coefficients from $t_i-t_{\rm peak}=16M$ which corresponds roughly with the minimum of the mismatch for fitting case $\{22,32,42*\}$ as seen in Fig.~\ref{fig:MM0305_h_o7C223242_Tol12Star}.  Table~\ref{table:h7223242_Tol12Star} compares the actual amplitudes of the robust expansion coefficients used in the fits displayed in Fig.~\ref{fig:h22_o7C223242_omit}.
\begin{table}
\begin{tabular}{c|d{5}|d{6}|d{5}}
 Mode & \multicolumn{3}{c}{Amplitude}   \\
  & \multicolumn{1}{c}{$N=7\#$} & \multicolumn{1}{c}{*(0)} & \multicolumn{1}{c}{*(16)}  \\
 \hline\hline
 $C^+_{220}$ & 0.971 & 0.973 & 0.972  \\
 $C^+_{221}$ & 4.16 & 2.67 & 4.21  \\
 $C^+_{222}$ & 11.6 & 1.53 & 10.5  \\
 $C^+_{320}$ & 0.0319 & 0.00927 & 0.0316  \\
 $C^+_{420}$ & 0.00265 & 0.000144 & 0.00226 
\end{tabular}
\caption{\label{table:h7223242_Tol12Star} The magnitude of the QNM amplitudes from fitting $h$ with fitting case $\{22,32,42*\}$ and using SVD tolerance $\tau=10^{-12}$.  The column labeled *(0) displays the results from $t_i=t_{\rm peak}$.  The column labeled *(16) displays the results from $t_i-t_{\rm peak}=16M$.  For reference, the column labeled $N=7\#$ displays the $5$ relevant modes from fitting case $\{22,32,42\}$ with $N=7$ at $t_i=t_{\rm peak}$.}
\end{table}

Figures~\ref{fig:RelAmp22c_h_o7223242_Tol12}, \ref{fig:MM0305_h_o7C223242_Tol12Star}, and \ref{fig:h22_o7C223242_omit} provide a great deal if insight into the relevance of the nonrobust higher overtones in ringdown fitting.  It is clear that the nonrobust modes have an impact on the quality of the fitting, even for fits that start late in the ring down (i.e.\ large $t_i$).  Certainly they have the largest impact for fits that start early in the ringdown.  Including the nonrobust higher overtones allows for the more slowly decaying modes to be fit robustly, even for fits that start before $t_{\rm peak}$.  GIST addressed the concern that including higher overtones may simply offer additional degrees of freedom that can be used to fit away nonlinearities that may exist in the waveform at times near $t_{\rm peak}$.  They showed that even small fractional changes to the QNMs with $n>0$ resulted in fits which did not agree as well with the remnant parameters.  This suggests that generic complex exponential modes do not improve the fit as well as the correct QNMs.

\section{Discussion}\label{sec:discussion}

In this paper, we have examined various aspects of multimode Kerr ringdown fitting.  In Sec.~\ref{sec:conventions}, we have given detailed descriptions of two approaches for fitting the gravitational ringdown signal $\psi_{\rm NR}$ from a numerical relativity simulation to a fitting function $\psi_{\rm fit}$ consisting of a superposition of QNMs.  The first approach, referred to as the Eigenvalue Method, is fundamentally based on the overlap of $\Psi_{\rm NR}$ and $\psi_{\rm fit}$ defined in Eq.~(\ref{eqn:rho2}).  Extremizing this overlap yields a unique maximum overlap eigenvalue (\ref{eqn:rho2max}), with its eigenvector giving the expansion coefficients associated with each QNM that make up $\psi_{\rm fit}$.  The eigenvector is, in fact, identical to the solution of a related least-squares problem, Eq.~(\ref{eqn:chi2full}).  Given a set of remnant parameters $\mathcal{R}$, and a range of times $[t_i,t_e]$ over which to perform the inner products, Eq.~(\ref{eqn:innerproduct}), the Eigenvalue Method provides a simple way to compute the coefficients of the optimum fit and the overlap associated with that fit.  The second approach, referred to simply as the Least-Squares Method, is based on a traditional minimization of $\chi_{ls}^2$ as defined in Eq.~(\ref{eqn:chi2ls}).  As discussed in Sec.~\ref{sec:waveformfitting}, the two methods are closely related but distinct.  The distinction is simply that the fitting function in the Least-Squares Method is limited in its angular dependence to the angular modes present in the data being fit.  The two methods can be put on an equal footing by defining a projected, or mode-limited fitting function $\tilde\psi_{\rm fit}$ via Eq.~(\ref{eqn:proj_psi_fit_k}).  Replacing $\psi_{\rm fit}$ with $\tilde\psi_{\rm fit}$ in the Eigenvalue Method results in a method we have called the mode-limited Eigenvalue Method which is essentially the same as the Least-Squares Method.

As we have discussed in Sec.~\ref{sec:waveformfitting}, and shown in Sec.~\ref{sec:C22-higher-order}, the EV method can incorporates more information about the overlap of the fitting modes because it makes use of all of the angular information available for each QNM.  In contrast, the LS and mlEV methods projects out, for each $C_{\ell{m}}$ only the corresponding angular behavior from each QNM used to fit it.  When enough simulation modes are used, the results from the EV and mlEV methods are comparable, but when the fitting modes contain significantly more angular information than is present in the simulation modes being fit, the approaches yield discernibly different results.

We reproduced the fitting results for the SXS data set SXS:BBH:0305 obtained by GIST\cite{giesler-etal-2019}, based on the restricted fitting model of Eq.~(\ref{eqn:GISTfitmodel}), and fitting to the gravitational strain $h$.  In a minor extension, we showed that we can obtain similar results by fitting the same restricted model to the Newman-Penrose scalar $\Psi_4$.  This was done primarily to test the {\em Mathematica} codes used to implement the various fitting methods but also to provide a point of comparison when exploring the full fitting model.

In exploring the behavior of the full fitting model, we found that with both the EV  and mlEV methods, we can successfully fit using a wide range of combinations of the simulation modes that are fit to, and the QNMs with which the fits are made.  In general, it proved beneficial to include higher $\ell$ modes in the set of QNMs used to perform the fits.  It also proved beneficial to include higher $\ell$ modes in the set of simulation modes that were being fit.  Given the inherent errors in the numerical simulation, it was not helpful to go beyond $\ell=4$.  And, as first described in GIST, the inclusion of overtones up to $n=7$ is essential in allowing the fits to be performed well at times as early as, or even earlier than, $t_{\rm peak}$.  For all practical purposes, the results for the $\{22,32,42\}$ fitting case with $N=7$ were as effective as the fits performed by GIST with the restricted model and $N=7$.  Given that the fitting model used by GIST ignored the spheroidal harmonic expansion amplitudes, perhaps it is more appropriate to say that their results are, for all practical purposes, as good as those we obtained.  It is remarkable how closely the amplitudes of the various QNMs agree when comparing the two approaches (see Table~\ref{table:h7noA} as compared to Table~\ref{table:h7223242}).  Of course, using multimode fitting with full fitting that correctly includes the spheroidal harmonic expansion coefficients allows us to obtain useful information about the $\ell=3$ and $4$, $m=2$ QNMs.  And, while details of the fitting were not included in this paper, simply trying to fit the $C_{32}$ simulation mode to the set $\{{\rm QNM}\}$ from fitting case $\{22\text{+3}\}$ yields rather poor results.  Much better results are obtained with the multimode approach and fitting several $C_{\ell{m}}$ simultaneously to the same set of QNMs.

When using the overlap (or mismatch) to gauge the quality of a fit, we have seen that it is important to clearly specify how the overlap is computed.  By default, the EV and the mlEV methods use Eq.~(\ref{eqn:rho2max}) to compute the overlap.  Note that Eq.~(\ref{eqn:rho2max}) is simply an alternative to Eq.~(\ref{eqn:rho2explicit}) for computing the overlap when $\mathbb{B}^{-1}$ has been computed and the expansion coefficients are computed via Eq.~(\ref{eqn:least-squares_coefs}).  The LS method computes the expansion coefficients in a way that is nearly identical to the mlEV method.  The overlap is then computed using Eq.~(\ref{eqn:rho2explicit}).  If the overlap is computed using $\mathbb{B}$ as in the EV method, then the results can seem quite poor (see Fig.~\ref{fig:h7_C22NEa}).  Instead, for the LS method, it is most appropriate to compute the overlap using $\mathbb{B}_{\rm ml}$.

In many of the examples we explored in Sec.~\ref{sec:sxsbbh0305}, the mismatch was used simply as a measure of the quality of a given fit.  In those cases, the option to recompute the mismatch of the EV method results using $\mathbb{B}_{\rm ml}$ allowed for a more direct comparison between approaches.  In particular, the mismatch used by GIST (and in fitting case $\{22\mbox{-$\mathcal{A}$}\}$) corresponds to the mode-limited approach, and comparison with the mode-limited mismatch for the fits made with the EV method shows that including the values of the spheroidal expansion coefficients does not degrade the fit as significantly as it seems when comparing with the default EV mismatch.  A reasonable question to ask is whether or not one of these two methods for computing the overlap is preferable.  If the mismatch were used only as a simple measure of the quality of a given fit, then the question is probably not terribly important.  One would simply use whichever version made for the most convenient comparison with other fits.  

However, the choice of which mismatch we use plays a much more important role when we consider the nonlinear fitting of model parameters because this involves minimizing the value of the mismatch.  The first evidence of the importance  of this choice was displayed in Figs.~\ref{fig:MMcp_h22_o7C22_0} and \ref{fig:MMcp_h22_mix_o7C22_0}.
In particular for the $n=7$ panel, the color-density plots show a much narrower and deeper minimum when the mode-limited mismatch is used.  The fit-series plots in Figs.~\ref{fig:FitMM_h_CvsN}, \ref{fig:Fitdelta_h_CvsN}, \ref{fig:Fitchi_h_CvsN}, and \ref{fig:FitAll_h_N7_mixml} show the same effect, and more, in much more detail.  Comparing the $N=7$ panels of Figs.~\ref{fig:FitMM_h_CvsN}, \ref{fig:Fitdelta_h_CvsN}, and \ref{fig:Fitchi_h_CvsN} with the left column of Fig.~\ref{fig:FitAll_h_N7_mixml} shows more quantitatively the same difference observed in the color-density plots.  In these cases, the expansion coefficients $C^\pm_{\ell{m}n}$ are computed for a given set of model parameters using $\mathbb{B}$.  The difference is that the model parameters are obtained by minimizing the mismatch computed using $\mathbb{B}$ for Figs.~\ref{fig:FitMM_h_CvsN}, \ref{fig:Fitdelta_h_CvsN}, and \ref{fig:Fitchi_h_CvsN}, but are obtained by minimizing the mismatch computed using $\mathbb{B}_{\rm ml}$ for Fig.~\ref{fig:FitAll_h_N7_mixml}.  The right column of Fig.~\ref{fig:FitAll_h_N7_mixml} shows yet another option.  In this case, the expansion coefficients $C^\pm_{\ell{m}n}$ are also computed using $\mathbb{B}_{\rm ml}$.

To make the differences of these various approaches as clear as possible, we present the equations used to compute the expansion coefficients $C^\pm_{\ell{m}n}$ and the overlap $\rho$ for each method:
\begin{enumerate}
	\item The Eigenvalue (EV) Method computes both $\rho$ and $C^\pm_{\ell{m}n}$ using the full $\mathbb{B}$ and $\mathbb{B}^{-1}$.  The relevant equations are
	\begin{equation}
	\rho^2 = \frac{\vec{A}^\dag\!\cdot\mathbb{B}^{{}^{-1}}\!\!\!\cdot\!\vec{A}}
	{\langle\psi_{\rm NR}|\psi_{\rm NR}\rangle} \quad:\quad \vec{C} = \mathbb{B}^{{}^{-1}}\!\!\!\cdot\!\vec{A}.
	\end{equation}
	\item The mixed Eigenvalue (mixEV) Method differs from the EV method by computing $\rho$ using the mode-limited $\mathbb{B}_{ml}$. The relevant equations are
	\begin{equation}
	\rho^2 = \frac{\left(\vec{A}^\dag\!\cdot\mathbb{B}^{{}^{-1}}\!\!\!\cdot\!\vec{A}\right)^2}
	{\langle\psi_{\rm NR}|\psi_{\rm NR}\rangle\vec{A}^\dag\!\cdot\!\mathbb{B}^{{}^{-1}}\!\!\cdot\mathbb{B}_{\rm ml}\!\cdot\mathbb{B}^{{}^{-1}}\!\!\!\cdot\!\vec{A}} \ :\ \vec{C} = \mathbb{B}^{{}^{-1}}\!\!\!\cdot\!\vec{A}.
	\end{equation}
	\item The mode-limited Eigenvalue (mlEV) Method computes both $\rho$ and $C^\pm_{\ell{m}n}$ using the mode-limited $\mathbb{B}_{\rm ml}$ and $\mathbb{B}^{-1}_{\rm ml}$.  The relevant equations are
	\begin{equation}
	\rho^2 = \frac{\vec{A}^\dag\!\cdot\mathbb{B}_{\rm ml}^{{}^{-1}}\!\!\cdot\!\vec{A}}
	{\langle\psi_{\rm NR}|\psi_{\rm NR}\rangle} \quad:\quad \vec{C} = \mathbb{B}_{\rm ml}^{{}^{-1}}\!\!\cdot\!\vec{A}.
	\end{equation}
\end{enumerate}
Clearly, a fourth permutation exists, but it is less well motivated and will not be considered further.
\begin{enumerate}
	\item[4.] The relevant equations of the fourth permutation are
	\begin{equation}
	\rho^2 = \frac{\left(\vec{A}^\dag\!\cdot\mathbb{B}_{\rm ml}^{{}^{-1}}\!\!\!\cdot\!\vec{A}\right)^2}
	{\langle\psi_{\rm NR}|\psi_{\rm NR}\rangle\vec{A}^\dag\!\cdot\!\mathbb{B}_{\rm ml}^{{}^{-1}}\!\!\cdot\mathbb{B}\!\cdot\mathbb{B}_{\rm ml}^{{}^{-1}}\!\!\!\cdot\!\vec{A}} \ :\ \vec{C} = \mathbb{B}_{\rm ml}^{{}^{-1}}\!\!\!\cdot\!\vec{A}.
	\end{equation}
\end{enumerate}

Clearly, there is very little difference between the approaches for the $\{22,32,42\}$ case.  For the $\{22,32\}$ case, the differences are still small but are easily visible in the figures.  But, for the $\{22\}$ case, the choice of method makes a substantial difference.  For a more quantitative comparison, Table~\ref{table:modelfitcmp} compares the three fitting methods for the three main fitting cases.  All fits are for $N=7$ and $\tau=0$.   For each fit, the top line gives the mismatch $\mathcal{M}$, dimensionless mass ratio $\delta$, and dimensionless spin $\chi_f$ obtained at $t_i=t_{\rm peak}$.  The second line gives the fit values averaged over the range $0\le t_i-t_{\rm peak}\le40M$.  The third line gives the root-mean-square (rms) deviation of the fit value from the corresponding value from $\mathcal{R}_{\rm NR}$.
Table~\ref{table:modelfitcmp} shows the results of fitting for the model parameters using the first three of these versions of the Eigenvalue Method.
\begin{table}
\begin{tabular}{lr|ld{4}d{4}}
 && \multicolumn{1}{c}{$\mathcal{M}$} & \multicolumn{1}{c}{$\delta$} & \multicolumn{1}{c}{$\chi_f$}  \\
 \hline\hline
 \multicolumn{2}{c|}{NR Simulation} && 0.9520 & 0.6921 \\ % NR
  &&& \pm0.0003 & \pm0.0007 \\ % NR
 \hline
 && $1.093\times10^{-3}$ & 0.8643 & 0.5461 \\ 
 & EV && 0.8023 & 0.4086 \\ 
 &&& \pm0.15 & \pm0.29 \\
\cline{2-5}
 && $5.977\times10^{-6}$ & 0.9490 & 0.6889 \\ 
 $\{22\},N=7$: & mixEV && 0.9419 & 0.6795 \\ 
 &&& \pm0.011 & \pm0.014 \\
\cline{2-5}
 && $2.521\times10^{-7}$ & 0.9518 & 0.6923 \\ 
 & mlEV && 0.9516 & 0.6917 \\ 
 &&& \pm0.0024 & \pm0.0031 \\
 \hline
 && $9.021\times10^{-6}$ & 0.9504 & 0.6897 \\ 
 & EV && 0.9481 & 0.6869 \\ 
 &&& \pm0.0042 & \pm0.0055 \\
\cline{2-5}
 && $1.312\times10^{-6}$ & 0.9515 & 0.6911 \\ 
 $\{22,32\},N=7$: & mixEV && 0.9517 & 0.6916 \\ 
 &&& \pm0.0008 & \pm0.0009 \\
\cline{2-5}
 && $1.261\times10^{-6}$ & 0.9515 & 0.6912 \\ 
 & mlEV && 0.9518 & 0.6917 \\ 
 &&& \pm0.0007 & \pm0.0009 \\
 \hline
 && $1.339\times10^{-6}$ & 0.9516 & 0.6912 \\ 
 & EV && 0.9518 & 0.6917 \\ 
 &&& \pm0.0007 & \pm0.0009 \\
\cline{2-5}
 && $1.299\times10^{-6}$ & 0.9516 & 0.6912 \\ 
 $\{22,32,42\},N=7$: & mixEV && 0.9518 & 0.6917 \\ 
 &&& \pm0.0007 & \pm0.0009 \\
\cline{2-5}
 && $1.297\times10^{-6}$ & 0.9516 & 0.6912 \\ 
 & mlEV && 0.9518 & 0.6918 \\ 
 &&& \pm0.0007 & \pm0.0009 \\
\end{tabular}
 \caption{\label{table:modelfitcmp} Comparison of minimum-$\mathcal{M}$ model-fit parameters for each of the three fitting cases, $\{22\}$, $\{22,32\}$, and $\{22,32,42\}$ with $N=7$ and $\tau=0$.  For each case, we compute the model parameters using the EV, mixEV, and mlEV methods.  Note that the results from the mlEV version of the $\{22\}$ case are equivalent to those from case $\{22\mbox{-$\mathcal{A}$}\}$.  For comparison, the values from the numerical simulation are on the top line.  For each case, the upper line gives the $\mathcal{M}$, $\delta$, and $\chi_f$ values extracted at $t_i=t_{\rm peak}$.  The second line presents the model parameters averaged over $0\le t_i-t_{\rm peak}\le40M$.  The bottom line gives the rms deviations of $\delta$ and $\chi_f$ from their corresponding simulation values over the same interval.}
\end{table}
It is clear from Table~\ref{table:modelfitcmp} that when a sufficient number of simulation modes are used, the results produced by the three versions of the Eigenvalue Method converge.  But the three methods are clearly different for the $\{22\}$ case.  Interestingly, we see that the best agreement for the minimum-$\mathcal{M}$ fit model parameters at $t_i=t_{\rm peak}$ with the remnant parameters $\mathcal{R}_{\rm NR}$ occurs in the mlEV version of $\{22\}$ which also displays the smallest mismatch $\mathcal{M}$.  On the other hand, when averaged over $0\le t_i-t_{\rm peak}\le40M$, all three methods for the $\{22,32,42\}$ case are nearly identical and are as good as, or better than, all other examples.

An interesting question would be: is there $1$ of the $3$ methods that can be considered best?  I do not think that the examples explored in this paper can satisfactorily answer this question.  If one is simply looking to get the best minimum-$\mathcal{M}$ fit, then the results we have presented suggest that the mlEV method, which is essentially the standard Least-Squares method, is best because it produced nearly identical results for all three fitting cases.  The uncertainties are simply smaller for the $\{22,32\}$ and $\{22,32,42\}$ cases than for the $\{22\}$ case.  But, is this better agreement just by chance for this data set?  

Clearly the question of which method is best only arises when the set of simulation modes $\{\rm NR\}$ being fit is not large enough to adequately match the angular-function space of the set of modes $\{\rm QNM\}$ that constitute the fit function.  In this case, one can argue that the small mismatch from the mlEV method could be giving false confidence in the results.  On the other hand, while the standard EV method may under perform when the set $\{\rm NR\}$ is too small, it does correctly indicate that more simulation and/or fitting modes are needed to have confidence in the results.

The work in this paper has been targeted at investigating the ringdown signal from numerical simulations where the waveform is known over all angles, as opposed to fitting detected gravitational wave signals.  Since numerical simulations can independently determine the mass and angular momentum of the remnant black hole, fitting these values serves primarily as a confidence test.  The real goal of fitting the ringdown signal is to explore the "spectroscopy" of the ringdown.  That is, to understand which modes are excited, and by how much, during a black hole binary collision or some other dynamic event that leaves behind a remnant black hole.  Exploring this question necessarily includes exploring the question of when the onset of ringdown occurs.

Figures~\ref{fig:RelAmp22a_h_o7223242_Tol12}, \ref{fig:RelAmp22c_h_o7223242_Tol12}, \ref{fig:MM0305_h_o7C223242_Tol12Star}, and \ref{fig:h22_o7C223242_omit} show that the dominant $m=2$ modes of the ringdown signal are well fit by five modes in $\{\rm QNN_{rob}\}$ for $t_i-t_{\rm peak}\gtrsim16M$ when the expansion coefficients are fixed by their values at $t_i-t_{\rm peak}=16M$.  Furthermore, Table~\ref{table:h7223242_Tol12Star} shows that values for the same expansion coefficients can be obtained at $t_i=t_{\rm peak}$ if many additional modes are used.  This suggests that the onset of ringdown has occurred by $t_i-t_{\rm peak}=16M$, but does not place a more stringent bound on the onset of ringdown.  A better understanding of the onset of ringdown will almost certainly benefit from quantitative information about the ``Kerrness'' of the remnant spacetime as discussed in Ref.~\cite{bhagwat-etal-2018}.

Another question that remains open regards the reliability of the nonrobust modes used in fits.  Their use clearly makes other modes appear more robust even though they are not robust themselves.  But, is it reasonable to think that these highly damped overtones can be robustly fit?  We have seen that, because they become so small late in the ringdown, including them in the fit can cause fitting errors.  So, clearly it seems that we should only consider their robustness over a smaller interval of time early in the fitting.  It seems likely that the time interval for the highest $n$ overtones will be too short to make any kind of determination of their robustness using continuously-fit relative amplitude plots.  Is there some other way to measure the reliability of these higher overtones?

In this work, we have only explored the dominant $m=2$ modes of one simulation.  Certainly, we can explore many additional sets of modes.  We have also not presented results that explore an off axis angular-momentum vector.  However, both have been explored at a preliminary level by Zalutskiy\cite{Thesis:Zalutskiy}.  Neither present any problems.  The most interesting aspect of exploring simulations in which the angular momentum points off axis is that it couples together multiple $m$-modes.  With multimode fitting, these can be successfully fit simultaneously.  However, it would also be possible to preprocess the waveform, effectively rotating the numerical relativity simulation frame to align the remnant angular momentum.  This would eliminate the coupling between $m$-modes and simplify the ringdown fitting.  The open question is whether or not this is computationally advantageous.

\acknowledgments Some computations were performed on the Wake Forest
University DEAC Cluster, a centrally managed resource with support
provided in part by the University.


\begin{thebibliography}{34}%
\makeatletter
\providecommand \@ifxundefined [1]{%
 \@ifx{#1\undefined}
}%
\providecommand \@ifnum [1]{%
 \ifnum #1\expandafter \@firstoftwo
 \else \expandafter \@secondoftwo
 \fi
}%
\providecommand \@ifx [1]{%
 \ifx #1\expandafter \@firstoftwo
 \else \expandafter \@secondoftwo
 \fi
}%
\providecommand \natexlab [1]{#1}%
\providecommand \enquote  [1]{``#1''}%
\providecommand \bibnamefont  [1]{#1}%
\providecommand \bibfnamefont [1]{#1}%
\providecommand \citenamefont [1]{#1}%
\providecommand \href@noop [0]{\@secondoftwo}%
\providecommand \href [0]{\begingroup \@sanitize@url \@href}%
\providecommand \@href[1]{\@@startlink{#1}\@@href}%
\providecommand \@@href[1]{\endgroup#1\@@endlink}%
\providecommand \@sanitize@url [0]{\catcode `\\12\catcode `\$12\catcode
  `\&12\catcode `\#12\catcode `\^12\catcode `\_12\catcode `\%12\relax}%
\providecommand \@@startlink[1]{}%
\providecommand \@@endlink[0]{}%
\providecommand \url  [0]{\begingroup\@sanitize@url \@url }%
\providecommand \@url [1]{\endgroup\@href {#1}{\urlprefix }}%
\providecommand \urlprefix  [0]{URL }%
\providecommand \Eprint [0]{\href }%
\providecommand \doibase [0]{https://doi.org/}%
\providecommand \selectlanguage [0]{\@gobble}%
\providecommand \bibinfo  [0]{\@secondoftwo}%
\providecommand \bibfield  [0]{\@secondoftwo}%
\providecommand \translation [1]{[#1]}%
\providecommand \BibitemOpen [0]{}%
\providecommand \bibitemStop [0]{}%
\providecommand \bibitemNoStop [0]{.\EOS\space}%
\providecommand \EOS [0]{\spacefactor3000\relax}%
\providecommand \BibitemShut  [1]{\csname bibitem#1\endcsname}%
\let\auto@bib@innerbib\@empty
%</preamble>
\bibitem [{\citenamefont {Abbott}\ \emph
  {et~al.}(2016{\natexlab{a}})\citenamefont {Abbott} \emph
  {et~al.}}]{LIGO-grtests-2016}%
  \BibitemOpen
  \bibfield  {author} {\bibinfo {author} {\bibfnamefont {B.~P.}\ \bibnamefont
  {Abbott}} \emph {et~al.} (\bibinfo {collaboration} {LIGO Scientific and Virgo
  Collaborations}),\ }\bibfield  {title} {\bibinfo {title} {Tests of general
  relativity with {GW150914}},\ }\href
  {https://doi.org/10.1103/PhysRevLett.116.221101} {\bibfield  {journal}
  {\bibinfo  {journal} {Phys. Rev. Lett.}\ }\textbf {\bibinfo {volume} {116}},\
  \bibinfo {pages} {221101} (\bibinfo {year} {2016}{\natexlab{a}})}\BibitemShut
  {NoStop}%
\bibitem [{\citenamefont {Isi}\ \emph {et~al.}(2019)\citenamefont {Isi},
  \citenamefont {Giesler}, \citenamefont {Farr}, \citenamefont {Scheel},\ and\
  \citenamefont {Teukolsky}}]{Isi-etal-2019}%
  \BibitemOpen
  \bibfield  {author} {\bibinfo {author} {\bibfnamefont {M.}~\bibnamefont
  {Isi}}, \bibinfo {author} {\bibfnamefont {M.}~\bibnamefont {Giesler}},
  \bibinfo {author} {\bibfnamefont {W.~M.}\ \bibnamefont {Farr}}, \bibinfo
  {author} {\bibfnamefont {M.~A.}\ \bibnamefont {Scheel}},\ and\ \bibinfo
  {author} {\bibfnamefont {S.~A.}\ \bibnamefont {Teukolsky}},\ }\bibfield
  {title} {\bibinfo {title} {Testing the no-hair theorem with {GW150914}},\
  }\href {https://doi.org/10.1103/PhysRevLett.123.111102} {\bibfield  {journal}
  {\bibinfo  {journal} {Phys. Rev. Lett.}\ }\textbf {\bibinfo {volume} {123}},\
  \bibinfo {pages} {111102} (\bibinfo {year} {2019})}\BibitemShut {NoStop}%
\bibitem [{\citenamefont {Carullo}\ \emph {et~al.}(2019)\citenamefont
  {Carullo}, \citenamefont {Del~Pozzo},\ and\ \citenamefont
  {Veitch}}]{carullo-pozzo-veitch-2019}%
  \BibitemOpen
  \bibfield  {author} {\bibinfo {author} {\bibfnamefont {G.}~\bibnamefont
  {Carullo}}, \bibinfo {author} {\bibfnamefont {W.}~\bibnamefont {Del~Pozzo}},\
  and\ \bibinfo {author} {\bibfnamefont {J.}~\bibnamefont {Veitch}},\
  }\bibfield  {title} {\bibinfo {title} {Observational black hole spectroscopy:
  A time-domain multimode analysis of {GW150914}},\ }\href
  {https://doi.org/10.1103/PhysRevD.99.123029} {\bibfield  {journal} {\bibinfo
  {journal} {Phys. Rev. D}\ }\textbf {\bibinfo {volume} {99}},\ \bibinfo
  {pages} {123029} (\bibinfo {year} {2019})}\BibitemShut {NoStop}%
\bibitem [{\citenamefont {Babak}\ \emph {et~al.}(2017)\citenamefont {Babak},
  \citenamefont {Taracchini},\ and\ \citenamefont
  {Buonanno}}]{babak-taracchini-buonanno-2017}%
  \BibitemOpen
  \bibfield  {author} {\bibinfo {author} {\bibfnamefont {S.}~\bibnamefont
  {Babak}}, \bibinfo {author} {\bibfnamefont {A.}~\bibnamefont {Taracchini}},\
  and\ \bibinfo {author} {\bibfnamefont {A.}~\bibnamefont {Buonanno}},\
  }\bibfield  {title} {\bibinfo {title} {Validating the effective-one-body
  model of spinning, precessing binary black holes against numerical
  relativity},\ }\href {https://doi.org/10.1103/PhysRevD.95.024010} {\bibfield
  {journal} {\bibinfo  {journal} {Phys. Rev. D}\ }\textbf {\bibinfo {volume}
  {95}},\ \bibinfo {pages} {024010} (\bibinfo {year} {2017})}\BibitemShut
  {NoStop}%
\bibitem [{\citenamefont {Bhagwat}\ \emph {et~al.}(2018)\citenamefont
  {Bhagwat}, \citenamefont {Okounkova}, \citenamefont {Ballmer}, \citenamefont
  {Brown}, \citenamefont {Giesler}, \citenamefont {Scheel},\ and\ \citenamefont
  {Teukolsky}}]{bhagwat-etal-2018}%
  \BibitemOpen
  \bibfield  {author} {\bibinfo {author} {\bibfnamefont {S.}~\bibnamefont
  {Bhagwat}}, \bibinfo {author} {\bibfnamefont {M.}~\bibnamefont {Okounkova}},
  \bibinfo {author} {\bibfnamefont {S.~W.}\ \bibnamefont {Ballmer}}, \bibinfo
  {author} {\bibfnamefont {D.~A.}\ \bibnamefont {Brown}}, \bibinfo {author}
  {\bibfnamefont {M.}~\bibnamefont {Giesler}}, \bibinfo {author} {\bibfnamefont
  {M.~A.}\ \bibnamefont {Scheel}},\ and\ \bibinfo {author} {\bibfnamefont
  {S.~A.}\ \bibnamefont {Teukolsky}},\ }\bibfield  {title} {\bibinfo {title}
  {On choosing the start time of binary black hole ringdowns},\ }\href
  {https://doi.org/10.1103/PhysRevD.97.104065} {\bibfield  {journal} {\bibinfo
  {journal} {Phys. Rev. D}\ }\textbf {\bibinfo {volume} {97}},\ \bibinfo
  {pages} {104065} (\bibinfo {year} {2018})}\BibitemShut {NoStop}%
\bibitem [{\citenamefont {Giesler}\ \emph {et~al.}(2019)\citenamefont
  {Giesler}, \citenamefont {Isi}, \citenamefont {Scheel},\ and\ \citenamefont
  {Teukolsky}}]{giesler-etal-2019}%
  \BibitemOpen
  \bibfield  {author} {\bibinfo {author} {\bibfnamefont {M.}~\bibnamefont
  {Giesler}}, \bibinfo {author} {\bibfnamefont {M.}~\bibnamefont {Isi}},
  \bibinfo {author} {\bibfnamefont {M.~A.}\ \bibnamefont {Scheel}},\ and\
  \bibinfo {author} {\bibfnamefont {S.~A.}\ \bibnamefont {Teukolsky}},\
  }\bibfield  {title} {\bibinfo {title} {Black hole ringdown: the importance of
  overtones},\ }\href {https://doi.org/10.1103/PhysRevX.9.041060} {\bibfield
  {journal} {\bibinfo  {journal} {Phys. Rev. X}\ }\textbf {\bibinfo {volume}
  {9}},\ \bibinfo {pages} {041060} (\bibinfo {year} {2019})}\BibitemShut
  {NoStop}%
\bibitem [{\citenamefont {Hughes}\ \emph {et~al.}(2019)\citenamefont {Hughes},
  \citenamefont {Apte}, \citenamefont {Khanna},\ and\ \citenamefont
  {Lim}}]{huges-etal-2019a}%
  \BibitemOpen
  \bibfield  {author} {\bibinfo {author} {\bibfnamefont {S.~A.}\ \bibnamefont
  {Hughes}}, \bibinfo {author} {\bibfnamefont {A.}~\bibnamefont {Apte}},
  \bibinfo {author} {\bibfnamefont {G.}~\bibnamefont {Khanna}},\ and\ \bibinfo
  {author} {\bibfnamefont {H.}~\bibnamefont {Lim}},\ }\bibfield  {title}
  {\bibinfo {title} {Learning about black hole binaries from their ringdown
  spectra},\ }\href {https://doi.org/10.1103/PhysRevLett.123.161101} {\bibfield
   {journal} {\bibinfo  {journal} {Phys. Rev. Lett.}\ }\textbf {\bibinfo
  {volume} {123}},\ \bibinfo {pages} {161101} (\bibinfo {year}
  {2019})}\BibitemShut {NoStop}%
\bibitem [{\citenamefont {Buonanno}\ \emph {et~al.}(2007)\citenamefont
  {Buonanno}, \citenamefont {Cook},\ and\ \citenamefont
  {Pretorius}}]{BCP-2007}%
  \BibitemOpen
  \bibfield  {author} {\bibinfo {author} {\bibfnamefont {A.}~\bibnamefont
  {Buonanno}}, \bibinfo {author} {\bibfnamefont {G.~B.}\ \bibnamefont {Cook}},\
  and\ \bibinfo {author} {\bibfnamefont {F.}~\bibnamefont {Pretorius}},\
  }\bibfield  {title} {\bibinfo {title} {Inspiral, merger, and ring-down of
  equal-mass black-hole binaries},\ }\href
  {https://doi.org/10.1103/PhysRevD.75.124018} {\bibfield  {journal} {\bibinfo
  {journal} {Phys. Rev. D}\ }\textbf {\bibinfo {volume} {75}},\ \bibinfo
  {pages} {124018} (\bibinfo {year} {2007})}\BibitemShut {NoStop}%
\bibitem [{\citenamefont {London}\ \emph {et~al.}(2014)\citenamefont {London},
  \citenamefont {Shoemaker},\ and\ \citenamefont
  {Healy}}]{londonShoemakerhealy-2014}%
  \BibitemOpen
  \bibfield  {author} {\bibinfo {author} {\bibfnamefont {L.}~\bibnamefont
  {London}}, \bibinfo {author} {\bibfnamefont {D.}~\bibnamefont {Shoemaker}},\
  and\ \bibinfo {author} {\bibfnamefont {J.}~\bibnamefont {Healy}},\ }\bibfield
   {title} {\bibinfo {title} {Modeling ringdown: Beyond the fundamental
  quasinormal modes},\ }\href {https://doi.org/10.1103/PhysRevD.90.124032}
  {\bibfield  {journal} {\bibinfo  {journal} {Phys. Rev. D}\ }\textbf {\bibinfo
  {volume} {90}},\ \bibinfo {pages} {124032} (\bibinfo {year}
  {2014})}\BibitemShut {NoStop}%
\bibitem [{\citenamefont {Kamaretsos}\ \emph
  {et~al.}(2012{\natexlab{a}})\citenamefont {Kamaretsos}, \citenamefont
  {Hannam}, \citenamefont {Husa},\ and\ \citenamefont
  {Sathyaprakash}}]{kamaretsos-etal-2012a}%
  \BibitemOpen
  \bibfield  {author} {\bibinfo {author} {\bibfnamefont {I.}~\bibnamefont
  {Kamaretsos}}, \bibinfo {author} {\bibfnamefont {M.}~\bibnamefont {Hannam}},
  \bibinfo {author} {\bibfnamefont {S.}~\bibnamefont {Husa}},\ and\ \bibinfo
  {author} {\bibfnamefont {B.~S.}\ \bibnamefont {Sathyaprakash}},\ }\bibfield
  {title} {\bibinfo {title} {Black-hole hair loss: Learning about binary
  progenitors from ringdown signals},\ }\href
  {https://doi.org/10.1103/PhysRevD.85.024018} {\bibfield  {journal} {\bibinfo
  {journal} {Phys. Rev. D}\ }\textbf {\bibinfo {volume} {85}},\ \bibinfo
  {pages} {024018} (\bibinfo {year} {2012}{\natexlab{a}})}\BibitemShut
  {NoStop}%
\bibitem [{\citenamefont {Lim}\ \emph {et~al.}(2019)\citenamefont {Lim},
  \citenamefont {Khanna}, \citenamefont {Apte},\ and\ \citenamefont
  {Hughes}}]{huges-etal-2019b}%
  \BibitemOpen
  \bibfield  {author} {\bibinfo {author} {\bibfnamefont {H.}~\bibnamefont
  {Lim}}, \bibinfo {author} {\bibfnamefont {G.}~\bibnamefont {Khanna}},
  \bibinfo {author} {\bibfnamefont {A.}~\bibnamefont {Apte}},\ and\ \bibinfo
  {author} {\bibfnamefont {S.~A.}\ \bibnamefont {Hughes}},\ }\bibfield  {title}
  {\bibinfo {title} {Exciting black hole modes via misaligned coalescences.
  {II}. the mode content of late-time coalescence waveforms},\ }\href
  {https://doi.org/10.1103/PhysRevD.100.084032} {\bibfield  {journal} {\bibinfo
   {journal} {Phys. Rev. D}\ }\textbf {\bibinfo {volume} {100}},\ \bibinfo
  {pages} {084032} (\bibinfo {year} {2019})}\BibitemShut {NoStop}%
\bibitem [{\citenamefont {Zimmerman}\ and\ \citenamefont
  {Chen}(2011)}]{Zimmerman-Chen-2011}%
  \BibitemOpen
  \bibfield  {author} {\bibinfo {author} {\bibfnamefont {A.}~\bibnamefont
  {Zimmerman}}\ and\ \bibinfo {author} {\bibfnamefont {Y.}~\bibnamefont
  {Chen}},\ }\bibfield  {title} {\bibinfo {title} {New generic ringdown
  frequencies at the birth of a {K}err black hole},\ }\href
  {https://doi.org/10.1103/PhysRevD.84.084012} {\bibfield  {journal} {\bibinfo
  {journal} {Phys. Rev. D}\ }\textbf {\bibinfo {volume} {84}},\ \bibinfo
  {pages} {084012} (\bibinfo {year} {2011})}\BibitemShut {NoStop}%
\bibitem [{\citenamefont {Zalutskiy}(2016)}]{Thesis:Zalutskiy}%
  \BibitemOpen
  \bibfield  {author} {\bibinfo {author} {\bibfnamefont {M.~P.}\ \bibnamefont
  {Zalutskiy}},\ }\emph {\bibinfo {title} {Investigations of Black-Hole
  Spectra: Purely-imaginary Modes and {K}err ringdown Radiation}},\ \href@noop
  {} {Ph.D. thesis},\ \bibinfo  {school} {Wake Forest University} (\bibinfo
  {year} {2016})\BibitemShut {NoStop}%
\bibitem [{\citenamefont {Kamaretsos}\ \emph
  {et~al.}(2012{\natexlab{b}})\citenamefont {Kamaretsos}, \citenamefont
  {Hannam},\ and\ \citenamefont {Sathyaprakash}}]{kamaretsos-et-al-2012b}%
  \BibitemOpen
  \bibfield  {author} {\bibinfo {author} {\bibfnamefont {I.}~\bibnamefont
  {Kamaretsos}}, \bibinfo {author} {\bibfnamefont {M.}~\bibnamefont {Hannam}},\
  and\ \bibinfo {author} {\bibfnamefont {B.~S.}\ \bibnamefont
  {Sathyaprakash}},\ }\bibfield  {title} {\bibinfo {title} {Is black-hole
  ringdown a memory of its progenitor?},\ }\href
  {https://doi.org/10.1103/PhysRevLett.109.141102} {\bibfield  {journal}
  {\bibinfo  {journal} {Phys. Rev. Lett.}\ }\textbf {\bibinfo {volume} {109}},\
  \bibinfo {pages} {141102} (\bibinfo {year} {2012}{\natexlab{b}})}\BibitemShut
  {NoStop}%
\bibitem [{\citenamefont {Varma}\ \emph {et~al.}(2019)\citenamefont {Varma},
  \citenamefont {Field}, \citenamefont {Scheel}, \citenamefont {Blackman},
  \citenamefont {Kidder},\ and\ \citenamefont {Pfeiffer}}]{varma-etal-2019}%
  \BibitemOpen
  \bibfield  {author} {\bibinfo {author} {\bibfnamefont {V.}~\bibnamefont
  {Varma}}, \bibinfo {author} {\bibfnamefont {S.~E.}\ \bibnamefont {Field}},
  \bibinfo {author} {\bibfnamefont {M.~A.}\ \bibnamefont {Scheel}}, \bibinfo
  {author} {\bibfnamefont {J.}~\bibnamefont {Blackman}}, \bibinfo {author}
  {\bibfnamefont {L.~E.}\ \bibnamefont {Kidder}},\ and\ \bibinfo {author}
  {\bibfnamefont {H.~P.}\ \bibnamefont {Pfeiffer}},\ }\bibfield  {title}
  {\bibinfo {title} {Surrogate model of hybridized numerical relativity binary
  black hole waveforms},\ }\href {https://doi.org/10.1103/PhysRevD.99.064045}
  {\bibfield  {journal} {\bibinfo  {journal} {Phys. Rev. D}\ }\textbf {\bibinfo
  {volume} {99}},\ \bibinfo {pages} {064045} (\bibinfo {year}
  {2019})}\BibitemShut {NoStop}%
\bibitem [{\citenamefont {Baibhav}\ \emph {et~al.}(2018)\citenamefont
  {Baibhav}, \citenamefont {Berti}, \citenamefont {Cardoso},\ and\
  \citenamefont {Khanna}}]{baibhav-etal-2018}%
  \BibitemOpen
  \bibfield  {author} {\bibinfo {author} {\bibfnamefont {V.}~\bibnamefont
  {Baibhav}}, \bibinfo {author} {\bibfnamefont {E.}~\bibnamefont {Berti}},
  \bibinfo {author} {\bibfnamefont {V.}~\bibnamefont {Cardoso}},\ and\ \bibinfo
  {author} {\bibfnamefont {G.}~\bibnamefont {Khanna}},\ }\bibfield  {title}
  {\bibinfo {title} {Black hole spectroscopy: Systematic errors and ringdown
  energy estimates},\ }\href {https://doi.org/10.1103/PhysRevD.97.044048}
  {\bibfield  {journal} {\bibinfo  {journal} {Phys. Rev. D}\ }\textbf {\bibinfo
  {volume} {97}},\ \bibinfo {pages} {044048} (\bibinfo {year}
  {2018})}\BibitemShut {NoStop}%
\bibitem [{SXS()}]{SXS-waveforms}%
  \BibitemOpen
  \href {http://www.black-holes.org/waveforms} {\bibinfo {title} {See
  http://www.black-holes.org/waveforms}}\BibitemShut {NoStop}%
\bibitem [{\citenamefont {Mrou\'e}\ \emph {et~al.}(2013)\citenamefont
  {Mrou\'e}, \citenamefont {Scheel}, \citenamefont {Szil\'agyi}, \citenamefont
  {Pfeiffer}, \citenamefont {Boyle}, \citenamefont {Hemberger}, \citenamefont
  {Kidder}, \citenamefont {Lovelace}, \citenamefont {Ossokine}, \citenamefont
  {Taylor}, \citenamefont {Zengino\u{g}lu}, \citenamefont {Buchman},
  \citenamefont {Chu}, \citenamefont {Foley}, \citenamefont {Giesler},
  \citenamefont {Owen},\ and\ \citenamefont {Teukolsky}}]{SXS-catalog-2013}%
  \BibitemOpen
  \bibfield  {author} {\bibinfo {author} {\bibfnamefont {A.~H.}\ \bibnamefont
  {Mrou\'e}}, \bibinfo {author} {\bibfnamefont {M.~A.}\ \bibnamefont {Scheel}},
  \bibinfo {author} {\bibfnamefont {B.}~\bibnamefont {Szil\'agyi}}, \bibinfo
  {author} {\bibfnamefont {H.~P.}\ \bibnamefont {Pfeiffer}}, \bibinfo {author}
  {\bibfnamefont {M.}~\bibnamefont {Boyle}}, \bibinfo {author} {\bibfnamefont
  {D.~A.}\ \bibnamefont {Hemberger}}, \bibinfo {author} {\bibfnamefont {L.~E.}\
  \bibnamefont {Kidder}}, \bibinfo {author} {\bibfnamefont {G.}~\bibnamefont
  {Lovelace}}, \bibinfo {author} {\bibfnamefont {S.}~\bibnamefont {Ossokine}},
  \bibinfo {author} {\bibfnamefont {N.~W.}\ \bibnamefont {Taylor}}, \bibinfo
  {author} {\bibfnamefont {A.}~\bibnamefont {Zengino\u{g}lu}}, \bibinfo
  {author} {\bibfnamefont {L.~T.}\ \bibnamefont {Buchman}}, \bibinfo {author}
  {\bibfnamefont {T.}~\bibnamefont {Chu}}, \bibinfo {author} {\bibfnamefont
  {E.}~\bibnamefont {Foley}}, \bibinfo {author} {\bibfnamefont
  {M.}~\bibnamefont {Giesler}}, \bibinfo {author} {\bibfnamefont
  {R.}~\bibnamefont {Owen}},\ and\ \bibinfo {author} {\bibfnamefont {S.~A.}\
  \bibnamefont {Teukolsky}},\ }\bibfield  {title} {\bibinfo {title} {Catalog of
  174 binary black hole simulations for gravitational wave astronomy},\ }\href
  {https://doi.org/10.1103/PhysRevLett.111.241104} {\bibfield  {journal}
  {\bibinfo  {journal} {Phys. Rev. Lett.}\ }\textbf {\bibinfo {volume} {111}},\
  \bibinfo {pages} {241104} (\bibinfo {year} {2013})}\BibitemShut {NoStop}%
\bibitem [{\citenamefont {Boyle}\ \emph {et~al.}(2019)\citenamefont {Boyle},
  \citenamefont {Hemberger}, \citenamefont {Iozzo}, \citenamefont {Lovelace},
  \citenamefont {Ossokine}, \citenamefont {Pfeiffer}, \citenamefont {Scheel},
  \citenamefont {Stein}, \citenamefont {Woodford}, \citenamefont {Zimmerman},
  \citenamefont {Afshari}, \citenamefont {Barkett}, \citenamefont {Blackman},
  \citenamefont {Chatziioannou}, \citenamefont {Chu}, \citenamefont {Demos},
  \citenamefont {Deppe}, \citenamefont {Field}, \citenamefont {Fischer},
  \citenamefont {Foley}, \citenamefont {Fong}, \citenamefont {Garcia},
  \citenamefont {Giesler}, \citenamefont {Hebert}, \citenamefont {Hinder},
  \citenamefont {Katebi}, \citenamefont {Khan}, \citenamefont {Kidder},
  \citenamefont {Kumar}, \citenamefont {Kuper}, \citenamefont {Lim},
  \citenamefont {Okounkova}, \citenamefont {Ramirez}, \citenamefont
  {Rodriguez}, \citenamefont {R\"uter}, \citenamefont {Schmidt}, \citenamefont
  {Szilagyi}, \citenamefont {Teukolsky}, \citenamefont {Varma},\ and\
  \citenamefont {Walker}}]{SXS-catalog-2019}%
  \BibitemOpen
  \bibfield  {author} {\bibinfo {author} {\bibfnamefont {M.}~\bibnamefont
  {Boyle}}, \bibinfo {author} {\bibfnamefont {D.}~\bibnamefont {Hemberger}},
  \bibinfo {author} {\bibfnamefont {D.~A.~B.}\ \bibnamefont {Iozzo}}, \bibinfo
  {author} {\bibfnamefont {G.}~\bibnamefont {Lovelace}}, \bibinfo {author}
  {\bibfnamefont {S.}~\bibnamefont {Ossokine}}, \bibinfo {author}
  {\bibfnamefont {H.~P.}\ \bibnamefont {Pfeiffer}}, \bibinfo {author}
  {\bibfnamefont {M.~A.}\ \bibnamefont {Scheel}}, \bibinfo {author}
  {\bibfnamefont {L.~C.}\ \bibnamefont {Stein}}, \bibinfo {author}
  {\bibfnamefont {C.~J.}\ \bibnamefont {Woodford}}, \bibinfo {author}
  {\bibfnamefont {A.~B.}\ \bibnamefont {Zimmerman}}, \bibinfo {author}
  {\bibfnamefont {N.}~\bibnamefont {Afshari}}, \bibinfo {author} {\bibfnamefont
  {K.}~\bibnamefont {Barkett}}, \bibinfo {author} {\bibfnamefont
  {J.}~\bibnamefont {Blackman}}, \bibinfo {author} {\bibfnamefont
  {K.}~\bibnamefont {Chatziioannou}}, \bibinfo {author} {\bibfnamefont
  {T.}~\bibnamefont {Chu}}, \bibinfo {author} {\bibfnamefont {N.}~\bibnamefont
  {Demos}}, \bibinfo {author} {\bibfnamefont {N.}~\bibnamefont {Deppe}},
  \bibinfo {author} {\bibfnamefont {S.~E.}\ \bibnamefont {Field}}, \bibinfo
  {author} {\bibfnamefont {N.~L.}\ \bibnamefont {Fischer}}, \bibinfo {author}
  {\bibfnamefont {E.}~\bibnamefont {Foley}}, \bibinfo {author} {\bibfnamefont
  {H.}~\bibnamefont {Fong}}, \bibinfo {author} {\bibfnamefont {A.}~\bibnamefont
  {Garcia}}, \bibinfo {author} {\bibfnamefont {M.}~\bibnamefont {Giesler}},
  \bibinfo {author} {\bibfnamefont {F.}~\bibnamefont {Hebert}}, \bibinfo
  {author} {\bibfnamefont {I.}~\bibnamefont {Hinder}}, \bibinfo {author}
  {\bibfnamefont {R.}~\bibnamefont {Katebi}}, \bibinfo {author} {\bibfnamefont
  {H.}~\bibnamefont {Khan}}, \bibinfo {author} {\bibfnamefont {L.~E.}\
  \bibnamefont {Kidder}}, \bibinfo {author} {\bibfnamefont {P.}~\bibnamefont
  {Kumar}}, \bibinfo {author} {\bibfnamefont {K.}~\bibnamefont {Kuper}},
  \bibinfo {author} {\bibfnamefont {H.}~\bibnamefont {Lim}}, \bibinfo {author}
  {\bibfnamefont {M.}~\bibnamefont {Okounkova}}, \bibinfo {author}
  {\bibfnamefont {T.}~\bibnamefont {Ramirez}}, \bibinfo {author} {\bibfnamefont
  {S.}~\bibnamefont {Rodriguez}}, \bibinfo {author} {\bibfnamefont {H.~R.}\
  \bibnamefont {R\"uter}}, \bibinfo {author} {\bibfnamefont {P.}~\bibnamefont
  {Schmidt}}, \bibinfo {author} {\bibfnamefont {B.}~\bibnamefont {Szilagyi}},
  \bibinfo {author} {\bibfnamefont {S.~A.}\ \bibnamefont {Teukolsky}}, \bibinfo
  {author} {\bibfnamefont {V.}~\bibnamefont {Varma}},\ and\ \bibinfo {author}
  {\bibfnamefont {M.}~\bibnamefont {Walker}},\ }\bibfield  {title} {\bibinfo
  {title} {The {SXS} collaboration catalog of binary black hole simulations},\
  }\href {https://doi.org/10.1088/1361-6382/ab34e2} {\bibfield  {journal}
  {\bibinfo  {journal} {Classical Quantum Gravity}\ }\textbf {\bibinfo {volume}
  {36}},\ \bibinfo {pages} {195006} (\bibinfo {year} {2019})}\BibitemShut
  {NoStop}%
\bibitem [{\citenamefont {Bondi}\ \emph {et~al.}(1962)\citenamefont {Bondi},
  \citenamefont {Van~der Burg},\ and\ \citenamefont {Metzner}}]{bondi-1962}%
  \BibitemOpen
  \bibfield  {author} {\bibinfo {author} {\bibfnamefont {H.}~\bibnamefont
  {Bondi}}, \bibinfo {author} {\bibfnamefont {M.~G.~J.}\ \bibnamefont {Van~der
  Burg}},\ and\ \bibinfo {author} {\bibfnamefont {A.}~\bibnamefont {Metzner}},\
  }\bibfield  {title} {\bibinfo {title} {Gravitational waves in general
  relativity, {VII}. waves from axi-symmetric isolated system},\ }\href
  {https://doi.org/10.1098/rspa.1962.0161} {\bibfield  {journal} {\bibinfo
  {journal} {Proceedings of the Royal Society of London. Series A. Mathematical
  and Physical Sciences}\ }\textbf {\bibinfo {volume} {269}},\ \bibinfo {pages}
  {21} (\bibinfo {year} {1962})}\BibitemShut {NoStop}%
\bibitem [{\citenamefont {Bishop}\ \emph {et~al.}(1996)\citenamefont {Bishop},
  \citenamefont {G\'omez}, \citenamefont {Lehner},\ and\ \citenamefont
  {Winicour}}]{bishop_etal96b}%
  \BibitemOpen
  \bibfield  {author} {\bibinfo {author} {\bibfnamefont {N.~T.}\ \bibnamefont
  {Bishop}}, \bibinfo {author} {\bibfnamefont {R.}~\bibnamefont {G\'omez}},
  \bibinfo {author} {\bibfnamefont {L.}~\bibnamefont {Lehner}},\ and\ \bibinfo
  {author} {\bibfnamefont {J.}~\bibnamefont {Winicour}},\ }\bibfield  {title}
  {\bibinfo {title} {Cauchy-characteristic extraction in numerical
  relativity},\ }\href {https://doi.org/10.1103/PhysRevD.54.6153} {\bibfield
  {journal} {\bibinfo  {journal} {Phys. Rev. D}\ }\textbf {\bibinfo {volume}
  {54}},\ \bibinfo {pages} {6153} (\bibinfo {year} {1996})}\BibitemShut
  {NoStop}%
\bibitem [{\citenamefont {Handmer}\ \emph {et~al.}(2016)\citenamefont
  {Handmer}, \citenamefont {Szil{\'{a}}gyi},\ and\ \citenamefont
  {Winicour}}]{Handmer-etal-2016}%
  \BibitemOpen
  \bibfield  {author} {\bibinfo {author} {\bibfnamefont {C.~J.}\ \bibnamefont
  {Handmer}}, \bibinfo {author} {\bibfnamefont {B.}~\bibnamefont
  {Szil{\'{a}}gyi}},\ and\ \bibinfo {author} {\bibfnamefont {J.}~\bibnamefont
  {Winicour}},\ }\bibfield  {title} {\bibinfo {title} {Spectral {C}auchy
  characteristic extraction of strain, news and gravitational radiation flux},\
  }\href {https://doi.org/10.1088/0264-9381/33/22/225007} {\bibfield  {journal}
  {\bibinfo  {journal} {Classical Quantum Gravity}\ }\textbf {\bibinfo {volume}
  {33}},\ \bibinfo {pages} {225007} (\bibinfo {year} {2016})}\BibitemShut
  {NoStop}%
\bibitem [{\citenamefont {Berti}\ \emph {et~al.}(2006)\citenamefont {Berti},
  \citenamefont {Cardoso},\ and\ \citenamefont {Will}}]{berticardosowill-2006}%
  \BibitemOpen
  \bibfield  {author} {\bibinfo {author} {\bibfnamefont {E.}~\bibnamefont
  {Berti}}, \bibinfo {author} {\bibfnamefont {V.}~\bibnamefont {Cardoso}},\
  and\ \bibinfo {author} {\bibfnamefont {C.~M.}\ \bibnamefont {Will}},\
  }\bibfield  {title} {\bibinfo {title} {Gravitational-wave spectroscopy of
  massive black holes with the space interferometer {LISA}},\ }\href
  {https://doi.org/10.1103/PhysRevD.73.064030} {\bibfield  {journal} {\bibinfo
  {journal} {Phys. Rev. D}\ }\textbf {\bibinfo {volume} {73}},\ \bibinfo
  {pages} {064030} (\bibinfo {year} {2006})}\BibitemShut {NoStop}%
\bibitem [{\citenamefont {Teukolsky}(1973)}]{teukolsky-1973}%
  \BibitemOpen
  \bibfield  {author} {\bibinfo {author} {\bibfnamefont {S.~A.}\ \bibnamefont
  {Teukolsky}},\ }\bibfield  {title} {\bibinfo {title} {Perturbations of a
  rotating black hole.\ {I}.\ {F}undamental equations for gravitational,
  electromagnetic, and neutrino-field perturbations},\ }\href
  {https://doi.org/10.1086/152444} {\bibfield  {journal} {\bibinfo  {journal}
  {Astrophys. J.}\ }\textbf {\bibinfo {volume} {185}},\ \bibinfo {pages} {635}
  (\bibinfo {year} {1973})}\BibitemShut {NoStop}%
\bibitem [{\citenamefont {Cook}\ and\ \citenamefont
  {Zalutskiy}(2014)}]{cook-zalutskiy-2014}%
  \BibitemOpen
  \bibfield  {author} {\bibinfo {author} {\bibfnamefont {G.~B.}\ \bibnamefont
  {Cook}}\ and\ \bibinfo {author} {\bibfnamefont {M.}~\bibnamefont
  {Zalutskiy}},\ }\bibfield  {title} {\bibinfo {title} {Gravitational
  perturnbations of the {K}err geometry: {H}igh-accuracy study},\ }\href
  {https://doi.org/10.1103/PhysRevD.90.124021} {\bibfield  {journal} {\bibinfo
  {journal} {Phys. Rev. D}\ }\textbf {\bibinfo {volume} {90}},\ \bibinfo
  {pages} {124021} (\bibinfo {year} {2014})}\BibitemShut {NoStop}%
\bibitem [{\citenamefont {Boyle}(2016)}]{boyle-2016}%
  \BibitemOpen
  \bibfield  {author} {\bibinfo {author} {\bibfnamefont {M.}~\bibnamefont
  {Boyle}},\ }\bibfield  {title} {\bibinfo {title} {Transformations of
  asymptotic gravitational-wave data},\ }\href
  {https://doi.org/10.1103/PhysRevD.93.084031} {\bibfield  {journal} {\bibinfo
  {journal} {Phys. Rev. D}\ }\textbf {\bibinfo {volume} {93}},\ \bibinfo
  {pages} {084031} (\bibinfo {year} {2016})}\BibitemShut {NoStop}%
\bibitem [{\citenamefont {Zalutskiy}(2014)}]{MSThesis:Zalutskiy}%
  \BibitemOpen
  \bibfield  {author} {\bibinfo {author} {\bibfnamefont {M.~P.}\ \bibnamefont
  {Zalutskiy}},\ }\emph {\bibinfo {title} {Efficient Iterative algorithm for
  computing quasinormal modes of black holes and information extraction from
  black hole ringdown signal}},\ \href@noop {} {Master's thesis},\ \bibinfo
  {school} {Wake Forest University} (\bibinfo {year} {2014})\BibitemShut
  {NoStop}%
\bibitem [{\citenamefont {Press}\ \emph {et~al.}(2007)\citenamefont {Press},
  \citenamefont {Teukolsky}, \citenamefont {Wetterling},\ and\ \citenamefont
  {Flannery}}]{numrec_c++}%
  \BibitemOpen
  \bibfield  {author} {\bibinfo {author} {\bibfnamefont {W.~H.}\ \bibnamefont
  {Press}}, \bibinfo {author} {\bibfnamefont {S.~A.}\ \bibnamefont
  {Teukolsky}}, \bibinfo {author} {\bibfnamefont {W.~T.}\ \bibnamefont
  {Wetterling}},\ and\ \bibinfo {author} {\bibfnamefont {B.~P.}\ \bibnamefont
  {Flannery}},\ }\href@noop {} {\emph {\bibinfo {title} {Numerical Recipes}}},\
  \bibinfo {edition} {3rd}\ ed.\ (\bibinfo  {publisher} {Cambridge University
  Press},\ \bibinfo {address} {Cambridge, England},\ \bibinfo {year}
  {2007})\BibitemShut {NoStop}%
\bibitem [{\citenamefont {Morrison}\ and\ \citenamefont
  {Parker}(1987)}]{Rotations-in-QM-1987}%
  \BibitemOpen
  \bibfield  {author} {\bibinfo {author} {\bibfnamefont {M.~A.}\ \bibnamefont
  {Morrison}}\ and\ \bibinfo {author} {\bibfnamefont {G.~A.}\ \bibnamefont
  {Parker}},\ }\bibfield  {title} {\bibinfo {title} {A guide to rotations in
  quantum mechanics},\ }\href {https://doi.org/10.1071/PH870465} {\bibfield
  {journal} {\bibinfo  {journal} {Aust. J. Phys.}\ }\textbf {\bibinfo {volume}
  {40}},\ \bibinfo {pages} {465} (\bibinfo {year} {1987})}\BibitemShut
  {NoStop}%
\bibitem [{\citenamefont {Cook}(2019)}]{QNMdata-cook-2019}%
  \BibitemOpen
  \bibfield  {author} {\bibinfo {author} {\bibfnamefont {G.~B.}\ \bibnamefont
  {Cook}},\ }\bibfield  {title} {\bibinfo {title} {Kerr quasinormal modes:
  s=-2, n=0--7},\ }\href {https://doi.org/10.5281/zenodo.2650358}
  {10.5281/zenodo.2650358} (\bibinfo {year} {2019}),\ \bibinfo {note}
  {{Z}enodo}\BibitemShut {NoStop}%
\bibitem [{\citenamefont {Abbott}\ \emph
  {et~al.}(2016{\natexlab{b}})\citenamefont {Abbott} \emph
  {et~al.}}]{GW150914-2016}%
  \BibitemOpen
  \bibfield  {author} {\bibinfo {author} {\bibfnamefont {B.~P.}\ \bibnamefont
  {Abbott}} \emph {et~al.} (\bibinfo {collaboration} {LIGO Scientific
  Collaboration and Virgo Collaboration}),\ }\bibfield  {title} {\bibinfo
  {title} {Observation of gravitational waves from a binary black hole
  merger},\ }\href {https://doi.org/10.1103/PhysRevLett.116.061102} {\bibfield
  {journal} {\bibinfo  {journal} {Phys. Rev. Lett.}\ }\textbf {\bibinfo
  {volume} {116}},\ \bibinfo {pages} {061102} (\bibinfo {year}
  {2016}{\natexlab{b}})}\BibitemShut {NoStop}%
\bibitem [{\citenamefont {Thrane}\ \emph {et~al.}(2017)\citenamefont {Thrane},
  \citenamefont {Lasky},\ and\ \citenamefont {Levin}}]{thranelaskyleven-2017}%
  \BibitemOpen
  \bibfield  {author} {\bibinfo {author} {\bibfnamefont {E.}~\bibnamefont
  {Thrane}}, \bibinfo {author} {\bibfnamefont {P.~D.}\ \bibnamefont {Lasky}},\
  and\ \bibinfo {author} {\bibfnamefont {Y.}~\bibnamefont {Levin}},\ }\bibfield
   {title} {\bibinfo {title} {Challenges for testing the no-hair theorem with
  current and planned gravitational-wave detectors},\ }\href
  {https://doi.org/10.1103/PhysRevD.96.102004} {\bibfield  {journal} {\bibinfo
  {journal} {Phys. Rev. D}\ }\textbf {\bibinfo {volume} {96}},\ \bibinfo
  {pages} {102004} (\bibinfo {year} {2017})}\BibitemShut {NoStop}%
\bibitem [{\citenamefont {Carullo}\ \emph {et~al.}(2018)\citenamefont
  {Carullo}, \citenamefont {van~der Schaaf}, \citenamefont {London},
  \citenamefont {Pang}, \citenamefont {Tsang}, \citenamefont {Hannuksela},
  \citenamefont {Meidam}, \citenamefont {Agathos}, \citenamefont {Samajdar},
  \citenamefont {Ghosh}, \citenamefont {Li}, \citenamefont {Del~Pozzo},\ and\
  \citenamefont {Van Den~Broeck}}]{carullo-etal-2018}%
  \BibitemOpen
  \bibfield  {author} {\bibinfo {author} {\bibfnamefont {G.}~\bibnamefont
  {Carullo}}, \bibinfo {author} {\bibfnamefont {L.}~\bibnamefont {van~der
  Schaaf}}, \bibinfo {author} {\bibfnamefont {L.}~\bibnamefont {London}},
  \bibinfo {author} {\bibfnamefont {P.~T.~H.}\ \bibnamefont {Pang}}, \bibinfo
  {author} {\bibfnamefont {K.~W.}\ \bibnamefont {Tsang}}, \bibinfo {author}
  {\bibfnamefont {O.~A.}\ \bibnamefont {Hannuksela}}, \bibinfo {author}
  {\bibfnamefont {J.}~\bibnamefont {Meidam}}, \bibinfo {author} {\bibfnamefont
  {M.}~\bibnamefont {Agathos}}, \bibinfo {author} {\bibfnamefont
  {A.}~\bibnamefont {Samajdar}}, \bibinfo {author} {\bibfnamefont
  {A.}~\bibnamefont {Ghosh}}, \bibinfo {author} {\bibfnamefont {T.~G.~F.}\
  \bibnamefont {Li}}, \bibinfo {author} {\bibfnamefont {W.}~\bibnamefont
  {Del~Pozzo}},\ and\ \bibinfo {author} {\bibfnamefont {C.}~\bibnamefont {Van
  Den~Broeck}},\ }\bibfield  {title} {\bibinfo {title} {Empirical tests of the
  black hole no-hair conjecture using gravitational-wave observations},\ }\href
  {https://doi.org/10.1103/PhysRevD.98.104020} {\bibfield  {journal} {\bibinfo
  {journal} {Phys. Rev. D}\ }\textbf {\bibinfo {volume} {98}},\ \bibinfo
  {pages} {104020} (\bibinfo {year} {2018})}\BibitemShut {NoStop}%
\bibitem [{\citenamefont {Baibhav}\ and\ \citenamefont
  {Berti}(2019)}]{baibhavberti-2019}%
  \BibitemOpen
  \bibfield  {author} {\bibinfo {author} {\bibfnamefont {V.}~\bibnamefont
  {Baibhav}}\ and\ \bibinfo {author} {\bibfnamefont {E.}~\bibnamefont
  {Berti}},\ }\bibfield  {title} {\bibinfo {title} {Multimode black hole
  spectroscopy},\ }\href {https://doi.org/10.1103/PhysRevD.99.024005}
  {\bibfield  {journal} {\bibinfo  {journal} {Phys. Rev. D}\ }\textbf {\bibinfo
  {volume} {99}},\ \bibinfo {pages} {024005} (\bibinfo {year}
  {2019})}\BibitemShut {NoStop}%
\end{thebibliography}
\end{document}